\documentclass[11pt]{article}
\usepackage{amsmath}
\usepackage{graphicx}
\usepackage{natbib}
\usepackage{url} 

\usepackage[ruled,vlined,noend]{algorithm2e}
\SetAlFnt{\small}
\usepackage{amsfonts}
\usepackage{amsmath}
\usepackage{bbm}
\usepackage{floatrow}
\usepackage{subcaption}
\usepackage{hyperref}
\usepackage{multibib}
\newcites{sup}{Supplementary Material References}

\usepackage{color}

\newcommand{\PP}{\mathrm{P}}
\newcommand{\EE}{\mathrm{E}}
\newcommand{\Cov}{\mathrm{Cov}}

\newcommand{\Var}{\mathrm{Var}}

\newcommand{\mO}{\mathcal{O}}

\newcommand{\bo}{\mathbf{0}}

\newcommand{\bx}{\mathbf{x}}
\newcommand{\by}{\mathbf{y}}
\newcommand{\bz}{\mathbf{z}}

\newtheorem{prop}{Proposition}
\newtheorem{lem}{Lemma}

\newtheorem{cor}{Corollary}

\def\T{ {\mathrm{\scriptscriptstyle T}} }
\def\dif{\mathrm d}
\def\me{\mathrm e}
\def\diag{\mathrm{diag}}

\begin{document}

\def\spacingset#1{\renewcommand{\baselinestretch}%
{#1}\small\normalsize} \spacingset{1.5}


\begin{titlepage}

\begin{center}
{\Large On Irreversible Metropolis Sampling Related to Langevin Dynamics}

\vspace{.1in} Zexi Song\footnotemark[1] \& Zhiqiang Tan\footnotemark[1]

\vspace{.1in}
\today
\end{center}

\footnotetext[1]{Department of Statistics, Rutgers University. Address: 110 Frelinghuysen Road,
Piscataway, NJ 08854. E-mails: zexisong@stat.rutgers.edu, ztan@stat.rutgers.edu.
The authors thank Kshitij Khare, Ron Levy, and Bin Zhang for helpful references.
}

\paragraph{Abstract.}
There has been considerable interest in designing Markov chain Monte Carlo algorithms by
exploiting numerical methods for Langevin dynamics, which includes Hamiltonian dynamics as a deterministic case.
A prominent approach is Hamiltonian Monte Carlo (HMC), where a leapfrog discretization of Hamiltonian dynamics is employed.
We investigate a recently proposed class of irreversible sampling algorithms, called Hamiltonian assisted Metropolis sampling (HAMS),
which uses an augmented target density similarly as in HMC, but involves a flexible proposal scheme and a carefully formulated acceptance-rejection scheme to
achieve generalized reversibility.
We show that as the step size tends to 0, the HAMS proposal satisfies a class of stochastic differential equations including Langevin dynamics as a special case.
We provide theoretical results for HAMS under the univariate Gaussian setting, including the stationary variance,
the expected acceptance rate, and the spectral radius. From these results, we derive default choices of tuning parameters for HAMS, such that only the step size needs to be tuned in applications.
Various relatively recent algorithms for Langevin dynamics
are also shown to fall in the class of HAMS proposals up to negligible differences.
Our numerical experiments on sampling high-dimensional latent variables
confirm that the HAMS algorithms consistently achieve superior performance, compared with several Metropolis-adjusted algorithms
based on popular integrators of Langevin dynamics.

\paragraph{Key words and phrases.} Generalized reversibility; Hamiltonian Monte Carlo;  Langevin dynamics; Markov chain Monte Carlo; Metropolis-Hastings sampling;

\end{titlepage}

\newpage
\spacingset{1.5} 


\section{Introduction}
\label{sec:intro}

Stochastic simulations are widely used in scientific computing across various fields \citep[e.g.,][]{Brooks2011}.
Examples include molecular dynamics (MD) simulations in physics and chemistry,
and posterior simulations in Bayesian statistical analysis.
As indicated by these applications, there are at least two distinct modes of stochastic simulations.


One mode of simulations involves generating Markov chains as numerical discretizations of continuous-time processes defined by stochastic differential equations (SDEs).
In particular, consider underdamped Langevin dynamics defined by the SDE:
\begin{align}
\dif x_t = u_t \,\dif t, \quad \dif u_t = - \eta\, u_t \,\dif t - \nabla U(x_t) \,\dif t + \sqrt{2\eta} \,\dif W_t, \label{eq:langevin-SDE}
\end{align}
where $x_t$ is a position variable, $u_t$ is a momentum variable, $U(x)$ is a potential function, $\eta\geq 0$ is a friction coefficient, and $W_t$ is the standard Brownian motion.
See the end of this section for our notation.
The stationary distribution of (\ref{eq:langevin-SDE}) is determined by the augmented density
\begin{align}
\pi(x,u) \propto \exp \{ - H(x,u) \} = \exp\left\{ - U(x) - u^\T u/2 \right \},  \label{eq:aug-density}
\end{align}
where $H(x,u) = U(x) + u^\T u / 2$ is called the Hamiltonian.
The marginal distribution of $x$ is called the Boltzmann distribution, with density  $\pi (x) \propto \exp \{ - U(x) \}$.
The momentum $u$ can be considered an auxiliary variable, with a standard Gaussian distribution.
For simplicity, unit mass and temperature are used, and the Boltzmann constant is set to 1.
Various algorithms have been proposed for Langevin dynamics in computational physics and related fields,
including early development \citep{vanGunsteren1982,Brunger1984};
and more recent contributions \citep{Mannella2004, Bussi2007, Melchionna2007, Goga2012, Leimkuhler2012, Farago2013}.
Typically, it is desired that the stationary distribution of $x$ obtained is close to
the Boltzmann distribution $\pi(x)$, while allowing large step sizes.
Then trajectory averages of functions of $x$ can be used to estimate the expectations under $\pi(x)$.

Another mode of simulations is Monte Carlo simulations, or known as Markov chain Monte Carlo (MCMC),
where a target distribution is defined with density $\pi(x) \propto \exp\{ -U(x) \}$
for some analytically tractable function $U(x)$, which can be interpreted as a potential function.
For MCMC, Markov chains are directly constructed in discrete time such that the associated stationary distribution
gives exactly the target distribution $\pi(x)$. One of the main workhorses in MCMC is Metropolis--Hastings sampling \citep{Metropolis1953,Hastings1970}.
Given the current variable $x_0$, the Metropolis--Hastings  algorithm generates $x^*$ from a proposal density
$ Q(x^*| x_0)$, and then accepts $x_1 = x^*$ as the next variable with probability
\begin{equation}
 \min\left\{1, \frac{\pi(x^*) Q(x_0|x^*) }{\pi(x_0) Q(x^*|x_0)} \right\}, \label{eq:MH-prob}
\end{equation}
or rejects $x^*$ and set $x_1=x_0$, where $\pi(x^*)/\pi(x_0)$ can be evaluated as $\exp\{- U(x^*) + U(x_0)\}$
without requiring the normalizing constant.
The update from $x_0$ to $x_1$ defines a Markov transition $K(x_1 | x_0)$, depending on both the proposal density and the acceptance-rejection step,
such that reversibility is satisfied: $\pi(x_0) K(x_1 | x_0) = \pi(x_1) K(x_0 | x_1)$. This condition is also
called detailed balance in physics.
As a result, the Markov chain generated by Metropolis--Hastings sampling is reversible and admits $\pi(x)$ as a stationary distribution.

Metropolis--Hastings sampling provides a versatile framework, where different choices of the proposal density $Q$ lead to different methods.
For example, random walk Metropolis (RWM) is obtained when the proposal $x^*$ is generated by adding a Gaussian noise to $x_0$.
For the Metropolis-adjusted Langevin algorithm (MALA) \citep{Besag1994, Roberts1996},
the proposal $x^*$ is defined as Euler's discretization of the overdamped Langevin process, which can be deduced from (\ref{eq:langevin-SDE}) in the high-friction limit
with the momentum dropped out. For these methods,
the proposal is accepted or rejected using the Metropolis--Hastings probability (\ref{eq:MH-prob}).
The Markov chain obtained is  reversible and may suffer slow mixing.

To induce irreversible sampling, the two modes of simulations are combined in Hamiltonian Monte Carlo (HMC),
which uses the leapfrog discretization of Hamiltonian dynamics corresponding to (\ref{eq:langevin-SDE}) with $\eta=0$,
in conjunction with momentum resampling \citep{Duane1987, Neal2011}.
The HMC algorithm can be described as follows, given the current $(x_0,u_0)$.
\vspace{-.05in}
\begin{itemize}\addtolength{\itemsep}{-.1in}
  \item Sample $u^+ \sim \mathcal N(\bo,I)$. 
  \item Compute  $(x^*, u^-)$  from $(x_0, u^+)$  by one or multiple leapfrog updates with a step size $\epsilon$.
  Set $(x_1,u_1) = (x^*, u^-)$ with probability $\alpha= \min\{1, \pi(x^*,u^-)/\pi(x_0,u^+)\}$
   or set $(x_1,u_1) = (x_0, -u^+)$ with the remaining probability.
\end{itemize} \vspace{-.05in}
To facilitate our subsequent discussion, the above description differs slightly from \cite{Neal2011}, Section 5.3.2.1.
For HMC, each iteration can be justified as a composition of two steps,
first $(x_0, u_0) \to (x_0, u^+)$
and then $(x_0, u^+) \to (x_1,u_1)$, where each step leaves the target $\pi(x,u)$ invariant.
For the second step, a subtle point is that the next momentum $u_1$ is defined as $u^-$ upon acceptance, but $- u^+$ in the case of rejection,
even though the acceptance probability $\alpha$ appears the same as the usual Metropolis probability for a symmetric proposal.

Recently, \cite{Song2020} proposed a new class of irreversible sampling algorithms, called Hamiltonian assisted Metropolis sampling (HAMS), using the augmented
target density (\ref{eq:aug-density}) similarly as in HMC.\ 
There are two main ingredients in HAMS, similar to those in Metropolis--Hastings sampling.\ Given the current variables $(x_0, u_0)$, HAMS
first generates $(x^*, u^*)$ from a proposal density $Q(x^*, u^* | x_0, u_0)$, as defined by (\ref{eq:proposal1-mat})--(\ref{eq:proposal3-mat}) later.
Then HAMS performs acceptance-rejection: set $(x_1,u_1)= (x^*,u^*)$ with acceptance probability $\alpha$ and
set $(x_1, u_1) = (x_0, -u_0)$ with the remaining probability, where
\begin{align}
 \alpha =\min\left\{ 1, \frac{\pi(x^*, u^*) Q( x_0, -u_0 | x^*, -u^*) }
  {\pi(x_0,u_0) Q ( x^*, u^* | x_0, u_0) }\right\} . \label{eq:gMH-prob}
\end{align}
In contrast with HMC, randomness can be directly incorporated into an irreversible proposal.
The Markov chain defined by the HAMS update is irreversible and satisfies the following generalized reversibility
(or generalized detailed balance): \vspace{-.08in}
\begin{align}
\pi(x_0, u_0) K(x_1,u_1 | x_0, u_0) = \pi(x_1, u_1) K(x_0, -u_0 | x_1, -u_1) , \label{eq:gDB}
\end{align}
where $K(x_1, u_1 | x_0, u_0)$ denotes the transition kernel, depending on both the proposal and acceptance-rejection schemes.
Compared with the usual formula (\ref{eq:MH-prob}), the momentum is negated in defining the backward proposal in acceptance probability (\ref{eq:gMH-prob}).
Such a generalized detailed balance is exploited in various algorithms related to Langevin dynamics \citep{Scemama2006, Bussi2007, fang-etal-2014}.

More broadly, a framework of generalized Metropolis--Hastings sampling is formulated in \cite{Song2020} to achieve generalized reversibility, where
the acceptance probability (\ref{eq:gMH-prob}) is used in conjunction with a general proposal density $Q$, including but not restricted to the HAMS proposal.
The second step in HMC can be obtained as a special case, where the proposal $Q$ is defined by the (deterministic) leapfrog discreteization of Hamiltonian dynamics
and the ratio in $Q$ in (\ref{eq:gMH-prob}) reduces to 1 due to time reversibility of leapfrog.
The proposal $Q$ can also be defined by a dicretization of Langevin dynamics (\ref{eq:langevin-SDE}),
which together with (\ref{eq:gMH-prob}) leads to various Metropolis-adjusted algorithms. See Supplement Section \ref{sec:MA-algrithms} for examples.

In this article, we further investigate HAMS in several directions, together with existing algorithms related to Langevin dynamics.
Our main findings can be summarized as follows.
First, we show that under an appropriate parametrization with a step size tending to 0,
the HAMS proposal satisfies a class of SDEs which include Langevin dynamics as a special case (Section~\ref{sec:sde}).
In fact, two specific versions of HAMS using one noise per iteration, called HAMS-A and HAMS-B in \cite{Song2020},
are associated with two extreme cases respectively, the underdamped Langevin SDE (\ref{eq:langevin-SDE})
and a different SDE in which the position $x$ appears to be dampened instead of the momentum $u$ as in (\ref{eq:langevin-SDE}).

Second, we study HAMS in general with two noise vectors used per iteration from various perspectives,
including algebraic properties of the acceptance probability (Section~\ref{sec:phi}),
the stationary variance implied by the HAMS proposal without acceptance-rejection
and the expected acceptance rate with acceptance-rejection under a univariate Gaussian target or harmonic oscillator (Sections~\ref{sec:variance}--\ref{sec:accept}),
and the convergence rate quantified by the spectral radius under a standard Gaussian target (Section~\ref{sec:eigen}).
From these studies, we derive default choices of tuning parameters for HAMS, such that only the step size needs to be tuned in applications.
Moreover, we identify a particular version of HAMS which exhibits a favorable tradeoff between the expected acceptance rate and the convergence rate.
This algorithm, called HAMS-$k$, is close to HAMS-A, but uses two noise vectors per iteration to incorporate a small amount of friction on the position as in HAMS-B in addition to dampening the momentum.

Third, we demonstrate that various relatively recent algorithms for Langevin dynamics
can all be put into the class of HAMS proposals, up to negligible differences compared with the associated leading terms of the step size (Section~\ref{sec:matching}).
Examples include Mannella's leapfrog \citep{Mannella2004,Burrage2007}, the BP algorithm \citep{Bussi2007},  stochastic position Verlet \citep{Melchionna2007},
impulsive Langevin leapfrog \citep{Goga2012}, BAOAB and ABOBA \citep{Leimkuhler2012}, and the GJF algorithm \citep{Farago2013}.
These results not only shed new light on the relationship between the existing algorithms,
but also attest to the broad scope of HAMS.

Finally, we conduct several numerical experiments (Section~\ref{sec:sim}), to compare different versions of HAMS and Metropolis-adjusted BP, BAOAB, and ABOBA algorithms,
which are derived by incorporating acceptance-rejection in the framework of generalized Metropolis--Hastings sampling mentioned earlier.
The newly identified HAMS-$k$ algorithms lead to the smallest errors in configurational sampling in our double well experiment.
Moreover, the HAMS algorithms consistently achieve superior performance in terms of two measures of effective sample sizes in  two experiments on sampling
high-dimensional latent variables.

\vspace{.1in}
\textbf{Notation.}
Assume that a target density $\pi(x)$ is defined on $\mathbb R^k$. The potential energy function $U(x)$ is defined
such that $\pi(x) \propto \exp\{-U(x)\}$. Denote the gradient of $U(x)$ as $\nabla U(x)$ and hessian $\nabla^2 U(x)$.
The (multivariate) normal distribution with mean $\mu$ and variance $\Sigma$ is denoted as $\mathcal N(\mu,\Sigma)$, and the density
function as $\mathcal N(\cdot |\mu,\Sigma)$.
Write $\bo$ for a vector or matrix with all $0$ entries, and $I$ for an identity matrix of appropriate dimensions.

\section{Review of HAMS}
\label{sec:review}
We give a description of HAMS, a class of irreversible sampling algorithms in \cite{Song2020}.
Throughout, we write the current variables as $(x_0,u_0)$, a proposal as $(x^*,u^*)$, and the next variables as $(x_1,u_1)$ after the acceptance-rejection step.

Given the current variables $(x_0,u_0)$, HAMS generates a proposal $(x^*,u^*)$ as follows.\vspace{-.1in}
\begin{itemize}\addtolength{\itemsep}{-.1in}
\item Sample \vspace{-.1in}
\begin{align}
&  Z_0 = \begin{pmatrix}
    Z_{0}^{(1)}\\
    Z_{0}^{(2)}
  \end{pmatrix}
    \sim \mathcal N (\bo,2A - A^2),
    \quad \text{with } A =
    \begin{pmatrix}
    a_1 I & a_2 I\\
    a_2 I & a_3 I
    \end{pmatrix} , \label{eq:proposal1-mat}
\end{align}
where each $I$ is a $k\times k$ identity matrix, with $k$ the dimension of $x$, and $a_1,a_2,a_3$ are scalar coefficients such
that  $\bo \leq A \leq 2 I$ (in the positive semi-definite sense),
hence ensuring that $2A -A^2$ is a variance matrix.
\item Compute \vspace{-.1in}
\begin{align}
  \begin{pmatrix}
    x^*\\
    u^\dag
  \end{pmatrix}
  & =
  \begin{pmatrix}
    x_0\\
    -u_0
  \end{pmatrix}
  -A
  \begin{pmatrix}
    \nabla U(x_0)\\
    -u_0
  \end{pmatrix}
  +
  \begin{pmatrix}
    Z_{0}^{(1)}\\
    Z_{0}^{(2)}
  \end{pmatrix}, \label{eq:proposal2-mat} \\
 u^* & = u^\dag + \phi( x^* -x_0 - \nabla U(x^*) + \nabla U(x_0) ) , \label{eq:proposal3-mat}
\end{align}
where $\phi$ is a scalar coefficient.
\end{itemize} \vspace{-.1in}
There are four tuning parameters, $a_1,a_2,a_3$ and $\phi$.
The proposal scheme (\ref{eq:proposal1-mat})--(\ref{eq:proposal3-mat}) is derived in several steps \citep{Song2020}: first applying
an auxiliary variable argument and an over-relaxation technique to obtain a reversible proposal,
introducing negation of the momentum to obtain (\ref{eq:proposal2-mat}), and incorporating the new gradient $\nabla U(x^*)$ in the momentum update (\ref{eq:proposal3-mat}).

To describe the acceptance-rejection scheme, it is helpful to rewrite the update formulas (\ref{eq:proposal2-mat}) and (\ref{eq:proposal3-mat}) as follows:
\begin{align}
& \tilde{Z}^{(1)}  = Z_{0}^{(1)}  - a_1\nabla U(x_0) + a_2 u_0,\qquad \tilde{Z}^{(2)}  = Z_{0}^{(2)}  - a_2\nabla U(x_0) + a_3 u_0, \label{eq:proposal1} \\
    & x^* = x_0 +  \tilde{Z}^{(1)} , \label{eq:proposal2-x} \\
    & u^* = -u_0 + \tilde{Z}^{(2)}  + \phi(\tilde{Z}^{(1)} + \nabla U(x_0) - \nabla U(x^*)). \label{eq:proposal2-u}
  \end{align}
Equations (\ref{eq:proposal2-x})--(\ref{eq:proposal2-u}) determine a forward transition from $(x_0,u_0)$ to $(x^*, u^*)$, depending on
the noise vector $Z_0$. For a backward transition, we compute the new noise vector
$Z^* = (\, (Z^{(1)*})^\T, (Z^{(2)*})^\T \, )^\T$, with
\begin{align}
 Z^{(1)*}  = \tilde{Z}^{(1)}  - a_1\nabla U(x^*) - a_2 u^*,\qquad  Z^{(2)*} = \tilde{Z}^{(2)} - a_2\nabla U(x^*) - a_3 u^*. \label{eq:proposal3}
\end{align}
Then (\ref{eq:proposal2-x}), (\ref{eq:proposal2-u}), and (\ref{eq:proposal3}) can be equivalently rearranged to
\begin{align}
& -\tilde Z^{(1)} = -Z^{(1)*} - a_1\nabla U(x^*)  + a_2 (-u^*),\quad -\tilde Z^{(2)} = -Z^{(2)*} -a_2 \nabla U(x^*) + a_3 (-u^*), \label{eq:backward-Z}\\
& x_0 = x^* + (-\tilde Z^{(1)}),   \label{eq:backward-x} \\
& -u_0 = u^* +(-\tilde Z^{(2)})  + \phi(-\tilde{Z}^{(1)} + \nabla U(x^*) - \nabla U(x_0) ) .   \label{eq:backward-u}
\end{align}
Importantly, equations (\ref{eq:backward-Z})--(\ref{eq:backward-u}) corresponds to the {\it same} mapping as (\ref{eq:proposal1})--(\ref{eq:proposal2-u}), but
applied from $(x^*, -u^*)$ to $(x_0, -u_0)$ using the new noise vector $-Z^*$.
Hence the forward and backward transitions in (\ref{eq:proposal1})--(\ref{eq:proposal2-u}) and (\ref{eq:backward-Z})--(\ref{eq:backward-u}) can be illustrated as
\begin{align}
  \begin{pmatrix}
    x_0 \\
    u_0
  \end{pmatrix}
  \stackrel{Z_0}{\longrightarrow}
  \begin{pmatrix}
    x^* \\
    u^*
  \end{pmatrix},
  \qquad
  \begin{pmatrix}
    x^* \\
    -u^*
  \end{pmatrix}
  \stackrel{-Z^*}{\longrightarrow}
  \begin{pmatrix}
    x_0 \\
    -u_0
  \end{pmatrix}, \label{eq:transition-diagram}
\end{align}
where the two arrows denote the {\it same} mapping, depending on $Z_0$ or $-Z^*$.

Once $(x^*,u^*)$ are obtained, HAMS sets the next variables $(x_1,u_1) = (x^*, u^*)$ with probability $\alpha$, and $(x_1,u_1) = (x_0, -u_0)$ with the remaining
probability $1-\alpha$, where
\begin{align}
\alpha & = \min \left[1,  \frac{\exp\{- H(x^*, u^*)\} \mathcal N( Z^*| \bo, 2A-A^2) }{ \exp\{ -H(x_0, u_0)\} \mathcal N( Z_0 | \bo, 2 A - A^2) } \right] \nonumber \\
  & = \min[1, \exp\{G(x_0,u_0,Z_0) - G(x^*,u^*,Z^*)\}] = \min[1, \exp(-\Delta G)], \label{eq:mr}
\end{align}
with  $\Delta G = G(x^*,u^*,Z^*) - G(x_0,u_0,Z_0)$ and
$  G(x,u,Z) = H(x,u) + \frac{1}{2} Z^\T (2A - A^2)^{-1} Z$.
We can view $G$ as a generalized Hamiltonian, being analogous to the Hamiltonian $H$ but also incorporating the noise vector $Z$.
Notice that in case of rejection, while the configuration $x$ remains the current $x_0$, the momentum is reset by negating the current $u_0$.

There are two desirable properties simultaneously achieved by HAMS.\ First, the HAMS algorithm produces irreversible Markov chains with the augmented density $\pi(x,u)$ as a stationary distribution.
In fact, HAMS can be understood as an example of generalized Metropolis--Hastings sampling as discussed in Section~\ref{sec:intro}.
The proposal densities in the forward and backward transitions are respectively
$Q ( x^*, u^* | x_0, u_0) = \mathcal N( Z_0 | \, \bo, 2A - A^2 )$
and $Q (x_0, -u_0 |x^*, -u^*) = \mathcal N( -Z^* | \, \bo, 2A - A^2 )$, as illustrated in (\ref{eq:transition-diagram}).
The acceptance probability (\ref{eq:gMH-prob}) reduces to (\ref{eq:mr}), due to the evenness of mean-zero Gaussian distributions.
Hence, by Proposition 3 in \cite{Song2020}, HAMS satisfies the generalized detailed balance (\ref{eq:gDB}) and admits $\pi(x,u)$ as a stationary distribution,
where $K(x_1, u_1 | x_0, u_0) $ denotes the transition kernel from $(x_0,u_0)$ to $(x_1,u_1)$, defined by both the
proposal and acceptance-rejection schemes.

Second, the HAMS algorithm becomes rejection-free, i.e., the proposal $(x^*,u^*)$ is always accepted, when the target density $\pi(x)$ is standard Gaussian.
In this case, the proposal scheme (\ref{eq:proposal2-mat})--(\ref{eq:proposal3-mat}), with $\nabla U(x) = x$, reduces to
\begin{align}
  \begin{pmatrix}
    x^*\\
    u^*
  \end{pmatrix}
  = (I - A)
  \begin{pmatrix}
    x_0\\
    -u_0
  \end{pmatrix}+
  \begin{pmatrix}
    Z_{0}^{(1)}\\
    Z_{0}^{(2)}
  \end{pmatrix},  \label{eq:VAR}
\end{align}
which, by definition of $Z_0$ in (\ref{eq:proposal1-mat}), yields an irreversible vector autoregressive (VAR) process in $(x,u)$ with $\mathcal N(\bo, I)$ as a stationary distribution.
Moreover, the acceptance probability in (\ref{eq:mr}) can be directly calculated to be $\alpha \equiv 1$.
A subtle point is that the rejection-free property of HAMS under a standard Gaussian target $\pi(x)$ depends on
using both the proposal scheme (\ref{eq:proposal2-mat})--(\ref{eq:proposal3-mat}) and the generalized Metropolis--Hastings probability (\ref{eq:mr}), derived from (\ref{eq:gMH-prob}).
If the standard Metropolis--Hastings probability similar to (\ref{eq:MH-prob}) were used,
HAMS would not achieve the rejection-free property under standard Gaussian; otherwise, the resulting Markov chain would be reversible, which contradicts the irreversibility of (\ref{eq:VAR}).

Two special cases of HAMS are further investigated in \cite{Song2020}, where the noise variance matrix $2A - A^2$ is singular (hence only a single noise vector is needed)
and a specific choice of $\phi$ is derived.
The first is called HAMS-A, where $A$ is singular with
\begin{align}
a_1 = a, \; a_3 = b, \; a_2 =  \sqrt{ab} , \; \phi = \sqrt{ab}/(2-a), \label{eq:HAMSA-ab}
\end{align}
subject to $a,b>0$ and $a+b\le 2$. The second is
called HAMS-B, where $2I - A$ is singular with
\begin{align}
a_1 = 2-\tilde a, \; a_3 = 2-\tilde b, \; a_2 = \sqrt{\tilde a \tilde b},\; \phi = \sqrt{\tilde b/\tilde a}, \label{eq:HAMSB-ab}
\end{align}
subject to $\tilde a, \tilde b>0$ and $\tilde a+ \tilde b \le 2$.
In addition, a concrete choice of $b$ given $a$ in HAMS-A and that of $\tilde b$ given $\tilde a$ in HAMS-B are identified by
studying the lag-1 stationary auto-covariance matrix of HAMS under a standard Gaussian target $\pi(x)$.
Extensions of these choices of $\phi$ and $(a,b)$ or $(\tilde a,\tilde b)$ are studied in Sections \ref{sec:phi} and \ref{sec:eigen}.

\section{Appropriate SDE limits}
\label{sec:sde}

The HAMS algorithm is derived from the perspective of Markov chain Monte Carlo as in Metropolis--Hastings sampling, where a discrete-time Markov chain is
simulated by generating and then accepting or rejecting a proposal.
In this section, we show that under an appropriate parametrization depending on a step size, HAMS leads to continuous-time limits
characterized by stochastic differential equations (SDEs) related to Langevin dynamics.

We introduce the following parametrization of $(a_1,a_2,a_3)$ for the matrix $A$ in (\ref{eq:proposal1-mat}):
\begin{align}
& a_1 = 2 - c_1(1 + \sqrt{1 - \epsilon^2}),\quad
  a_3 = c_2(1 + \sqrt{1-\epsilon^2}),\quad
  a_2 = \epsilon\sqrt{c_1c_2}, \label{eq:sdeparam1}
\end{align}
where $\epsilon \in [0,1]$ is a step size and $c_1 \ge 0$ and $c_2 \ge 0$ are carryover coefficients.
For $c_1>0$ and $c_2 >0$, the constraint $\bo \le A \le 2I$ is satisfied if and only if $0< c_1 \le 1$ and $0 < c_2 \le 1$.
To derive an SDE limit, we further impose the relationship
\begin{align}
  & c_1 = \me^{-\eta_1\epsilon/2},\quad c_2 = \me^{-\eta_2\epsilon/2}, \quad \phi = \mO(\epsilon), \label{eq:sdeparam2}
\end{align}
where $\eta_1 \ge 0$ and $\eta_2 \ge 0$ are friction coefficients similarly as in Langevin dynamics,
and the form of $\phi$ can be flexible, subject to being of order $\mO(\epsilon)$.
For any fixed $\epsilon >0$, the preceding relationship about $(c_1,c_2)$ and $(\eta_1,\eta_2)$ is one-to-one and hence unrestricted.
However, as $\epsilon \to 0$, the relationship (\ref{eq:sdeparam2}) implies that $c_1\to 1$ and $c_2 \to 1$ for any fixed $(\eta_1,\eta_2)$.

\begin{prop}
    \label{prop:sde}
    For a target density $\pi(x)$, suppose that $\nabla^2 U(x)$ exists and the spectral norm $\|\nabla^2 U(x)\|$ is bounded in $x$ by a constant. Then as $\epsilon\to 0$,
    the HAMS proposal $(x^*,u^*)$ in (\ref{eq:proposal1-mat})--(\ref{eq:proposal3-mat}) with
    the parametrization (\ref{eq:sdeparam1})--(\ref{eq:sdeparam2}) and fixed $(\eta_1,\eta_2)$ satisfies the following SDE:
    \begin{equation}
        \label{eq:sde1}
        \begin{pmatrix}
          \dif x_t \\
          \dif u_t
        \end{pmatrix}
        = - \begin{pmatrix}
          \eta_1 & -1 \\
          1 & \eta_2
        \end{pmatrix}
        \begin{pmatrix}
          \nabla U(x_t) \\
          u_t
        \end{pmatrix}
        \dif t + \begin{pmatrix}
          \sqrt{2\eta_1}\, \dif W_t^{(1)} \\
          \sqrt{2\eta_2}\, \dif W_t^{(2)}
        \end{pmatrix},
      \end{equation}
    where $W_t^{(1)}$ and $W_t^{(2)}$ are independent, standard Brownian motions.
\end{prop}

The preceding result can be generalized under weaker conditions on the spectral norm $\|\nabla^2 U(x)\|$.
Nevertheless, the present form already gives several informative implications.

First, the SDE (\ref{eq:sde1}) falls into a general class of SDEs which admit the augmented density $\pi(x,u) \propto \exp\{ -H(x,u)\}$ as a stationary distribution in \cite{Ma2018},
where $H(x,u) = U(x) + u^\T u/2$. In fact, (\ref{eq:sde1}) can be put into the form
    \begin{align*}
    \dif z_t = - (D+Q) \nabla H(z_t) \dif t + (2D)^{1/2} \dif  W_t ,
    \end{align*}
where $z_t = (x_t^\T, u_t^\T)^\T$, $W_t= (W_t^{(1)\T}, W_t^{(2)\T})^\T$, $D$ is a positive semi-definite matrix, and $Q$ is a skew-symmetric matrix, defined as follows:
\begin{align*}
D = \begin{pmatrix}
          \eta_1 & 0 \\
          0 & \eta_2
        \end{pmatrix}, \quad
Q = \begin{pmatrix}
          0 & -1 \\
          1 & 0
        \end{pmatrix}.
\end{align*}
Moreover, by Theorem 2 in \cite{Ma2018}, the Markov process (\ref{eq:sde1}) satisfies generalized reversibility with respect to $\pi(x,u)$, where the backward process is defined by the SDE
    \begin{equation}
        \label{eq:sde1-backward}
        \begin{pmatrix}
          \dif x_t \\
          \dif u_t
        \end{pmatrix}
        = - \begin{pmatrix}
          \eta_1 & 1 \\
          -1 & \eta_2
        \end{pmatrix}
        \begin{pmatrix}
          \nabla U(x_t) \\
          u_t
        \end{pmatrix}
        \dif t + \begin{pmatrix}
          \sqrt{2\eta_1}\, \dif W_t^{(1)} \\
          \sqrt{2\eta_2}\, \dif W_t^{(2)}
        \end{pmatrix}.
      \end{equation}
Interestingly, the Markov process defined by (\ref{eq:sde1-backward}) is equivalent to that defined by (\ref{eq:sde1}) but with $u_t$ and $W_t^{(2)}$ replaced by $-u_t$ and $-W_t^{(2)}$, that is,
(\ref{eq:sde1-backward}) is equivalent to
    \begin{equation*}
        \begin{pmatrix}
          \dif x_t \\
          - \dif u_t
        \end{pmatrix}
        = - \begin{pmatrix}
          \eta_1 & -1 \\
          1 & \eta_2
        \end{pmatrix}
        \begin{pmatrix}
          \nabla U(x_t) \\
          -u_t
        \end{pmatrix}
        \dif t + \begin{pmatrix}
          \sqrt{2\eta_1}\, \dif W_t^{(1)} \\
          -\sqrt{2\eta_2}\, \dif W_t^{(2)}
        \end{pmatrix}.
      \end{equation*}
Hence the backward process (\ref{eq:sde1-backward}) is stochastically the same as the forward process (\ref{eq:sde1}) except for the sign reversal of the momentum.
This relationship between the forward and backward processes represents a continuous-time limit of that between the forward and backward transitions of the proposals in HAMS, as illustrated
in (\ref{eq:transition-diagram}).
The generalized reversibility of (\ref{eq:sde1}) as a limit of the HAMS proposal implies that HAMS with the parametrization  (\ref{eq:sdeparam1})--(\ref{eq:sdeparam2})
leads to an acceptance rate which tends to 1 in the limit of $\epsilon \to 0$.
Similarly as emphasized in \cite{Ma2018}, the limiting acceptance rate of 1 would not be obtained if the HAMS proposal were plugged into standard Metropolis--Hastings sampling.
This point also echoes the discussion after (\ref{eq:VAR}) about the rejection-free property of HAMS under standard Gaussian.

Second, it is interesting to examine two special cases of the SDE (\ref{eq:sde1}). On one hand, setting $\eta_1 = 0$ in (\ref{eq:sde1}) yields
the underdamped Langevin dynamics, i.e., (\ref{eq:langevin-SDE}) with $\eta=\eta_2$:
\begin{equation}
    \begin{pmatrix}
      \dif  x_t \\
      \dif  u_t
    \end{pmatrix}
    = - \begin{pmatrix}
      0& -1 \\
      1 & \eta_2
    \end{pmatrix}
    \begin{pmatrix}
      \nabla U(x_t) \\
      u_t
    \end{pmatrix}
    \dif t + \begin{pmatrix}
      0 \\
      \sqrt{2\eta_2} \dif W_t^{(2)}
    \end{pmatrix}. \label{eq:Langevin}
  \end{equation}
On the other hand, setting $\eta_2 = 0$ in (\ref{eq:sde1}) leads to the SDE,
\begin{equation}
    \begin{pmatrix}
      \dif  x_t \\
      \dif  u_t
    \end{pmatrix}
    = - \begin{pmatrix}
      \eta_1 & -1 \\
      1 & 0
    \end{pmatrix}
    \begin{pmatrix}
      \nabla U(x_t) \\
      u_t
    \end{pmatrix}
    \dif t + \begin{pmatrix}
      \sqrt{2\eta_1} \dif W_t^{(1)} \\
      0
    \end{pmatrix}. \label{eq:Langevin-new}
  \end{equation}
The intersection of the two cases obtained by $\eta_1=\eta_2=0$ corresponds to the
deterministic Hamiltonian dynamics: $\dif x_t = u_t \,\dif t$ and $\dif u_t =- \nabla U(x_t) \,\dif t$.
There are interesting differences between (\ref{eq:Langevin}) and (\ref{eq:Langevin-new}), in relation to the Hamiltonian dynamics.
The underdamped Langevin dynamics (\ref{eq:Langevin}) is widely used to describe the motions of particles in the presence of frictions,
where the momentum $u$ is dampened by a friction force and a Gaussian white noise.
By comparison, the SDE (\ref{eq:Langevin-new}) is mathematically well defined, indicating that the configuration $x$ is affected by the force field as in overdamped Langevin, in addition to the momentum.
To our knowledge, the physical meaning of (\ref{eq:Langevin-new}) remains to be studied.

Third, the two special cases of $\eta_1=0$ or $\eta_2=0$ are directly related to HAMS-A or HAMS-B respectively.
In fact, HAMS-A can be obtained by taking $\eta_1 = 0$ or equivalently $c_1 = 1$ in (\ref{eq:sdeparam1}) and the choice of $\phi$ as described in (\ref{eq:HAMSA-ab}),
\begin{align}
  a_1 = 1 - \sqrt{1 - \epsilon^2},\quad a_2  = \epsilon \sqrt{c_2},
  \quad a_3 = c_2(1 + \sqrt{1-\epsilon^2}),\quad \phi = \frac{\epsilon\sqrt{c_2}}{1 + \sqrt{1-\epsilon^2}} . \label{eq:HAMSA-param}
\end{align}
Then the proposal scheme (\ref{eq:proposal1-mat})--(\ref{eq:proposal3-mat}) for HAMS-A can be simplified to
\begin{align}
    \tilde u & = \sqrt{c_2}u_0 - \frac{\epsilon}{1 + \sqrt{1 - \epsilon^2}} \nabla U(x_0) + Z,  \label{eq:HAMSA-prop1} \\
    x^* & = x_0 + \epsilon \tilde u, \label{eq:HAMSA-prop2} \\
    u^* & = -u_0 + 2\sqrt{c_2} \tilde u + \frac{\epsilon\sqrt{c_2}}{1 + \sqrt{1 - \epsilon^2}}\left(\nabla U(x_0) - \nabla U(x^*)\right), \label{eq:HAMSA-prop3}
\end{align}
with $Z \sim \mathcal N(\bo, (1 - c_2) I)$.
Moreover, HAMS-B can be obtained by taking $\eta_2 = 0$ or equivalently $c_2 = 1$ and the choice of $\phi$ as described in in (\ref{eq:HAMSB-ab}),
\begin{align}
  a_1 = 2 - c_1(1+\sqrt{1 -\epsilon^2}),\quad a_2  = \epsilon \sqrt{c_1},
  \quad a_3 = 1 + \sqrt{1-\epsilon^2},\quad \phi = \frac{\epsilon}{\sqrt{c_1} (1+\sqrt{1-\epsilon^2})} . \label{eq:HAMSB-param}
\end{align}
Then the proposal scheme (\ref{eq:proposal1-mat})--(\ref{eq:proposal3-mat}) for HAMS-B can be simplified to
\begin{align}
    \tilde u & = \sqrt{c_1}u_0 - \frac{2 - c_1(1 + \sqrt{1-\epsilon^2})}{\epsilon}\nabla U(x_0) + \frac{\sqrt{c_1}(1 + \sqrt{1 - \epsilon^2})}{\epsilon} Z,  \label{eq:HAMSB-prop1} \\
    x^* & = x_0 + \epsilon \tilde u, \label{eq:HAMSB-prop2} \\
    u^* & = u_0 - \frac{\epsilon}{\sqrt{c_1}(1 + \sqrt{1 - \epsilon^2})}\left(\nabla U(x_0) + \nabla U(x^*)\right), \label{eq:HAMSB-prop3}
\end{align}
with $Z \sim \mathcal N(\bo, (1 - c_1) I)$.
Taking $c_2=1$ in (\ref{eq:HAMSA-prop1})--(\ref{eq:HAMSA-prop3}) or $c_1=1$ in (\ref{eq:HAMSB-prop1})--(\ref{eq:HAMSB-prop3}) yields
\begin{align*}
    \tilde u & = u_0 - \frac{\epsilon}{1 + \sqrt{1 - \epsilon^2}} \nabla U(x_0)  ,   \quad
    x^* = x_0 + \epsilon \tilde u, \quad
    u^* =  \tilde u - \frac{\epsilon}{1 + \sqrt{1 - \epsilon^2}} \nabla U(x^*),
\end{align*}
which is the same as the leapfrog discretization of the Hamiltonian dynamics, except with step size $\epsilon/(1+\sqrt{1-\epsilon^2})$ instead of $\epsilon/2$ for momentum updates.
We record the following consequence of Proposition~\ref{prop:sde} for HAMS-A and HAMS-B.

\begin{cor} \label{cor:sde-HAMS-AB}
Under the setting of Proposition~\ref{prop:sde}, in the limit of $\epsilon \to 0$, the HAMS-A proposal $(x^*, u^*)$ defined in (\ref{eq:HAMSA-prop1})--(\ref{eq:HAMSA-prop3}) with $c_2= \me^{-\eta_2\epsilon/2}$ and fixed $\eta_2$ satisfies the underdamped Langevin SDE (\ref{eq:Langevin}),
and the HAMS-B proposal $(x^*, u^*)$ defined in (\ref{eq:HAMSB-prop1})--(\ref{eq:HAMSB-prop3}) with $c_1= \me^{-\eta_1\epsilon/2}$ and fixed $\eta_1$
satisfies the SDE (\ref{eq:Langevin-new}).
\end{cor}

The preceding result sheds new light on differences between HAMS-A and HAMS-B.
The parametrization (\ref{eq:HAMSA-param}) for HAMS-A is the same as used in \cite{Song2020}.
But the parametrization (\ref{eq:HAMSB-param}) for HAMS-B differs slightly from that used in \cite{Song2020}, and has the conceptual
advantage of inducing the SDE (\ref{eq:Langevin-new}) with $c_1= \me^{-\eta_1\epsilon/2}$ and any fixed $\eta_1$ as $\epsilon\to 0$.
See the Supplement (Section~\ref{subsec:proof:param-HAMSB}) for further discussion.

\section{Default choice of $\phi$}
\label{sec:phi}

The SDE (\ref{eq:sde1}) is informative about the behavior of HAMS with the parametrization (\ref{eq:sdeparam1})--(\ref{eq:sdeparam2})
in the limit of $\epsilon \to 0$, where $\phi = \mO(\epsilon)$ can be of a flexible form.
To derive a specific choice of $\phi$ with fixed $\epsilon$, we study the generalized Hamiltonian difference, $\Delta G$,
in the acceptance probability (\ref{eq:mr}) under a univariate Gaussian target $\mathcal N(0 ,\gamma^{-1})$, which is called harmonic oscillators in physics.
As discussed in Section~\ref{sec:review},  when the target density $\pi(x)$ is standard Gaussian,
HAMS is rejection-free: the acceptance probability (\ref{eq:mr}) is always 1 or equivalently $\Delta G$ is always 0.
But when the target density is $\mathcal N(0 ,\gamma^{-1})$ with $\gamma\not=1$, HAMS is no longer rejection-free.
We seek a choice of $\phi$ such that $\Delta G$ is minimally affected as $\gamma$ deviates from 1.

\begin{prop}
  \label{prop:phichoice}
Suppose that the target density $\pi(x)$ is $\mathcal N(0,\gamma^{-1})$. Then $\Delta G$ defined in (\ref{eq:mr}) can be
expressed as a quadratic form,
\[
\Delta G =  (x_0, u_0, Z_{0}^{(1)}, Z_{0}^{(2)})\, D(\gamma) \, (x_0, u_0, Z_{0}^{(1)}, Z_{0}^{(2)})^\T,
\]
where $D(\gamma)$ is a $4 \times 4$ matrix. For $i, j = 1,2,3,4$, the $(i,j)$th entry of $D(\gamma)$ is
$d_{ij}(\gamma)$, a polynomial of $\gamma$, with coefficients depending on $(a_1,a_2,a_3,\phi)$. The
coefficient of the leading term of $d_{44}(\gamma)$ is always $0$. Furthermore, for any $a_1,a_2,a_3$ such that $\bo\leq A \leq 2 I$,
the coefficients of the leading terms of $d_{11}(\gamma),d_{22}(\gamma),d_{33}(\gamma)$ are simultaneously minimized in absolute values
by the choice $\phi = a_2/(2-a_1)$.
\end{prop}

The preceding result gives a default choice of $\phi$ for HAMS in a unified manner.
In the special cases of HAMS-A and HAMS-B, the choice $\phi = a_2/(2-a_1)$ is easily seen to
agree with those derived in \cite{Song2020}, as stated in (\ref{eq:HAMSA-param}) and (\ref{eq:HAMSB-param}).
The derivation of $\phi$ in \cite{Song2020} is similar as above for HAMS-A, but involves a seemingly different angle for HAMS-B, where
the choice of $\phi$ in (\ref{eq:HAMSB-param}) is unique in ensuring that the two backward noise vectors $Z_1^{(1)*}$ and $Z_2^{(2)*}$ in (\ref{eq:proposal3}) are proportional to each other.

As another interesting consequence of the default choice of $\phi$, the quantity $\Delta G$ can be substantially simplified even for a general target density $\pi(x)$.

\begin{cor}
  \label{cor:phichoice}
 Suppose that $\phi = a_2/(2-a_1)$ is used. Then for a general target density $\pi(x)$ on $\mathbb R^k$, $\Delta G$ in the acceptance probability (\ref{eq:mr}) reduces to
  \begin{equation}
   \label{eq:dG1}
    \Delta G = U(x^*) - U(x_0) + \frac{\{\nabla U(x_0) + \nabla U(x^*)\}^\T\left[a_1 \{\nabla U(x_0) + \nabla U(x^*)\} - 2(a_2u_0 +Z_{0}^{(1)} ) \right] }{2(2-a_1)}.
\end{equation}
Particularly, for a Gaussian target $\mathcal N(0,\gamma^{-1} I)$, $\Delta G$ reduces to
  \begin{equation}
    \label{eq:dG2}
    \Delta G = \frac{a_1\gamma(\gamma-1)}{2(2-a_1)}(a_2u_0 + Z_{0}^{(1)} - a_1\gamma x_0)^{\T}(a_2u_0 + Z_{0}^{(1)} + (2 - a_1\gamma)x_0).
  \end{equation}
\end{cor}

We remark that the expressions (\ref{eq:dG1}) and (\ref{eq:dG2}) depend on only $(a_1,a_2)$ and $(x_0,u_0, Z_0^{(1)})$, but not $a_3$ or $Z_0^{(2)}$,
even though the proposal $u^*$ depends on $a_3$ and $Z_0^{(2)}$.

\section{Stationary variance under univariate Gaussian}
\label{sec:variance}

In Section~\ref{sec:phi}, we derive a default choice of $\phi$ by exploiting an algebraic property of $\Delta G$ under a univariate Gaussian target.
In this and next two sections, we also consider a univariate Gaussian target density, but turn to study
stochastic properties of HAMS and understand impacts of different choices for the tuning parameters $(a_1,a_2,a_3)$.

We study the stationary variance of the HAMS proposal (\ref{eq:proposal1-mat})--(\ref{eq:proposal3-mat}), applied iteratively without performing acceptance-rejection,
under a univariate Gaussian target $\mathcal N(0,\gamma^{-1})$.
A similar approach can be found in \cite{Burrage2007} in comparing various methods for solving the underdamped Langevin SDE (\ref{eq:Langevin}).
In this setting, the updates (\ref{eq:proposal1-mat})--(\ref{eq:proposal3-mat}) can be equivalently written as an order-1 VAR process,
\begin{equation}
  \label{eq:proposal4}
   \begin{pmatrix}
    x^* \\
    u^*
  \end{pmatrix} = \Phi \begin{pmatrix}
    x_0 \\
    u_0
  \end{pmatrix} +
  \mathbf{\zeta}. \\
\end{equation}
where $\Phi$ is the coefficient matrix and $\zeta$ represents noise terms independent of $(x_0,u_0)$. Detailed expressions are
given in the Supplement (Section~\ref{subsec:proof:prop:variance}).

\begin{prop}
  \label{prop:variance}
Suppose that the target density $\pi(x)$ is $\mathcal N(0,\gamma^{-1})$, and the choice $\phi = a_2/(2-a_1)$ is used in HAMS.\
Then the stationary variance of the HAMS proposal (\ref{eq:proposal1-mat})--(\ref{eq:proposal3-mat}) or equivalently the VAR process (\ref{eq:proposal4}) is
\begin{equation}
  \label{eq:var1}
  \Var\begin{pmatrix}
  x \\
  u
\end{pmatrix} = \begin{pmatrix}
    \frac{a_1 - 2}{\gamma(a_1\gamma - 2)} & 0 \\
    0 & 1
  \end{pmatrix}.
\end{equation}
\end{prop}

\vspace{.1in}
It is interesting to compare the stationary variance matrix (\ref{eq:var1}) with the variance matrix from the augmented density $\pi(x,u)$, that is,
\begin{equation}
  \label{eq:var2}
\Gamma =
\begin{pmatrix}
  \gamma^{-1} & 0 \\
  0 & 1
\end{pmatrix}.
\end{equation}
At stationarity, the HAMS proposal leads to  $\Var(u)$ and $\Cov(x,u)$ which are the same as the target values in (\ref{eq:var2}). But $\Var(x)$ differs from $\gamma^{-1}$ unless $\gamma = 1$ or $a_1 = 0$.
The first case $\gamma=1$ confirms the rejection-free property of HAMS under standard Gaussian, as discussed in Section~\ref{sec:review}.
The latter case $a_1=0$ is degenerate, where $a_2$ must also be 0 by the positive semi-definiteness of $A$ and hence the update of $x$ becomes non-ergodic, $x^*=x_0$.

To study the order of error in $\Var(x)$, we use the parameterization of $a_1$ in (\ref{eq:sdeparam1})--(\ref{eq:sdeparam2})
and take a Taylor expansion with respect to the step size $\epsilon$:
\begin{equation}
  \label{eq:var3}
  \Var(x) = \frac{a_1 - 2}{\gamma(a_1\gamma - 2)} = \frac{1}{\gamma} + \frac{\gamma - 1}{\gamma} \cdot \frac{\eta_1}{2}\epsilon +
  \frac{\gamma - 1}{\gamma}\cdot \left(\frac{1+(\gamma - 1/2)\eta_1^2}{4}\right)\epsilon^2 + \mO(\epsilon^3),
\end{equation}
The leading error term in $\Var(x)$ is then
\[
  \frac{\gamma - 1}{\gamma} \cdot \frac{\eta_1}{2}\epsilon.
\]
For HAMS-A with $\eta_1 = 0$ as in (\ref{eq:HAMSA-param}), this term vanishes and the overall error is $\mO(\epsilon^2)$.
More generally, if we set $\eta_1 = k\epsilon$ for some $k\geq 0$, then the expansion (\ref{eq:var3}) becomes
\begin{equation}
  \label{eq:var4}
 \Var(x)= \frac{a_1 - 2}{\gamma(a_1\gamma - 2)} = \frac{1}{\gamma} + \frac{\gamma -1}{\gamma}\left(\frac{1}{4} + \frac{k}{2}\right)\epsilon^2 + \mO(\epsilon^3),
\end{equation}
and hence the overall error is also $\mO(\epsilon^2)$.
Nevertheless, for any fixed $\gamma$, the coefficient of the leading error term in (\ref{eq:var4}), $ \frac{\gamma -1}{\gamma} (\frac{1}{4} + \frac{k}{2})$, is minimized in absolute values
by taking $k = 0$, corresponding to HAMS-A. Therefore HAMS-A is the best within the HAMS class when considering the stationary variance in
the univariate Gaussian setting.

We remark that under univariate Gaussian (or harmonic oscillator), several existing algorithms for simulating Langevin dynamics are known to yield the correct variance $\gamma^{-1}$ for $x$, including
Mannella's leapfrog \citep{Mannella2004, Burrage2007}, GJF algorithm \citep{Farago2013}, and BAOAB and ABOBA  \citep{Leimkuhler2012, Leimkuhler2013}.
But the variance for $u$ obtained from these algorithms is shown to be $(1-\gamma^{-1} \epsilon^2/4)^{-1}$, hence with error of order $\mO(\epsilon^2)$.
For harmonic oscillators, the variances of $x$ and $u$ are related to the configurational and kinetic temperatures \citep{Farago2019}; see the Supplement, Section \ref{sec:sim-detail-DW}.
Nevertheless, these results are primarily of interest in the molecular dynamics settings where no acceptance-rejection is performed.
When using HAMS or Metropolis-adjusted versions of these algorithms for sampling from the augmented density $\pi(x,u)$,
the acceptance-rejection step is defined, depending on both $x$ and $u$.
We provide further discussion from the sampling perspective in Section~\ref{sec:accept}.

\section{Expected acceptance rate under univariate Gaussian}
\label{sec:accept}

Section 5 investigates the stationary variance under a univariate Gaussian target when using the HAMS proposal without acceptance-rejection.
In this section, we study the expected acceptance rate of HAMS under univariate Gaussian while incorporating the acceptance-rejection step.
We show that HAMS-A achieves an expected acceptance rate which is closest to 1 by the leading error as $\epsilon\to 0$ among the HAMS class,
in agreement with the best accuracy of the stationary variance obtained by HAMS-A without acceptance-rejection.

First, we provide a useful result about the expected acceptance rate of HAMS under a general target density $\pi(x)$.
A similar result is discussed in \cite{Neal2011} and \cite{Calvo2019} for Hamiltonian Monte Carlo (HMC),
which uses, as a proposal, a deterministic integrator such as the leapfrog integrator for the Hamiltonian dynamics.
By comparison, the HAMS proposal (\ref{eq:proposal1-mat})--(\ref{eq:proposal3-mat}) is a stochastic mapping, depending on a noise vector $Z$.

\begin{lem}
  \label{lem:accept}
  Assume that the HAMS chain is stationary, with $(x_0,u_0)\sim \pi(x,u)$. Then the expected acceptance rate is
\[
  \EE[\alpha] = 2 \PP[\Delta G <0] + \PP[\Delta G = 0],
\]
where $\alpha$ is the acceptance probability defined in (\ref{eq:mr}).
\end{lem}

The preceding result includes the term $\PP[\Delta G = 0]$, which is important for HAMS because $\PP[\Delta G = 0] = 1$ for HAMS under a standard Gaussian target.
Such a term is absent in the related result for HMC in \citet{Calvo2019}, where the probability that the change in the Hamiltonian, $\Delta H$, equals zero is assumed to be zero.

For a univariate Gaussian target, the expected acceptance rate of HAMS can be monotonically linked to the
expectation of the generalized Hamiltonian difference $\Delta G$.
Interestingly, \cite{Calvo2019} shows that under univariate Gaussian, the expected acceptance rate of HMC satisfies a similar formula as (\ref{eq:accept1}) below,
with $\Delta G$ replaced by $\Delta H$.
However, the rescaling argument used in \cite{Calvo2019} to directly transfer the formula from standard to non-standard Gaussian for HMC is not applicable to HAMS,
partly because (\ref{eq:accept1}) holds trivially for HAMS with $\alpha\equiv 1$ and $\Delta G \equiv 0$ under standard Gaussian.

\begin{prop}
  \label{prop:accept}
Suppose that the target density $\pi(x)$ is $\mathcal N(0,\gamma^{-1})$, and the choice $\phi = a_2/(2-a_1)$ is used in HAMS.\
Then the expected acceptance rate of HAMS  in stationarity is
\begin{equation}
  \label{eq:accept1}
  \EE[\alpha] = 1 - \frac{2}{\pi} \arctan\left(\sqrt{\frac{\EE[\Delta G]}{2}}\right).
\end{equation}
\end{prop}

\vspace{.1in}
To further assess the expected acceptance rate, we expand $\EE[\alpha]$ in terms of $\EE[\Delta G]$,
\[
  \EE[\alpha] = 1 - \frac{\sqrt{2}}{\pi}\sqrt{\EE[\Delta G]} + \mO((\EE[\Delta G])^{3/2}),
\]
and use the following formula obtained in the proof of Proposition~\ref{prop:accept},
\begin{equation*}
  \EE[\Delta G] = \frac{a_1^3(\gamma - 1)^2\gamma}{2(2-a_1)}.
\end{equation*}
For the parameterization of $a_1$ in (\ref{eq:sdeparam1})--(\ref{eq:sdeparam2}), direct calculation shows that
$ a_1 = \eta_1 \epsilon + (\frac{1}{2}-\frac{\eta_1^2}{4})\epsilon^2 + \mO (\epsilon^3)$ and
$ \EE [\alpha ] = 1- \sqrt{(\gamma-1)^2 \gamma} /(\sqrt{2}\pi) \cdot \eta_1^{3/2} \epsilon^{3/2} + \mO (\epsilon^3)$.
For HAMS-A with $\eta_1=0$, the leading error term $\mO(\epsilon^{3/2})$ vanishes in $\EE[\alpha]$, so that  $\EE [\alpha] = 1- \mO(\epsilon^3)$.
If we set $\eta_1 = k\epsilon$ for some $k\geq 0$, then the expected acceptance rate is also $1-\mO(\epsilon^3)$, satisfying
\begin{equation}
  \label{eq:accept2}
  \EE[\alpha] = 1 - \frac{(1+ 2k)^{3/2}}{4\pi}\sqrt{\gamma(\gamma - 1)^2}\cdot \epsilon^3 + \mO(\epsilon^5).
\end{equation}
For any fixed $\gamma$, a smaller $k$ leads to a higher expected acceptance rate as $\epsilon\to 0$ under univariate Gaussian,
with $k = 0$ (i.e., HAMS-A) being optimal in the HAMS class.

As mentioned in Section~\ref{sec:variance}, several existing algorithms lead to the correct variance for $x$, but incur errors in the variance for $u$ under univariate Gaussian.
In the Supplement, Section~\ref{sec:MA-algrithms}, we derive Metropolized versions of BAOAB and ABOBA \citep{Leimkuhler2012} and BP \citep{Bussi2007},
and then study the corresponding expected acceptance rates under univariate Gaussian. Interestingly,
a similar identity to (\ref{eq:accept1}) is obtained for each method with appropriately defined $\Delta G$,
and the expected acceptance rate satisfies
\begin{align}
    \EE[\alpha] = 1 - \frac{\sqrt{2}}{4\pi}\gamma\sqrt{\eta} \cdot \epsilon^{5/2} + \mO(\epsilon^{7/2}), \label{eq:accept-BAO}
\end{align}
for Metropolis-adjusted BAOAB and ABOBA, and
\begin{align}
     \EE[\alpha] = 1 - \frac{\gamma^{3/2 }}{4\pi} \cdot \epsilon^3 + \mO(\epsilon^9) , \label{eq:accept-BP}
\end{align}
for Metropolis-adjusted BP.\
The expected acceptance rate (\ref{eq:accept-BAO}) for Metropolis-adjusted BAOAB or ABOBA deviates from 1 by a lower order $\mO(\epsilon^{5/2})$, compared with $\mO(\epsilon^3)$ for HAMS-A or
HAMS with $\eta_1 = k\epsilon$.
The expected acceptance rate (\ref{eq:accept-BP}) for Metropolis-adjusted BP deviates from 1 by a leading term which is of same order $\mO(\epsilon^3)$ as in HAMS with $\eta_1=k\epsilon$, but
does not reduce to 0 in the special case of standard Gaussian ($\gamma=1$).

\section{Convergence rate under standard Gaussian}
\label{sec:eigen}

In Sections~\ref{sec:variance}--\ref{sec:accept}, we mainly study stationary properties of HAMS under univariate Gaussian, where the HAMS chain is assumed to be stationary
without or with acceptance-rejection. In this section, we examine how the convergence rate can be optimized within the HAMS class
under the standard Gaussian setting.
The investigation reveals an interesting trade-off between different versions of HAMS, and leads to
a specific choice of $c_2$ given $(\epsilon,c_1)$ or $c_1$ given $(\epsilon,c_2)$, which can be used as the default choices
when the target distribution can be transformed into roughly standard Gaussian after preconditioning.

Consider the standard Gaussian setting, where HAMS is rejection-free. In this case, the HAMS proposal reduces to the order-1 VAR process (\ref{eq:VAR})
or equivalently
\begin{align}
\begin{pmatrix}
    x^* \\
    u^*
  \end{pmatrix} = \Phi \begin{pmatrix}
    x_0 \\
    u_0
  \end{pmatrix} + Z_0, \label{eq:VAR2}
\end{align}
where $Z_0 \sim \mathcal N (\bo, 2A - A^2)$ and
$\Phi = \begin{pmatrix}
    1- a_1 & a_2 \\
    - a_2 & a_3-1
\end{pmatrix}$,
depending on $A = \begin{pmatrix}
    a_1 & a_2 \\
    a_2 & a_3
\end{pmatrix}$
such that $\bo \le A \le 2 I$ as stated in (\ref{eq:proposal1-mat}).
The convergence rate of the VAR process (\ref{eq:VAR2}) is known to be captured by the spectral radius of
the coefficient matrix $\Phi$, which is defined as the maximum modulus of its eigenvalues \citep[see][Theorem 1]{Roberts1997}.
A smaller spectral radius of $\Phi$ leads to faster convergence for the VAR process (\ref{eq:VAR2}).

If the coefficients $a_1,a_2,a_3$ are all free to choose,  subject to $\bo \le A \le 2I$, then the spectral radius of $\Phi$ can be made equal to 0, by
setting $a_3 = a_1$ and $a_2 = \pm(1-a_1)$ for $a_1 \in [.5, 1.5]$. The corresponding VAR process (\ref{eq:VAR2}) can be shown to converge to $\mathcal N(\bo,I)$ after 2 iterations for any initial value.
However, such choices of $A$ are incompatible with the parametrization (\ref{eq:sdeparam1})--(\ref{eq:sdeparam2}), where
$a_1 \to 0$ and $a_3 \to 2$ as $\epsilon \to 0$ for appropriate SDE limits.

To obtain a meaningful solution, we seek to minimize the spectral radius of $\Phi$ over possible choices of $(a_2,a_3)$ while fixing $a_1$ and $\nu \equiv a_2^2/a_3$.
For the parametrization (\ref{eq:sdeparam1}), this program corresponds to optimizing the choice of $c_2$ while fixing $(\epsilon, c_1)$,
which indicates that both $a_1 = 2 - c_1 (1 +\sqrt{1 + \epsilon^2})$ and $\nu = a_2^2 / a_3 = c_1 (1 -\sqrt{1-\epsilon^2})$ are fixed.

\begin{prop}
 \label{prop:eigen1}
 Suppose that the target density $\pi(x)$ is $\mathcal N(0,1)$.
 For any fixed $0 < a_1 <2$ and $\nu \equiv a_2^2 /a_3 \ge 0 $ such that $\nu \le a_1 \le 1 + \nu$,
 the convergence rate of the HAMS process (\ref{eq:VAR2}) is optimized or the spectral radius of $\Phi$ is minimized over $(a_2,a_3)$ by the choices
 \begin{equation}
  \label{eq:eigen1}
   a_3^*  = ( \sqrt{ \nu + 2- a_1} - \sqrt{\nu} )^2, \quad a_2^* = \pm\sqrt{ \nu a_3^*},
\end{equation}
with the minimum spectral radius given by  $ \frac{ |a_3^*-a_1|}{2}$.
\end{prop}

There are two constraints on the fixed values of $(a_1,\nu)$ in the preceding result. The first constraint $\nu \le a_1$ is needed to ensure
$ a_1 a_3 \ge a_2^2 = \nu a_3 $ and hence $A \ge \bo$.
The second constraint $a_1 \le 1 + \nu$ is equivalent to requiring $c_1 \ge 1/2$ in the parametrization (\ref{eq:sdeparam1}), by the expression
of $(a_1,\nu)$ mentioned above.
If this constraint were relaxed, then the optimal choice of $(a_2,a_3)$ would be of a different form than (\ref{eq:eigen1}), by extending
the proof of Proposition~\ref{prop:eigen1}. Nevertheless,
the constraint $a_1 \le 1 + \nu$ or equivalently $c_1 \ge 1/2$ is automatically satisfied in HAMS-A, with $\nu=a_1$ and $c_1=1$.
For HAMS-A, the optimal choice (\ref{eq:eigen1}) given $a_1$ reduces to $a_3^* = (\sqrt{2}-\sqrt{a_1})^2$ and $a_2^* = \pm\sqrt{a_1 a_3^*}$,
in agreement with \cite{Song2020}, Lemma 3.
Moreover, for $c_1 = \me^{-\eta_1 \epsilon/2}$ in the parametrization (\ref{eq:sdeparam2}) with $\eta_1$ bounded, the constraint $c_1 \ge 1/2$
is also satisfied in the practical situation where the step size $\epsilon$ is relatively small.

As motivated by the discussions in Sections~\ref{sec:variance}--\ref{sec:accept}, we study HAMS
with $\eta_1= k\epsilon$ in (\ref{eq:sdeparam2}) for some constant $k \ge 0$.
Given $0 < \epsilon <1$ and $  1/2 \le c_1 \le 1$ in the parametrization (\ref{eq:sdeparam1}), $(a_1,\nu)$ are fixed as mentioned above, and
the optimal choice of $a_3$ in (\ref{eq:eigen1}) translates into
\begin{align}
a_3^* = \left\{ 3 - \sqrt{1-\epsilon^2} - 2 \sqrt{2} \epsilon ( 1+ \sqrt{1-\epsilon^2} )^{-1/2} \right\} c_1 . \label{eq:eigen2}
\end{align}
For $\eta_1=k \epsilon$ and $c_1 = \exp(-k\epsilon^2 /2)$,
by the expressions of $a_1$ in (\ref{eq:sdeparam1}) and $a_3^*$ in (\ref{eq:eigen2}),
the minimum spectral radius of $\Phi$ can be shown as $\epsilon \to 0$ to be
\begin{align}
  \label{eq:eigen5}
  \frac{|a_3^* - a_1|}{2} = 1 - \epsilon - k\epsilon^2 + \mO(\epsilon^3) , 
\end{align}
This means that a larger $k$ corresponds to a smaller spectral radius for $\Phi$, hence faster convergence for HAMS under standard Gaussian. By comparison,
as seen from (\ref{eq:var4}) and (\ref{eq:accept2}),
a smaller $k$ corresponds to a more accurate stationary variance without acceptance-rejection
and a higher expected acceptance rate with acceptance-rejection for HAMS when the target density is non-standard Gaussian.
Hence there exists a tradeoff in the behavior of HAMS when using $\eta_1=k\epsilon$ and the associated choice $a_3^*$ for different values of $k$.

The preceding scheme of optimizing over $(a_2,a_3)$ given $(a_1, \nu)$ extends the corresponding scheme for HAMS-A, where $\nu$ is identical to $a_1$ by definition.
A similar extension can be obtained for HAMS-B, by minimizing the spectral radius of $\Phi$ over possible choices of $(a_1,a_2)$ while fixing
$a_3$ and $\tilde \nu \equiv a_2^2 / (2-a_1)$. For the parametrization (\ref{eq:sdeparam1}), this scheme corresponds to
optimizing the choice of $c_1$ while fixing $(\epsilon, c_2)$,
which leads to fixed $a_3 = c_2 (1 +\sqrt{1 + \epsilon^2})$ and $\tilde \nu = a_2^2 / (2-a_1) = c_2 (1 -\sqrt{1-\epsilon^2})$.

\begin{prop}
 \label{prop:eigen1-b}
 Suppose that the target density $\pi(x)$ is $\mathcal N(0,1)$.
 For any fixed $0 < a_3 <2$ and $\tilde\nu \equiv a_2^2 /(2-a_1) \ge 0 $ such that $\tilde \nu \le 2-a_3 \le 1 + \tilde\nu$,
  the convergence rate of the HAMS process (\ref{eq:VAR2}) is optimized or the spectral radius of $\Phi$ is minimized over $(a_1,a_2)$ by the choices
 \begin{equation}
  \label{eq:eigen1-b}
   2-a_1^*  = ( \sqrt{ \tilde \nu + a_3} - \sqrt{\tilde \nu} )^2, \quad a_2^* = \pm\sqrt{ \tilde \nu (2-a_1^*) },
\end{equation}
with the minimum spectral radius given by  $ \frac{ |a_3-a_1^*|}{2}$.
\end{prop}

For HAMS-B, defined with $c_2=1$ as in (\ref{eq:HAMSB-param}), simple calculation shows that $\tilde\nu $ is identical to $2-a_3$,
and hence the optimal choice (\ref{eq:eigen1-b}) given $a_3$ reduces to
$ 2 -a_1^* = ( \sqrt{ 2} - \sqrt{2-a_3} )^2$ and $ a_2^* = \pm\sqrt{ (2-a_3) (2-a_1^*) }$.
This result is symmetric to \cite{Song2020}, Lemma 3, where
the optimal choices of $(a_2,a_3)$ given $a_1$  are determined for HAMS-B by the relationship
$2 - a_3 = ( \sqrt{2} - \sqrt{2-a_1})^2$.
The change that $a_1$ is tuned given $a_3$, instead of $a_3$ given $a_1$, is due to the parametrization (\ref{eq:HAMSB-param})
used here for HAMS-B, which is slightly different from that in \cite{Song2020} as mentioned at the end of Section~\ref{sec:sde}.

In connection with the SDEs in Section~\ref{sec:sde}, we record the implied choice of $\eta_2$ by (\ref{eq:eigen1}) for HAMS-A with $\eta_1=0$
and that of $\eta_1$ by (\ref{eq:eigen1-b}) for HAMS-B with $\eta_2=0$.


\begin{cor}
  \label{cor:eta-ab}
For HAMS-A, the choice of $\eta_2$ based on (\ref{eq:eigen1}) is of the order
$ 2 + \mO(\epsilon^2)$, and the associated HAMS-A proposal satisfies the underdamped Langevin SDE (\ref{eq:Langevin}) with $\eta_2=2$
as $\epsilon \to 0$.
For HAMS-B, the choice of $\eta_1$ based on (\ref{eq:eigen1-b}) is of the order
$ 2 + \mO(\epsilon^2)$ and the associated HAMS-B proposal satisfies the SDE (\ref{eq:Langevin}) with $\eta_1=2$ as $\epsilon \to 0$.
\end{cor}

To facilitate tuning, the formulas in Propositions~\ref{prop:eigen1} and \ref{prop:eigen1-b} can be used as the default choices given $(\epsilon, c_1)$ as in HAMS-A
or given $(\epsilon, c_2)$ as in HAMS-B.
Whenever possible, it is helpful to exploit preconditioning, that is,
applying a linear transformation of $x$ based on an approximate variance matrix such that the target density can be roughly aligned with standard Gaussian.
Further discussion about preconditioning is provided in the Appendix.

\section{Matching with existing algorithms}
\label{sec:matching}

The HAMS class is related to a class of SDEs (\ref{eq:sde1}) including underdamped Langevin in Section~\ref{sec:sde} in the limit of a small step size.
In this section, we show that various popular algorithms for simulating Langevin dynamics (\ref{eq:Langevin}) can be put in the HAMS class
up to negligible differences which are of higher orders of the step size than the associated leading terms.
For the purpose of matching, all physical quantities such as the Boltzmann constant, the temperature and the mass are set to 1 in the existing algorithms.
In addition, to simplify the notation, the target density $\pi(x)$ or the potential function $U(x)$ is assumed to be univariate.

As the existing algorithms are conventionally used for molecular dynamics simulations without acceptance-rejection, we only discuss how the proposal $(x^*, u^*)$
is defined, given the current variables $(x_0,u_0)$. Nevertheless, as described in the Supplement (Section~\ref{sec:MA-algrithms}),
an acceptance-rejection step can be incorporated into these algorithms in the framework of generalized Metropolis--Hastings sampling \citep{Song2020}.
The resulting sampling algorithms are used in our numerical experiments (Section~\ref{sec:sim}).

\vspace{.1in}
\textbf{GJF, BAOAB, and IL.}\;
First, we study three relatively recent algorithms, where a single noise is used in each update.
The GJF algorithm \citep{Farago2013} is defined as follows:
\begin{align}
    x^* & = x_0 - \frac{\epsilon^2}{2+\eta\epsilon}\nabla U(x_0) +
    \frac{2\epsilon}{2+\eta\epsilon}  u_0  + \frac{\epsilon}{2+\eta\epsilon} W, \label{eq:GJF1} \\
    u^* & = \frac{2-\eta\epsilon}{2+\eta\epsilon} u_0 +
    \frac{\eta\epsilon^2 - 2\epsilon}{2(2+\eta\epsilon)} \nabla U(x_0) - \frac{\epsilon}{2}\nabla U(x^*)
    +\frac{2}{2 + \eta \epsilon} W, \label{eq:GJF2}
\end{align}
where $W \sim \mathcal N(0, 2\eta \epsilon)$.
Throughout, $\eta \geq 0$ is the friction coefficient playing the role of $\eta_2$ in (\ref{eq:Langevin}).
The BAOAB algorithm \citep{Leimkuhler2012} is given by
\begin{align}
 \tilde{u} & = u_0 - \frac{\epsilon}{2}\nabla U(x_0),   \quad
 \tilde{x}  = x_0 + \frac{\epsilon}{2}\tilde{u}, \label{eq:BAO1} \\
 \tilde{\tilde u} & = e^{-\eta\epsilon}\tilde{u} + \sqrt{1-e^{-2\eta\epsilon}} W,  \label{eq:BAO2} \\
 x^* & = \tilde{x} + \frac{\epsilon}{2}\tilde{\tilde u}, \quad
 u^*  = \tilde{\tilde u} - \frac{\epsilon}{2}\nabla U(x^*), \label{eq:BAO3}
\end{align}
where $W \sim \mathcal N(0,1)$.
The impulsive Langevin leapfrog (IL) algorithm \citep{Goga2012} is expressed in terms of half-step momentums, which are denoted as $u_{-\frac{1}{2}}$ and $u_{\frac{1}{2}}$. The
update is given as follows:
\begin{align}
    \tilde{u} & = u_{-\frac{1}{2}} - \epsilon \nabla U(x_0), \quad
    \tilde{\tilde u}  = - \tilde{c}\tilde{u} + \sqrt{\tilde{c}(2 - \tilde{c})} W, \label{eq:IL1} \\
    x^* & = x_0 + \epsilon\left(\tilde{u} + \frac{1}{2} \tilde{\tilde u}\right), \quad
    u_{\frac{1}{2}}  = \tilde{u} + \tilde{\tilde u}, \label{eq:IL2}
\end{align}
where $W $ is a $\mathcal N(0,1)$ noise and $0\leq \tilde{c} \leq 1$ represents the fraction of momentum lost due to friction with
$\tilde{c} = 1 - e^{-\eta \epsilon}$.

\begin{prop}
  \label{prop:matching}
  Suppose that we rescale the momentum in GJF and BAOAB by
\[
  u^* \gets \frac{\sqrt{4 - \epsilon^2}}{2} u^*, \quad u_0 \gets \frac{\sqrt{4 - \epsilon^2}}{2} u_0,
\]
and define the full-step momentum in IL by
\[
  u_0 = \frac{2}{\sqrt{4-\epsilon^2}}\left(u_{-\frac{1}{2}} - \frac{\epsilon}{2}\nabla U(x_0)\right),\quad u^* = \frac{2}{\sqrt{4-\epsilon^2}}\left(u_{\frac{1}{2}} - \frac{\epsilon}{2}\nabla U(x^*)\right).
\]
See the Appendix for explicit expressions. Then the following results hold.
\begin{itemize}\addtolength{\itemsep}{-.1in}
  \item BAOAB and IL are equivalent to each other.
  \item GJF, BAOAB and IL can be put exactly into the HAMS form (\ref{eq:proposal1-mat})--(\ref{eq:proposal3-mat}), with $(a_1,a_2,a_3,\phi)$ satisfying (\ref{eq:HAMSA-ab}) in HAMS-A except for
  a difference of the order $\mO(\epsilon^2)$ in $\phi$.
\end{itemize}
\end{prop}

From the proof of Proposition~\ref{prop:matching}, the choices of  $(a_1,a_2,a_3,\phi)$ used to match rescaled GJF, BAO, and IL with HAMS-A are of the order
\begin{align}
& a_1 = \frac{\epsilon^2}{2} (1- \frac{\eta \epsilon}{2}) + \mO (\epsilon^4), \quad
 a_2 = \epsilon (1- \frac{\eta \epsilon}{2}) + \mO (\epsilon^3) , \quad a_3 = 2 (1- \frac{\eta \epsilon}{2}) + \mO (\epsilon^2) ,  \label{eq:matching-choice1} \\
& \phi = \frac{\epsilon}{2} + \mO(\epsilon^2).   \label{eq:matching-choice2}
\end{align}
By comparison, the SDE parametrization (\ref{eq:HAMSA-param}) for HAMS-A with $\eta_2= \eta$ satisfies
\begin{align}
& a_1 = \frac{\epsilon^2}{2} + \mO (\epsilon^4), \quad
 a_2 = \epsilon (1- \frac{\eta \epsilon}{4}) + \mO (\epsilon^3) ,  \quad a_3 = 2 (1- \frac{\eta \epsilon}{2}) + \mO (\epsilon^2) , \label{eq:HAMSA-param-expan1} \\
& \phi = \frac{\epsilon}{2} + \mO(\epsilon^2).   \label{eq:HAMSA-param-expan2}
\end{align}
Interestingly,  the choices of $(a_1,a_2,a_3)$ in (\ref{eq:matching-choice1})
and (\ref{eq:HAMSA-param-expan1}) agree in the (first) leading terms, but not in the second leading terms.
This difference does not affect the convergence of all these algorithms to underdamped Langevin SDE (\ref{eq:Langevin}) as $\epsilon \to 0$.

\vspace{.1in}
\textbf{BP and VEC.}\; Next we study two algorithms where two noise variables are used in each update. The BP algorithm \citep{Bussi2007} is given by
\begin{align}
     u^+  & = \sqrt{c}u_0 + \sqrt{1-c} W_1 , \label{eq:BP1} \\
     \tilde{u} & = u^+ - \frac{\epsilon}{2} \nabla U(x_0), \quad x^* = x_0 + \epsilon\tilde{u}, \quad u^- = \tilde{u} - \frac{\epsilon}{2} \nabla U(x^*), \label{eq:BP2} \\
     u^* & = \sqrt{c} u^- + \sqrt{1-c} W_2, \label{eq:BP3}
\end{align}
where $0 \leq  c \leq 1$ is the amount of momentum carryover defined as $c=e^{-\eta \epsilon}, $ and $ W_1,W_2 \sim \mathcal N(0,1)$ independently.
The VEC integrator described by Equation 21 in \citet{Vanden2006} is
\begin{align}
  x^* & = x_0 - \frac{\epsilon^2}{2} \nabla U(x_0) + \frac{2\epsilon - \eta\epsilon^2}{2} u_0
  + \frac{\sqrt{2\eta}\epsilon^{3/2}}{2} W_1 + \frac{\sqrt{6\eta}\epsilon^{3/2}}{6} W_2,  \label{eq:Vanden1} \\
  u^* & = \left(1-\eta\epsilon + \frac{\eta^2\epsilon^2}{2}\right) u_0 + \frac{\eta\epsilon^2 - \epsilon}{2}\nabla U(x_0)
  -\frac{\epsilon}{2}\nabla U(x^*) + \frac{\sqrt{2\eta\epsilon}}{2}(2-\eta\epsilon) W_1 - \frac{\sqrt{6}}{6}(\eta\epsilon)^{3/2}W_2,  \label{eq:Vanden2}
\end{align}
where $W_1, W_2 \sim \mathcal N(0,1)$, independently.

\begin{prop}
  \label{prop:matching2}
If the coefficient $\epsilon/2$ is replaced by $\epsilon/(1 + \sqrt{1 - \epsilon^2})$ for  $\nabla U(x_0)$ and $\nabla U(x^*)$ in (\ref{eq:BP2}),
then BP can be put exactly into the HAMS form (\ref{eq:proposal1-mat})--(\ref{eq:proposal3-mat}), with the default choice $\phi$ in Proposition~\ref{prop:phichoice}.
If the coefficient $\frac{\eta\epsilon^2 - \epsilon}{2}$ is replaced by $\frac{\eta\epsilon^2 - \epsilon}{2} - \frac{\epsilon^3}{4}$ for $\nabla U(x_0)$ in (\ref{eq:Vanden2}),
then VEC can be matched with HAMS with default $\phi$, except for differences of order $\mO(\epsilon^2)$ in the $\phi$ choice and $\mO(\epsilon^3)$ in the variances and covariance of $(x^*,u^*)$ given $(x_0,u_0)$.
See the Appendix for explicit expressions of the modified algorithms.
\end{prop}

From the proof of Proposition~\ref{prop:matching2}, the choices of  $(a_1,a_2,a_3,\phi)$ used to match modified BP and VEC with HAMS
satisfy the same expansions as in (\ref{eq:matching-choice1})--(\ref{eq:matching-choice2}).
Even though two noise variables are used per iteration, the leading terms of $(a_1, a_2, a_3)$ satisfy $a_1 a_3 = a_2^2$, characteristic of HAMS-A, which
explicitly uses only one noise variable per iteration.

\vspace{.1in}
\textbf{Shifted HAMS.}\; For all the methods discussed so far, including HAMS, $x^*$ is determined using the
gradient  $\nabla U(x_0)$  evaluated at the current step. There exist other methods, however, which first update $x_0$ to
some intermediate value $\tilde{x}$, and then use $\nabla U(\tilde{x})$ in the expression of $x^*$. In an
attempt to match such methods, we introduce a variation of HAMS, called shifted HAMS, with the following update:
\begin{equation}
  \label{eq:sHAMS1}
  \begin{pmatrix}
    x^* \\
    u^*
  \end{pmatrix} =
  \begin{pmatrix}
    x_0 \\
    -u_0
  \end{pmatrix} - \tilde{A}
  \begin{pmatrix}
    \nabla U(\tilde{x}) \\
    -u_0
  \end{pmatrix}
  +   \begin{pmatrix}
    Z_0^{(1)} \\
    Z_0^{(2)}
  \end{pmatrix},
\end{equation}
where $\tilde{x} = x_0 + b u_0$, $(Z_0^{(1)}, Z_0^{(2)})^\T \sim \mathcal N(\bo, 2A - A^2)$, and
\[
A = \begin{pmatrix}
  a_1 & a_2 \\
  a_2 & a_3
\end{pmatrix}, \quad
\tilde{A} = A
\begin{pmatrix}
  1 & b \\
  0 & 1
\end{pmatrix} =
\begin{pmatrix}
  a_1 & ba_1 + a_2 \\
  a_2 & ba_2 + a_3
\end{pmatrix}.
\]
Here $b$ is a scalar tuning parameter involved to define the shifted value $\tilde x$, such that the gradient $\nabla U(\tilde x)$
is used in the update, instead of $\nabla U (x_0)$.
Taking $b=0$ in (\ref{eq:sHAMS1}) leads back to the original HAMS update (\ref{eq:proposal2-mat}) before the $u^*$ update.
The coefficient matrix $\tilde A$ in (\ref{eq:sHAMS1}) is derived to achieve the property that
when the target density $\pi(x)$ is standard Gaussian with $\nabla U(x) =x$, the update (\ref{eq:sHAMS1}) reduces to
the same VAR process (\ref{eq:VAR}) as the original HAMS.

\vspace{.1in}
\textbf{ABOBA, SPV, and Mannella's leapfrog.}\;
The ABOBA algorithm \citep{Leimkuhler2012} is defined by the following update
\begin{align}
  \tilde{x} & = x_0 + \frac{\epsilon}{2} u_0 ,  \label{eq:ABO1} \\
  \tilde{u}  &= u_0 - \frac{\epsilon}{2} \nabla U(\tilde{x}), \quad \tilde{\tilde u} = e^{-\eta\epsilon}\tilde{u} + \sqrt{1-e^{-2\eta\epsilon}} W,
  \quad  u^* = \tilde{\tilde u} -\frac{\epsilon}{2}\nabla U(\tilde{x}) ,  \label{eq:ABO2} \\
   x^* & = \tilde{x} + \frac{\epsilon}{2}u^*, \label{eq:ABO3}
\end{align}
where $W \sim \mathcal N(0,1)$. The update in the stochastic position Verlet algorithm (SPV) \citep{Melchionna2007} is given by
\begin{align}
    \tilde{x} &= x_0 + \frac{\epsilon}{2} u_0, \label{eq:SPV1} \\
    u^* & = e^{-\eta\epsilon}u_0 - \frac{1-e^{-\eta\epsilon}}{\eta}\nabla U(\tilde{x}) + \sqrt{1 - e^{-2\eta\epsilon}} W, \label{eq:SPV2} \\
    x^* &= \tilde{x} + \frac{\epsilon}{2}u^*, \label{eq:SPV3}
\end{align}
where $W \sim \mathcal N(0,1)$. Mannella's leapfrog \citep{Mannella2004} is given by
\begin{align}
  \tilde{x} & = x_0 + \frac{\epsilon}{2} u_0, \label{eq:M-leapfrog1}\\
  u^* & = c_2(c_1u_0 - \epsilon\nabla U(\tilde{x}) + \sqrt{2\eta} W),  \label{eq:M-leapfrog2}\\
  x^* & = \tilde{x} + \frac{\epsilon}{2} u^*, \label{eq:M-leapfrog3}
\end{align}
where $W\sim \mathcal N(0,\epsilon)$, $c_1 = \frac{2-\eta\epsilon}{2}$, and $ c_2 = \frac{2}{2+\eta\epsilon}$.

\begin{prop}
  \label{prop:matching3}
Suppose that the coefficient $\epsilon/2$  for $u_0$ and $u^*$  is replaced by $\epsilon/(1 + \sqrt{1 - \epsilon^2})$ in (\ref{eq:ABO1}) and (\ref{eq:ABO3}) for ABOBA
and in (\ref{eq:M-leapfrog1}) and (\ref{eq:M-leapfrog3}) for Mannella's leapfrog, and replaced by
$$
b = \frac{2(1-e^{-\eta\epsilon}) /\eta} { 1 + e^{-\eta\epsilon} + \sqrt{(1+e^{-\eta\epsilon})^2 - 4(1-e^{-\eta\epsilon})^2 / \eta^2 } }
$$
in (\ref{eq:SPV1}) and (\ref{eq:SPV3}) for SPV.
Then ABOBA, SPV, and Mannella's leapfrog can be put into the form of (\ref{eq:sHAMS1}) in shifted HAMS with suitable choices $(a_1,a_2,a_3,b)$, except for
differences of order $\mO(\epsilon^3)$ in the variances and covariance of $(x^*,u^*)$ given $(x_0,u_0)$.
\end{prop}

From the proof of Proposition~\ref{prop:matching3}, the choices of $(a_1,a_2,a_3)$ used to match the three algorithms with shifted HAMS are
the same as in (\ref{eq:matching-choice1}) for matching other algorithms with HAMS. Moreover, the two choices of $b$ stated in Proposition~\ref{prop:matching3}
are both of the order
\begin{align}
b = \frac{ \epsilon}{2} + \mO (\epsilon^3), \label{eq:matching-choice3}
\end{align}
which shares the same leading term, $\epsilon/2$, as the $\phi$ choice in (\ref{eq:matching-choice2}), although the $\epsilon^2$ term vanishes in (\ref{eq:matching-choice3}).
These observations shed interesting light on the relationship among the existing algorithms, in addition to their connections with HAMS.

\section{Numerical Experiments} \label{sec:sim}
We conduct numerical experiments to compare HAMS-A, HAMS-B, HAMS with $\eta_1 = k\epsilon$, henceforth labeled as HAMS-$k$, for $k = 1,2,3$, and Metropolized versions
of BAOAB, ABOBA and BP, which are derived in the framework of generalized Metropolis--Hastings sampling as described in the Supplement, Section~\ref{sec:MA-algrithms}.
The target densities include a one-dimensional double well potential, and two higher-dimensional latent variable distributions.

\subsection{Sampling from a double well} \label{sec:sim-dw}
Consider the one-dimensional double well  as in \cite{Leimkuhler2013},
\[
\pi(x)\propto \exp(-T^{-1} U(x)),\quad U(x)  = (x^2 - 1)^2 + x,
\]
where both the temperature $T$ and the Boltzmann's constant are set to $1$. See the Supplement for a plot of $\pi(x)$.
For a unit mass, the augmented density with the momentum $u$ is
\begin{equation}
  \label{eq:temp1}
  \pi(x, u) \propto \exp\left\{- T^{-1}\left(U(x) + \frac{1}{2}u^2\right)\right\}.
\end{equation}
We compare different algorithms for sampling from the augmented density, where acceptance-rejection is included at each iteration.
In contrast, the experiment in \cite{Leimkuhler2013} is conducted in the molecular dynamics setting where
every proposal is accepted. In our setting, the estimation error is in general non-monotonic in the step size.

We follow \cite{Leimkuhler2013} and set the friction in underdamped Langevin to one.
Thus for HAMS-A and HAMS-1/2/3, we fix $\eta_2 = 1$. For BAOAB, ABOBA and BP, we fix $\eta = 1$.
For HAMS-B, which is associated with SDEs with fixed $\eta_2=0$ (Section~\ref{sec:sde}), we set $\eta_1 = 1$ to reciprocate.
We use $8$ different step sizes starting from $\epsilon = 0.04$ and increase by $0.04$ until $\epsilon = 0.32$. For each $\epsilon$, we collect $10000$ draws and repeat this
process $3000$ times. The starting values of $x$ and $u$ are randomly drawn from Uniform$[-1,1]$.

We assess the performance using the accuracy in temperature estimates.
In fact, the temperature $T$ can be equivalently expressed as the configurational temperatures,
\begin{equation}
  \label{eq:temp2}
  T_{C1} = \EE[x\cdot \nabla U(x)],\qquad T_{C2} = \frac{\EE[(\nabla U(x))^2]}{\EE[\nabla^2 U(x)]},
\end{equation}
or as the kinetic temperature,
\begin{equation}
  \label{eq:temp3}
  T_{K} = \EE[u^2].
\end{equation}
The expression $T_{C1}$ is used in \cite{Leimkuhler2013}, whereas $T_{C2}$ is used in \cite{Farago2019}.
It can be directly shown that the theoretical values of these expressions are the same as $T$, $T_{C1} = T_{C2} = T_{K} = T$ (see Supplement Section~\ref{sec:sim-detail-DW}).
However due to sampling errors, the empirical estimates of these temperatures can be different. We use root mean squared errors of $T_{C1}, T_{C2}$
and $T_{K}$ from repeated experiments as metrics.
In the Supplement, we also report density estimation and details of error calculation.

\begin{figure}[!tph]  
  \centering
  \begin{subfigure}[b]{.48\linewidth}
      \centering
      \includegraphics[width = 0.82\textwidth]{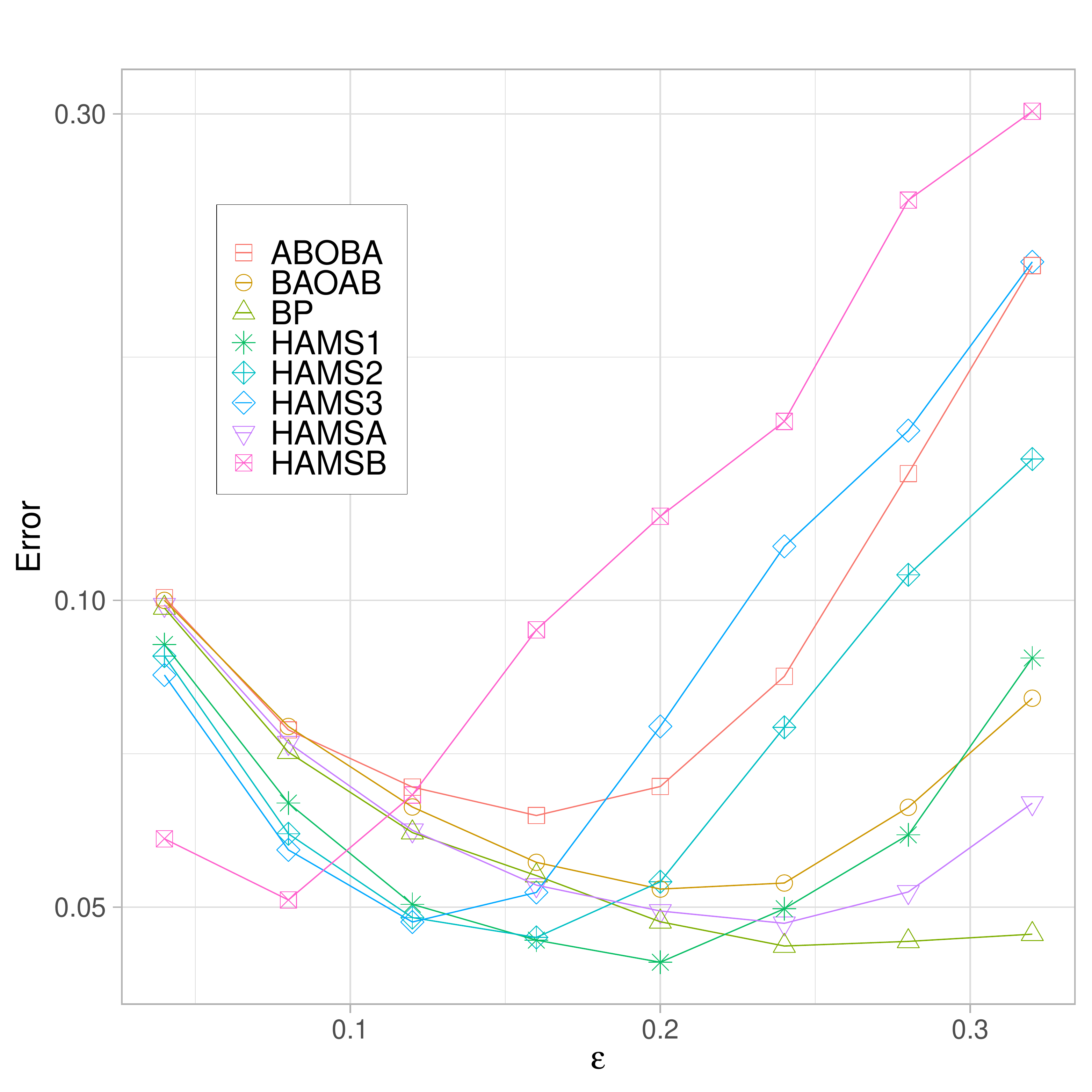}
      \caption{$T_{C1}$}\label{fig:DWTc1}
  \end{subfigure}
      \hfill
  \begin{subfigure}[b]{.48\linewidth}
      \centering
      \includegraphics[width = 0.82\textwidth]{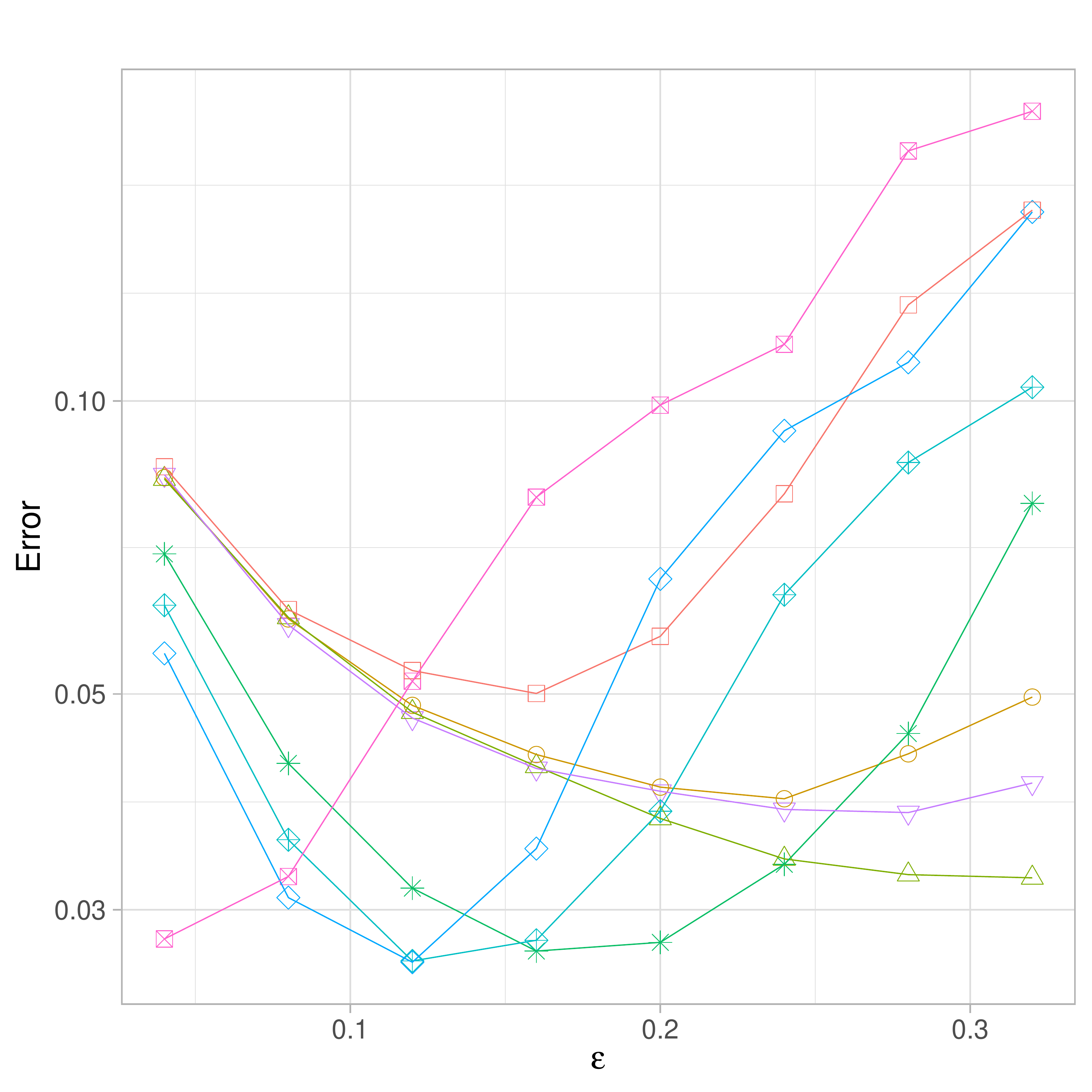}
      \caption{$T_{C2}$}\label{fig:DWTc2}
  \end{subfigure}
  \begin{subfigure}[b]{.48\linewidth}
    \centering
    \includegraphics[width = 0.82\textwidth]{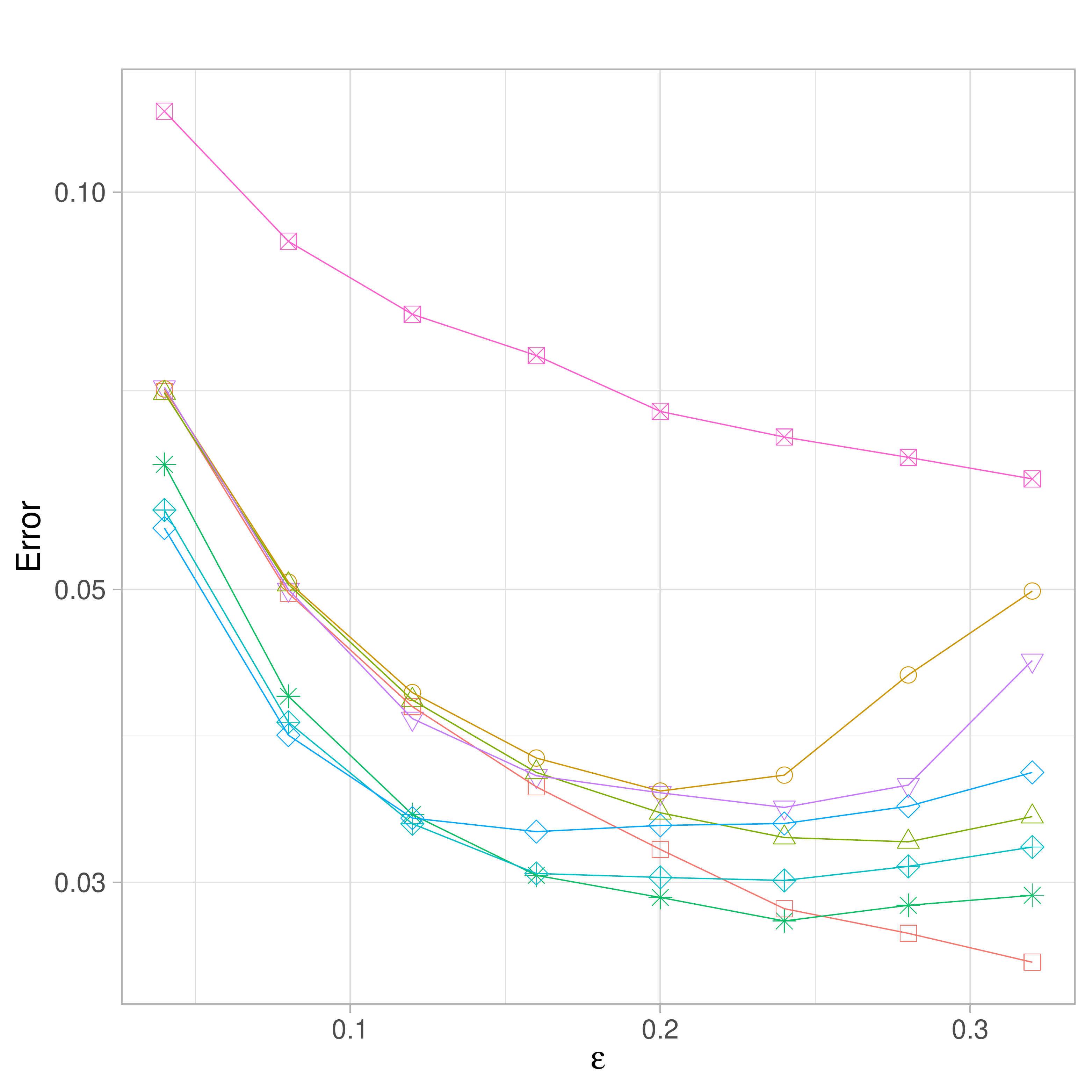}
    \caption{$T_K$}\label{fig:DWTk}
\end{subfigure}
    \hfill
\begin{subfigure}[b]{.48\linewidth}
    \centering
    \includegraphics[width = 0.82\textwidth]{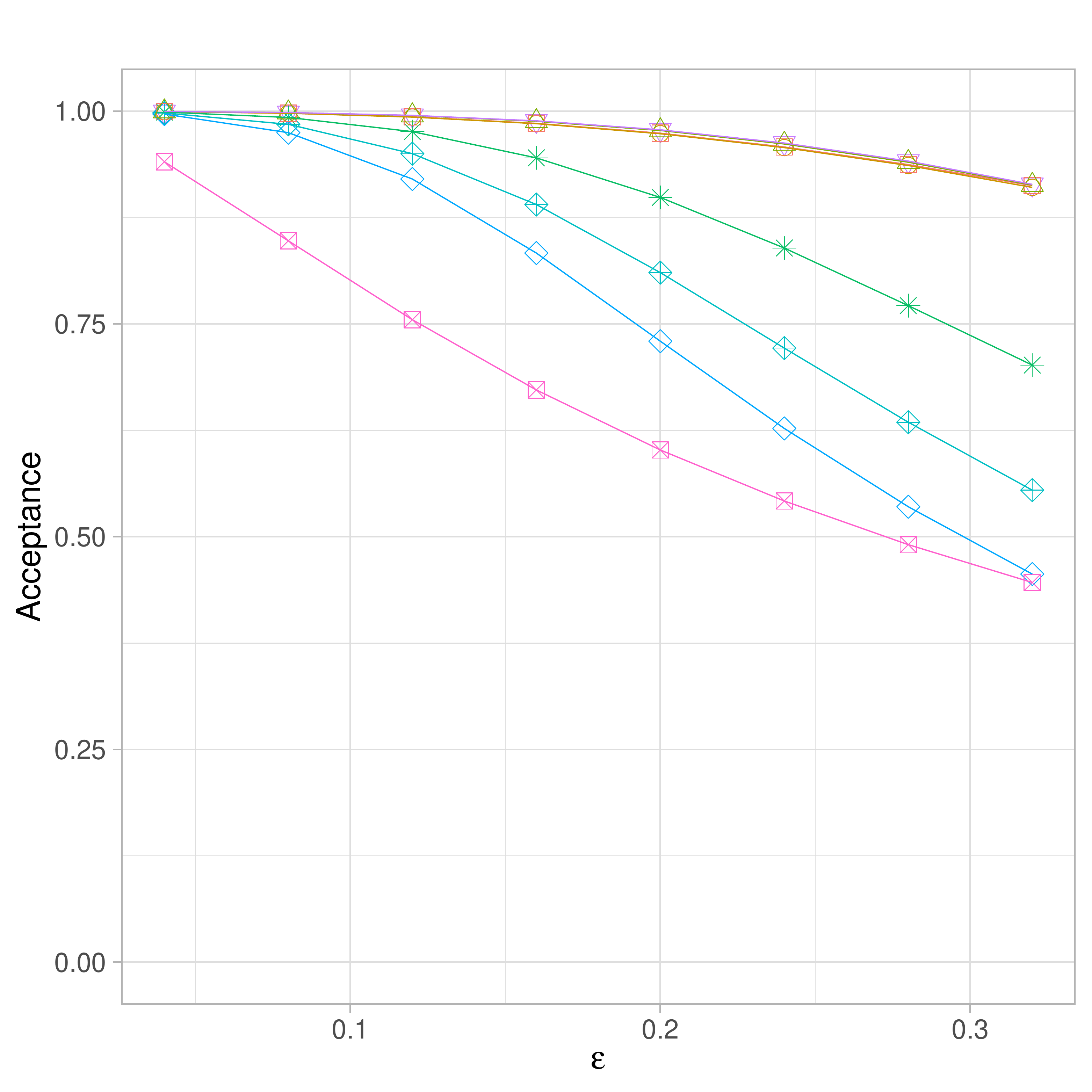}
    \caption{Acceptance rates}\label{fig:DWAccept}
\end{subfigure}
 \vspace{-.2in}  
  \RawCaption{\caption{Root mean squared errors in configurational temperatures and kinetic temperature,
  average acceptance rates for the double well. Results are based on $3000$ repeated experiments.}
  \label{fig:DWError}}
\end{figure}

In Figure~\ref{fig:DWError}, the errors in $T_{C1},T_{C2}$ and $T_{K}$ are plotted on the log scale. First, we examine estimates of the configurational temperatures.
There appears to be three groups among the algorithms studied.
The first group is HAMS-1/2/3, which lead to smallest errors in both $T_{C1}$ and $T_{C2}$ among all algorithms, when $\epsilon \le 0.16$.
The performance of HAMS-$k$ improves as $k$ increases from 1 to 3, when $\epsilon$ is small.
In the second group, the error curves of HAMS-A, BAOAB and BP are comparable and those of ABOBA consistently higher.
For both $T_{C1}$ and $T_{C2}$, HAMS-B, in its own group, is the best for the smallest $\epsilon$, but as $\epsilon$ increases its performance quickly deteriorates.
Over the whole range of $\epsilon$, HAMS-1 has the smallest $T_{C1}$ error, whereas both HAMS-2 and HAMS-3 reach the smallest error in $T_{C2}$.

For the kinetic temperature $T_K$, the same three groups of algorithms emerge as above.
When $\epsilon\leq 0.16$, we see two groups each with comparable performance: the first group is HAMS-1/2/3, which outperform the second group including HAMS-A, BAOBA, ABOBA and BP.
As $\epsilon$ increases beyond $0.2$ the two groups diverge with ABOBA achieving the smallest error overall while BAOAB producing larger errors.
In its own group, HAMS-B leads to large errors in $T_K$, which are decreasing as $\epsilon$ increases in the range studied.

We also present the average acceptance rates in Figure~\ref{fig:DWError}.
The algorithms can also be divided into three groups as above.
For the group HAMS-A, BAOAB, ABOBA and BP, which are directly related to underdamped Langevin,
the acceptance rates are relatively high across the range of $\epsilon$. For a fixed $\epsilon$, when $k$ increases (with $k=0$ corresponding to HAMS-A),
the acceptance rates of HAMS-$k$ drops, which is consistent with the discussion in Section~\ref{sec:accept}. When achieving the smallest $T_{C1}$ and $T_{C2}$ errors as remarked above, HAMS-1/2/3 have much lower acceptance rates compared with the HAMS-A group at the same step sizes.
The step sizes leading to the best accuracy in $T_{C1}$ and $T_{C2}$ are higher within the HAMS-A group than within HAMS-1/2/3.
This illustrates interesting behavior of HAMS-$k$, distinct from existing algorithms purely based on underdamped Langevin.

\subsection{Latent variable sampling} \label{sec:latent}

We compare the methods by sampling latent variables in a stochastic volatility model and  a log-Gaussian Cox model. In order to improve sampling efficiency, we perform preconditioning on the target densities,
as described in the Appendix for HAMS and in the Supplement for other algorithms studied. This allows us to employ the default tuning suggested by Propositions~\ref{prop:eigen1} and \ref{prop:eigen1-b}. Consequently each method only depends on a single step size parameter $\epsilon$. We tune $\epsilon$ during a burn-in period to achieve roughly $70\%$ acceptance rates for all methods. All simulation details are provided in Supplement Section~\ref{sec:sim-detail}.\par

To evaluate MCMC samples, a useful metric is the effective sample size,
$\mbox{ESS} = n /\{ 1+2\sum_{l=1}^{\infty}\rho(l)\}$,
where $n$ is the total number of draws and $\rho(l)$ is the lag-$l$ autocorrelation.
We report two estimators of ESS which are both suitable for irreversible Markov chains.
The first one is the Bartlett window estimator (labelled as ESS$_1$): \vspace{-.05in}
\begin{equation} \label{eq:ess}
\mbox{ESS}_1 = \frac{n}{1+2\sum_{l=1}^L \left(1-\frac{l}{L}\right)\rho(l)},
\end{equation}
where the cutoff value $L$ is a large number (taken to be $3000$ in our results). The second one (labelled as ESS$_2$) is based
on the within and between variances from multiple chains in repeated simulations. Suppose that we have $m$ Markov chains
each with $n$ draws, denoted as $\{x_{ij}: i = 1,\ldots,n, j = 1,\ldots,m\}$. Then ESS can be estimated by
\begin{equation} \label{eq:ess2}
  \mbox{ESS}_2 = n\,\frac{W}{B},\quad  W = \frac{1}{m(n-1)}\sum_{i,j} (x_{ij} - \bar{x}_{. j})^2,\quad  B = \frac{n}{m-1}\sum_{j}(\bar{x}_{. j} - \bar{x})^2,
\end{equation}
where $\bar{x}_{. j} =n^{-1} \sum_{i=1}^n x_{ij}$ and $\bar{x} = m^{-1} \sum_{j=1}^m \bar{x}_{. j}$. In fact, $B/n$ is an estimator of the variance of the average of $n$ draws, whereas $W$ is an estimator of the marginal variance of $x$.
For relatively large $m$ ($50$ in our experiments), the estimator $\mbox{ESS}_2$ can be more reliable than $\mbox{ESS}_1$ based on within-chain autocorrelations in directly measuring
consistency between repeated simulations.
Both ESS estimators are computed from each coordinate for a multi-dimensional distribution.
Following \cite{Girolami2011}, we report the minimum ESS over all coordinates, adjusted by runtime,
as a measure of computational efficiency.\par

\vspace{.1in}
\textbf{Stochastic volatility model.} First, consider a stochastic volatility model studied in \cite{Kim1998},
where latent volatilities are generated as
\begin{equation}
  \label{eq:sveq1}
x_t = \varphi x_{t-1} + \theta_t, \quad \theta_t \sim\mathcal N(0,\sigma^2),\quad t= 2,3,...,T ,
\end{equation}
with $x_1 \sim\mathcal N(0,\sigma^2/(1-\varphi^2))$, and the observations are generated as
\begin{equation}
  \label{eq:sveq2}
  y_t = z_t \beta \exp(x_t/2),\quad z_t\sim\mathcal N(0,1),\qquad t = 1,...,T.
\end{equation}

\begin{table}[h]
  \caption{Runtime and ESS comparison for sampling latent variables in the stochastic volatility model.
  Results are averaged over 50 repetitions. \label{tab:sv}}
\begin{tabular}{|cccccc|}
\hline
Method & Time (s) & \begin{tabular}[c]{@{}c@{}}ESS$_1$\\[-.1in] (min, median, max)\end{tabular}  & $\frac{\mbox{minESS}_1}{\mbox{Time}}$& \begin{tabular}[c]{@{}c@{}}ESS$_2$\\[-.1in] (min, median, max)\end{tabular}  & $\frac{\mbox{minESS}_2}{\mbox{Time}}$ \\ \hline
HAMS-A & 33.0     & (2000, 3728, 7034) & 60.56  & (563, 1093, 2619)    & 17.05  \\
HAMS-1 & 32.1     & (2117, 3461, 6349) & 65.99  & (505, 1032, 2101)    & 15.73  \\
HAMS-2 & 32.2     & (1936, 3276, 5754) & 60.05  & (496, 1029, 2247)    & 15.37   \\
HAMS-3 & 32.3     & (2199, 3221, 6014) & 68.11  & (461, 988,  2301)    & 14.27   \\
HAMS-B & 33.4     & (2301, 3487, 6890) & 68.84  & (501, 1058, 2997)    & 14.99    \\
BAOAB  & 33.8     & (466,  801,  1188) & 13.79  & (128, 235,  481)     & 3.81   \\
ABOBA  & 34.1     & (443,  756,  1143) & 13.00  & (132, 224,  538)     & 3.88   \\
BP     & 32.8     & (667,  1050, 1624) & 20.31  & (141, 318,  709)     & 4.29   \\ \hline
\end{tabular}
\end{table}

Let $\bx = (x_1,...,x_T)^\T $ and $ \by = (y_1,...,y_T)^\T $.
We generate $T = 1000$ observations from (\ref{eq:sveq1})--(\ref{eq:sveq2}) using parameter values $\beta = 0.65, \sigma = 0.15$ and $\varphi = 0.98$.
We fix $\by$ and the parameters, and then sample latent variables from $p(\bx|\by, \beta,\sigma,\varphi)$. See Supplement
for expressions of gradients and preconditioning matrices used. All algorithms are run for $5000$ burn-in iterations,
and then $5000$ draws are collected. Initial values of latent variables are drawn from standard normal distribution.
The simulation process is repeated for $50$ times.\par

Table~\ref{tab:sv} shows the runtime and ESS comparison. In terms of $\mbox{ESS}_1$, HAMS-B is the best, followed closely by HAMS-3 and then HAMS-1, which
are slightly better than HAMS-A and HAMS-2. On the other hand, in terms of $\mbox{ESS}_2$, HAMS-A clearly leads all other methods. We also observe
that the performance of HAMS-k improves as $k$ decreases, whereas HAMS-B is in between HAMS-2 and HAMS-3. Using either ESS metric, we see that all HAMS
variants are superior to BAOAB, ABOBA and BP.\par

\begin{figure}[t] 
  \begin{center}
  \includegraphics[height = 0.5\textheight]{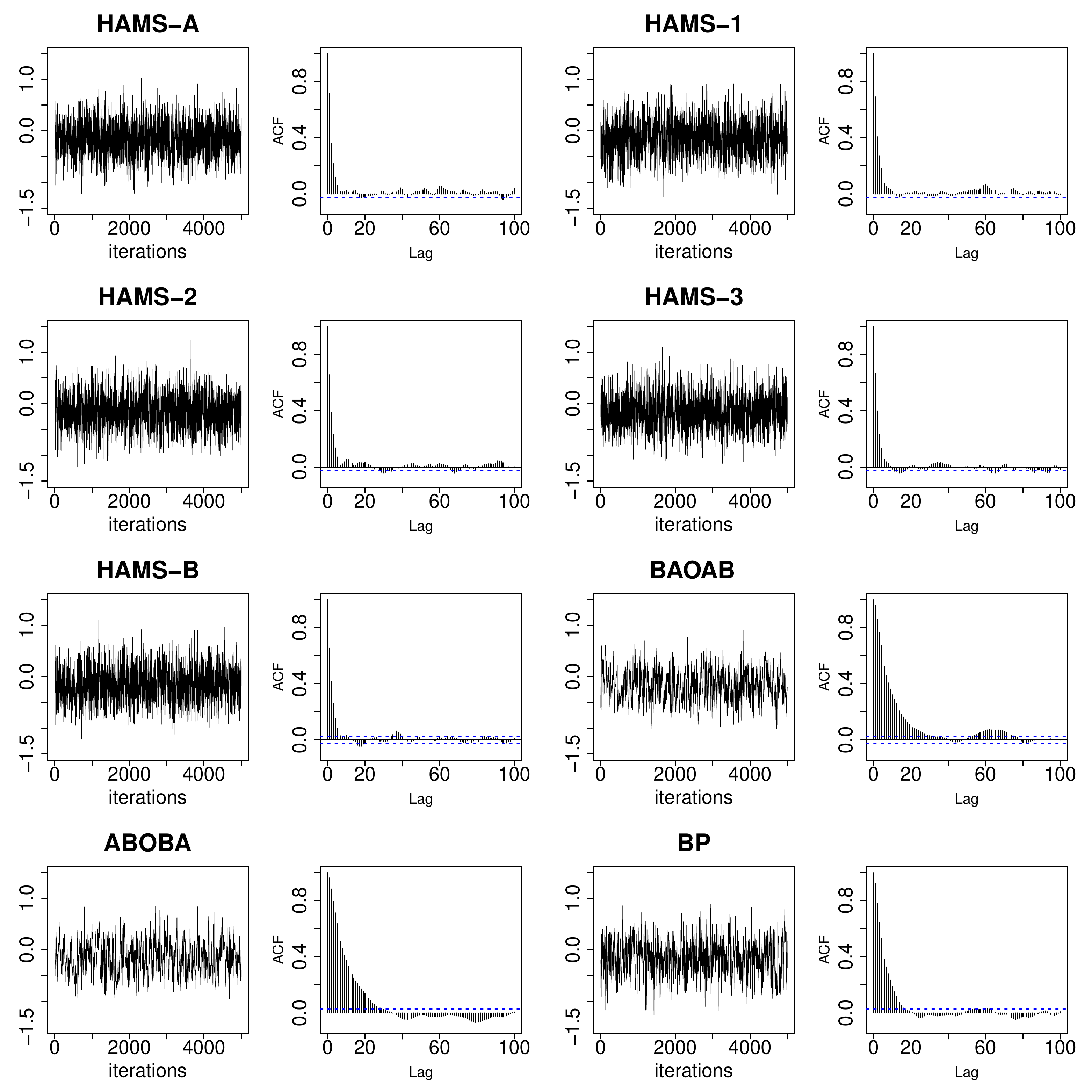}
  \end{center}
  \caption{Trace and ACF plots of a latent variable from an individual run
  for sampling latent variables in the stochastic volatility model.}
  \label{fig:svtrace} \vspace{-.1in}
\end{figure}

\begin{figure}[H] 
  \begin{center}
  \includegraphics[height = 0.3\textheight]{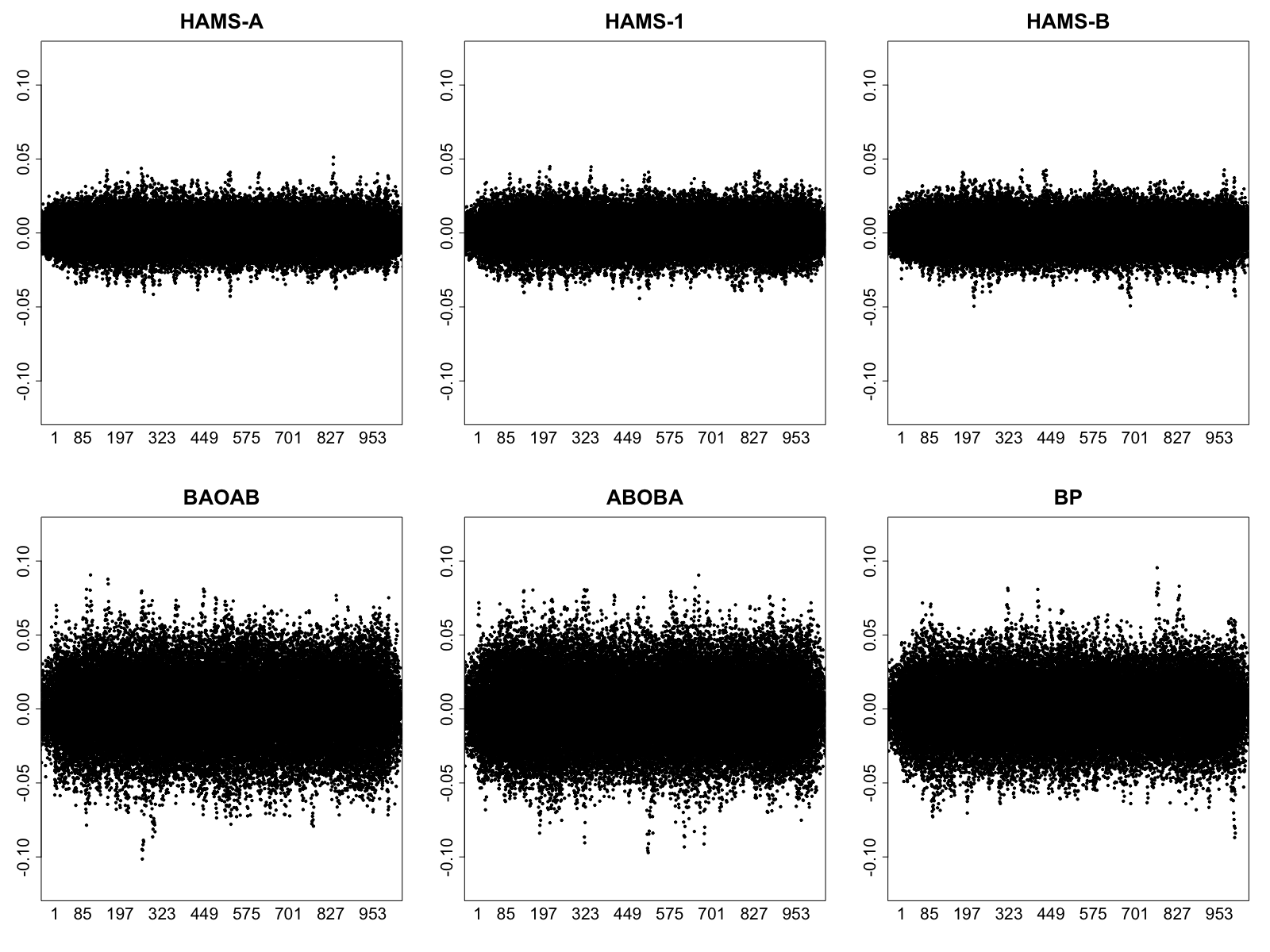}
  \end{center}
  \caption{Centered plots of sample means of all 1000 latent variables over $50$ repetitions
  for sampling latent variables in the stochastic volatility model.}
  \label{fig:svscatter} \vspace{-.1in}
\end{figure}

Trace plots in Figure~\ref{fig:svtrace} show that HAMS methods have much better mixing than the rest. The average sample means of latent variables are similar
across all methods, as shown by Figure~\ref{fig:svmeanvar} in the Supplement. Hence it is more informative to compare the variation among repeated experiments.
Figure~\ref{fig:svscatter} shows the sample means of latent variables after centering.
A thinner spread indicates better consistency.
We see that HAMS-A, HAMS-1 and HAMS-B have comparable spread, while BAOAB, ABOBA and BP show much larger variation.
The performance of HAMS-2/3 (omitted in Figure~\ref{fig:svscatter}) is similar to that of HAMS-1.\par

\vspace{.1in}
\textbf{Log-Gaussian Cox model.} Next consider a log-Gaussian Cox model, where the latent variables $\bx = (x_{ij})_{i,j = 1,...,m}$ are associated with an $m\times m$ grid
\citep{Christensen2005}. Assume that $x_{ij}$'s are normal with means $0$ and a covariance function
$ C[(i,j),(i',j')] = \sigma^2\exp(-\sqrt{(i-i')^2 + (j - j')^2}/(m\beta))$. By abuse of notation, we denote
$\bx \sim \mathcal N(\bo, C)$ of dimension $n=m^2$. The observations $(y_{ij})_{i,j = 1,...,m}$
are independently Poisson, where the mean of $y_{ij}$ is
$\lambda_{ij} = n^{-1}\exp(x_{ij} + \mu)$, with $\mu$ treated as known.
The density of latent variables given response $\by$ is
\begin{equation}
  \label{eq:coxeq1}
  p(\bx|\by,\beta,\sigma^2,\mu ) \propto \exp\left\{-\frac{1}{2}x^\T C^{-1}x\right\} \exp\left\{\sum_{i,j}(y_{ij}(x_{ij}+\mu) -\lambda_{ij} )\right\}.
\end{equation}

\begin{table}[t]
  \caption{Runtime and ESS comparison for sampling latent variables in the log-Gaussian Cox model.
  Results are averaged over 50 repetitions. \label{tab:cox}}
\begin{tabular}{|cccccc|}
\hline
Method & Time (s) & \begin{tabular}[c]{@{}c@{}}ESS$_1$\\[-.1in] (min, median, max)\end{tabular}  & $\frac{\mbox{minESS}_1}{\mbox{Time}}$& \begin{tabular}[c]{@{}c@{}}ESS$_2$\\[-.1in] (min, median, max)\end{tabular}  & $\frac{\mbox{minESS}_2}{\mbox{Time}}$ \\ \hline
HAMS-A & 528.4     & (968, 1467, 4607) & 1.83  & (218, 444, 1406)   & 0.41  \\
HAMS-1 & 530.1     & (665, 1142, 3118) & 1.25  & (175, 344, 937)    & 0.33  \\
HAMS-2 & 530.3     & (700, 1080, 2740) & 1.32  & (174, 323, 1011)   & 0.33   \\
HAMS-3 & 530.3     & (656, 1019, 2546) & 1.24  & (159, 308, 800)    & 0.30   \\
HAMS-B & 529.8     & (606, 938,  2680) & 1.14  & (142, 279, 804)    & 0.27    \\
BAOAB  & 530.5     & (316, 494,  972)  & 0.60  & (68,  144, 401)    & 0.13   \\
ABOBA  & 536.2     & (324, 478,  1080) & 0.60  & (76,  143, 348)    & 0.14   \\
BP     & 529.4     & (348, 555,  1215) & 0.66  & (75,  166, 420)    & 0.14   \\ \hline
\end{tabular}
\end{table}

We take $m = 64$ and generate $n = 64^2 = 4096$ observations using the parameter values $\sigma^2 = 1.91$, $\beta = 1/33$ and $\mu = \log(126) - 0.955$.
We fix the simulated $\by$ values and the parameters, and then sample latent variables $\bx$ from the density (\ref{eq:coxeq1}). All algorithms are run
for $5000$ burn-in iterations, and then $5000$ draws are collected. We initialize the latent variables from a standard normal distribution.
The simulation process is repeated for $50$ times.\par

From Table~\ref{tab:cox}, we see that for the Cox model, HAMS-A is the best in both $\mbox{ESS}_1$ and $\mbox{ESS}_2$. In terms of $\mbox{ESS}_2$, we observe that
similarly as in Table~\ref{tab:sv}, HAMS-k becomes worse as $k$ increases. However, this does not hold for $\mbox{ESS}_1$, where HAMS-2 is slightly better than HAMS-1. Among the
HAMS methods, HAMS-B has the lowest ESS in this case. Similarly to the stochastic volatility results, all three non-HAMS methods show inferior performance.\par

\begin{figure}[H] 
  \begin{center}
  \includegraphics[height = 0.5\textheight]{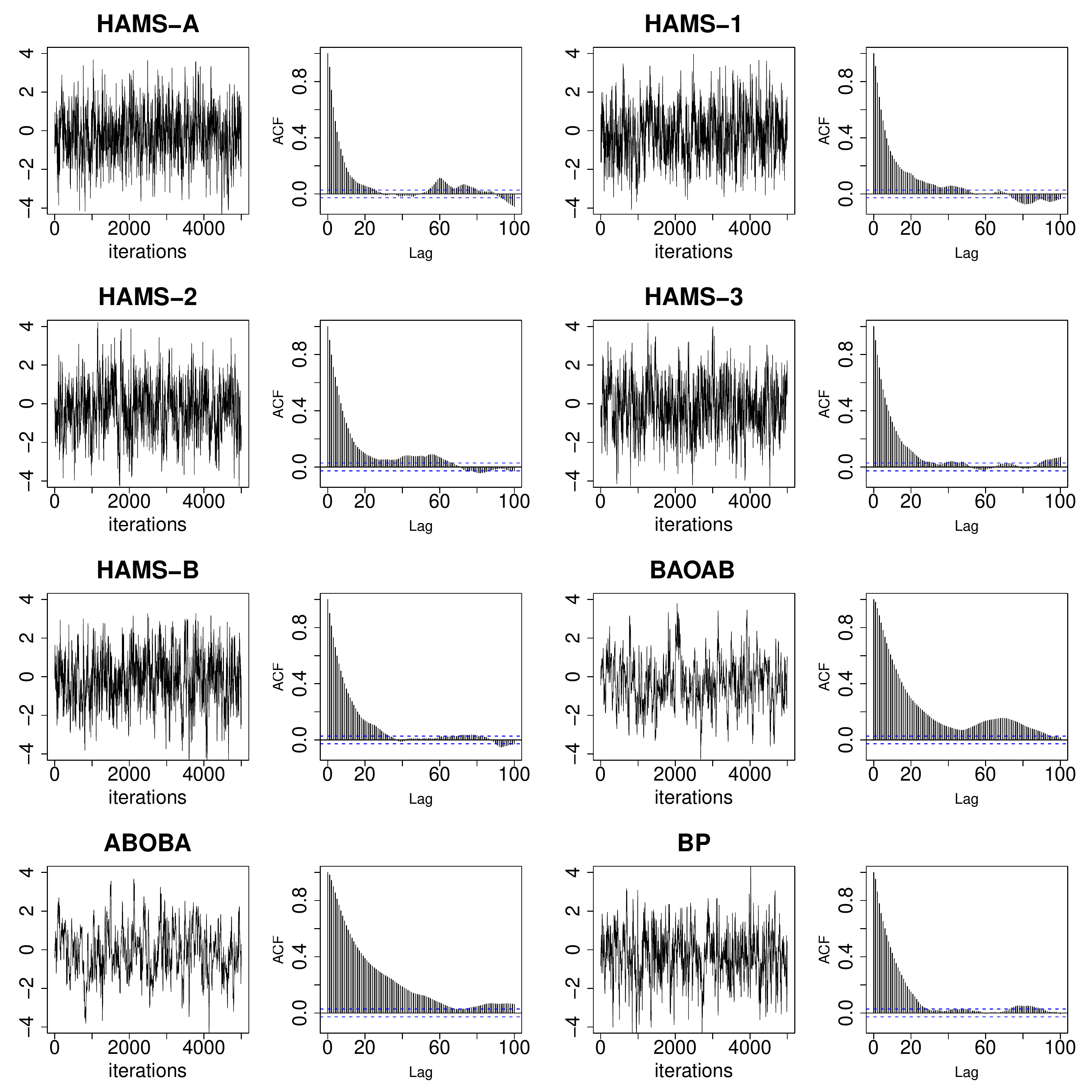}
  \end{center}
  \caption{Trace and ACF plots of a latent variable from an individual run
  for sampling latent variables in the log-Gaussian Cox model.}
  \label{fig:coxtrace}  \vspace{-.15in}
\end{figure}

\begin{figure}[H] 
  \begin{center}
  \includegraphics[height = 0.3\textheight]{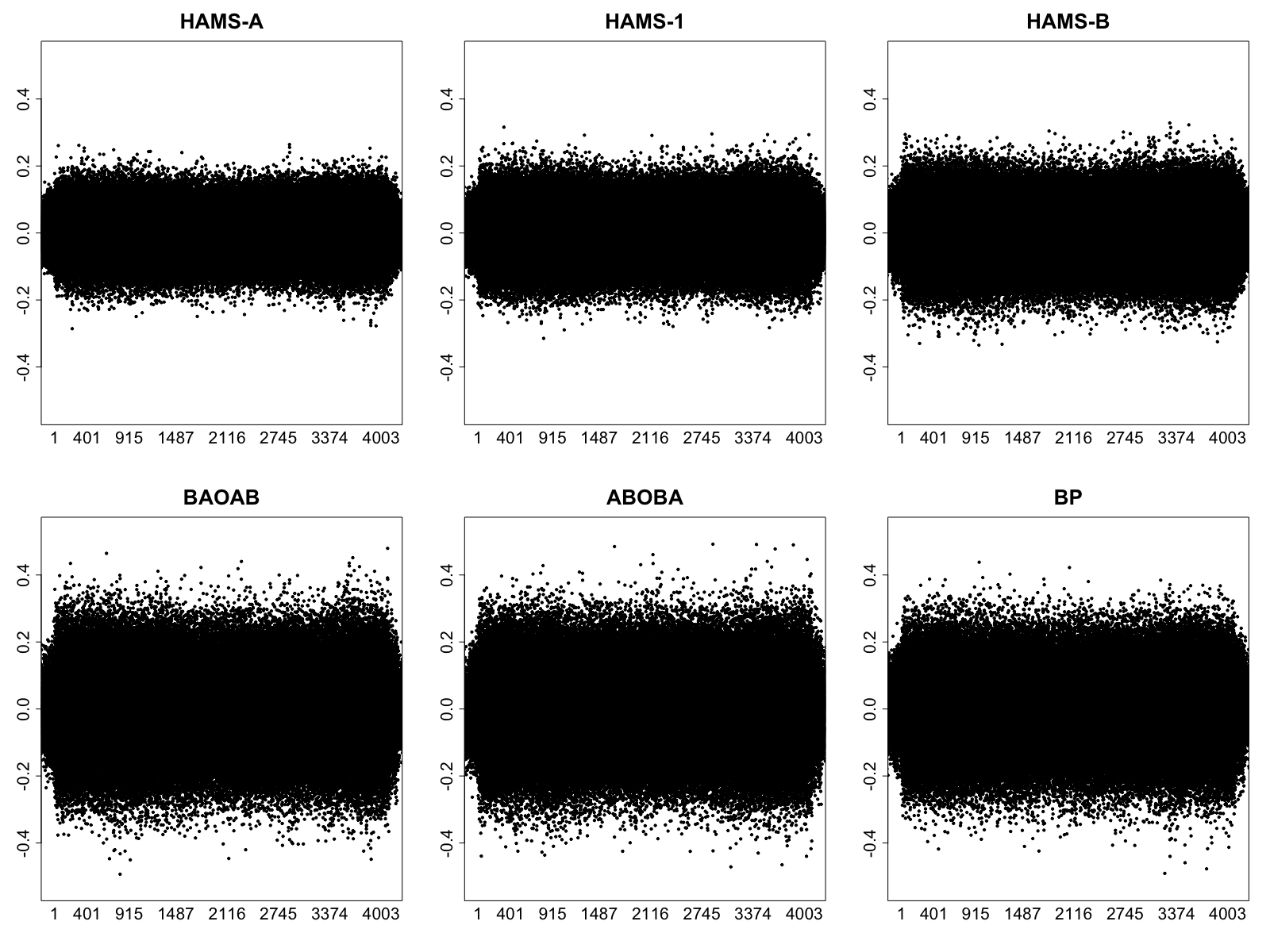}
  \end{center}
  \caption{Centered plots of sample means of all 4096 latent variables over $50$ repetitions
  for sampling latent variables in the stochastic volatility model.}
  \label{fig:coxscatter} \vspace{-.15in}
\end{figure}

According to trace plots in Figure~\ref{fig:coxtrace}, HAMS methods mix better than the other methods. Furthermore, the ACF of HAMS-A has the fastest decay.
The average sample means of latent variables are also aligned across different methods (see Figure~\ref{fig:coxmeanvar} in the Supplement).
From the centered sample means in Figure~\ref{fig:coxscatter}, we see that HAMS-A has a slight advantage over HAMS-1 and HAMS-B.
The three remaining methods are clearly less consistent than HAMS. The performance of HAMS-2/3 (omitted in Figure~\ref{fig:coxscatter}) is similar to that of HAMS-1.

\section{Conclusion}

We investigate HAMS in several directions, including deriving SDE limits,
studying theoretical properties under the univariate Gaussian setting, and establishing connections to existing algorithms for Langevin dynamics.
Recently, convergence properties of underdamped Langevin dynamics and Euler's discretizations have been obtained under fairly general settings
\citep{Cheng2018, Dalalyan2020,Cao2020}.
It is interesting to study the impact of using improved discretizations including HAMS and those in Section \ref{sec:matching} and that of incorporating acceptance-rejection.
Moreover, investigation of HAMS and related methods is desired in simulation settings more diverse than our current experiments.

\section{Appendix} \label{sec:appendix}

  \begin{algorithm}[!tph] 
    \SetAlgoNoLine
    Initialize $x_0,u_0,\hat{x}_0 = L^\T x_0$ and $\nabla U(\hat{x}_0) = L^{-1}\nabla U(x_0)$.\\
     \For{ $t = 0,1,2,..., N_{iter}$}{
       Sample $w \sim \text{Uniform} [0,1]$ and $ \begin{pmatrix}
         Z^{(1)}_0 \\
         Z^{(2)}_0
       \end{pmatrix} \sim \mathcal N(\bo,2A - A^2)$ \\
       $\xi = a_2 u_t + Z^{(1)}_0, \quad \hat{x}^* = \hat{x}_t - a_1 \nabla U(\hat{x}_t) + \xi$\\
       Propose $x^* = (L^\T )^{-1}\hat{x}^*$\\
       $\nabla U(\hat{x}^*) = L^{-1} \nabla U(x^*), \quad \tilde{\xi} = \nabla U(\hat{x}^*) + \nabla U(\hat{x}_t)$\\
       $\rho = \exp\left\{U(x_t) - U(x^*) + \frac{1}{2-a_1}(\tilde{\xi})^\T (\xi - \frac{a_1}{2}\tilde{\xi} )\right\}$ \\
      \eIf{$w < \min(1,\rho)$}{
        $x_{t+1} = x^*$,\quad $\hat{x}_{t+1} = \hat{x}^*$,\quad $\nabla U(\hat{x}_{t+1}) =  \nabla U(\hat{x}^*) $\qquad\# Accept \\
        $u_{t+1} = \frac{a_1 + a_2^2 + 2a_3 - a_1a_3 - 2}{2 - a_1}u_{t} - \frac{a_2}{2-a_1}\tilde{\xi} + \frac{a_2}{2-a_1}Z^{(1)}_0 + Z^{(2)}_0 $
       }{
        $x_{t+1} = x_t, u_{t+1} = -u_t, \hat{x}_{t+1} = \hat{x}_t,\nabla U(\hat{x}_{t+1}) = \nabla U(\hat{x}_t) $ \qquad\# Reject
      }
     }
     \caption{Preconditioned HAMS}   \label{alg:HAMSG}
    \end{algorithm}

\subsection{Preconditioning}

We present a preconditioned HAMS algorithm in Algorithm~\ref{alg:HAMSG}.
Let $\hat{\Sigma}$ be an approximation of $\Var(x)$, and take the Cholesky decomposition $\hat{\Sigma}^{-1} = L L^{\T}$ where $L$ is lower triangular.
By preconditioning, we apply HAMS to the target density of the transformed variable $\hat{x} = L^{\T} x$,
while keeping the momentum variable $ u\sim \mathcal N(\bo,I)$.
Algorithm~\ref{alg:HAMSG} is formulated similarly as the preconditioned HAMS-A/B in \cite{Song2020}, to
minimize the number of matrix-by-vector manipulations per iteration for efficient implementation.

\subsection{Modified algorithms for matching} \label{sec:modified-alg}

We first state the modified algorithms studied in Propositions~\ref{prop:matching}--\ref{prop:matching2} for matching with HAMS.\
The rescaled GJF update is
\begin{align*}
  x^* & = x_0 - \frac{\epsilon^2}{2+\eta\epsilon}\nabla U(x_0) +
  \frac{\epsilon\sqrt{4-\epsilon^2}}{2+\eta\epsilon}  u_0  + \frac{\epsilon}{2+\eta\epsilon} W,   \\
  u^* & = \frac{2-\eta\epsilon}{2+\eta\epsilon} u_0 +
  \frac{\eta\epsilon^2 - 2\epsilon}{\sqrt{4 - \epsilon^2}(2+\eta\epsilon)} \nabla U(x_0) - \frac{\epsilon}{\sqrt{4-\epsilon^2}}\nabla U(x^*)
  +\frac{4}{\sqrt{4 - \epsilon^2}(2 + \eta \epsilon )} W,
\end{align*}
where $W \sim \mathcal N(0, 2\eta \epsilon)$.
The rescaled BAOAB update is
\begin{align*}
 \tilde{u} & = u_0 - \frac{\epsilon}{\sqrt{4-\epsilon^2}}\nabla U(x_0),   \quad
 \tilde{x}  = x_0 + \frac{\epsilon\sqrt{4-\epsilon^2}}{4}\tilde{u},  \\
 \tilde{\tilde u} & = e^{-\eta\epsilon}\tilde{u} + \sqrt{\frac{1-e^{-2\eta\epsilon}}{4-\epsilon^2}} W,  \\
 x^* & = \tilde{x} + \frac{\epsilon\sqrt{4-\epsilon^2}}{4}\tilde{\tilde u}, \quad
 u^*  = \tilde{\tilde u} - \frac{\epsilon}{\sqrt{4-\epsilon^2}}\nabla U(x^*),
\end{align*}
which can be written more succinctly as
\begin{align*}
  x^* &= x_0 - \frac{\epsilon^2}{4}(1+e^{-\eta\epsilon})\nabla U(x_0) + \frac{\epsilon\sqrt{4-\epsilon^2}}{4}(1+e^{-\eta\epsilon}) u_0
  + \frac{\epsilon \sqrt{1-e^{-2\eta\epsilon}}}{2} W,   \\
  u^* &= e^{-\eta\epsilon}u_0 - \frac{\epsilon e^{-\eta\epsilon}}{\sqrt{4-\epsilon^2}}\nabla U(x_0) - \frac{\epsilon}{\sqrt{4-\epsilon^2}}\nabla U(x^*)
  + 2\sqrt{\frac{1-e^{-2\eta\epsilon}}{4-\epsilon^2}} W,
\end{align*}
where $W \sim \mathcal N(0, 1)$. With the full-step momentum in Proposition~\ref{prop:matching}, the IL update can be shown to be equivalent to rescaled BAOAB (see the Supplement, Section~\ref{sec:det-matching}).
The rescaled BP update is
\begin{align*}
     u^+  & = \sqrt{c}u_0 + \sqrt{1-c} W_1 ,  \\
     \tilde{u} & = u^+ - \frac{\epsilon}{1 + \sqrt{1 - \epsilon^2}} \nabla U(x_0), \quad x^* = x_0 + \epsilon\tilde{u}, \quad u^- = \tilde{u} - \frac{\epsilon}{1 + \sqrt{1 - \epsilon^2}} \nabla U(x^*),  \\
     u^* & = \sqrt{c} u^- + \sqrt{1-c} W_2,
\end{align*}
where $ W_1,W_2 \sim \mathcal N(0,1)$ independently. The modified VEC update is
\begin{align*}
  x^* & = x_0 - \frac{\epsilon^2}{2} \nabla U(x_0) + \frac{2\epsilon - \eta\epsilon^2}{2} u_0
  + \frac{\sqrt{2\eta}\epsilon^{3/2}}{2} W_1 + \frac{\sqrt{6\eta}\epsilon^{3/2}}{6} W_2 , \\
  u^* & = \left(1-\eta\epsilon + \frac{\eta^2\epsilon^2}{2}\right) u_0 + \left(\frac{\eta\epsilon^2 - \epsilon}{2} - \frac{\epsilon^3}{4} \right) \nabla U(x_0)
  -\frac{\epsilon}{2}\nabla U(x^*) \nonumber \\
  & \quad + \frac{\sqrt{2\eta\epsilon}}{2}(2-\eta\epsilon) W_1 - \frac{\sqrt{6}}{6}(\eta\epsilon)^{3/2}W_2,
\end{align*}
where $W_1, W_2 \sim \mathcal N(0,1)$, independently.

Next we state the modified algorithms studied in Proposition~\ref{prop:matching3} for matching with shifted HAMS.\
The modified ABOBA update is
\begin{align*}
  \tilde{x} & = x_0 + b u_0 ,  \\
  \tilde{u}  &= u_0 - \frac{\epsilon}{2} \nabla U(\tilde{x}), \quad \tilde{\tilde u} = e^{-\eta\epsilon}\tilde{u} + \sqrt{1-e^{-2\eta\epsilon}} W,
  \quad  u^* = \tilde{\tilde u} -\frac{\epsilon}{2}\nabla U(\tilde{x}) ,\\
  x^* & = \tilde{x} + b u^*,
\end{align*}
where $W \sim \mathcal N(0,1)$ and $b= \frac{\epsilon}{1 + \sqrt{1 - \epsilon^2}}$. The modified SPV update is
\begin{align*}
    \tilde{x} &= x_0 + b u_0,  \\
    u^* & = e^{-\eta\epsilon}u_0 - \frac{1-e^{-\eta\epsilon}}{\eta}\nabla U(\tilde{x}) + \sqrt{1 - e^{-2\eta\epsilon}} W,  \\
    x^* &= \tilde{x} + b u^*,
\end{align*}
where $W \sim \mathcal N(0,1)$ and $b$ is defined in Proposition~\ref{prop:matching3}. The modified Mannella's leapfrog is
\begin{align*}
  \tilde{x} & = x_0 + b u_0,  \\
  u^* & = c_2(c_1u_0 - \epsilon\nabla U(\tilde{x}) + \sqrt{2\eta} W), \\
  x^* & = \tilde{x} + b u^*,
\end{align*}
where $W\sim \mathcal N(0,\epsilon)$, $c_1 = \frac{2-\eta\epsilon}{2}$,  $ c_2 = \frac{2}{2+\eta\epsilon}$, and $b=\frac{\epsilon}{1 + \sqrt{1 - \epsilon^2}}$.

\bibliographystyle{apalike}
\bibliography{hams2ref}


\clearpage

\setcounter{page}{1}

\setcounter{section}{0}
\setcounter{equation}{0}

\setcounter{figure}{0}
\setcounter{table}{0}

\renewcommand{\theequation}{S\arabic{equation}}
\renewcommand{\thesection}{\Roman{section}}

\renewcommand\thefigure{S\arabic{figure}}
\renewcommand\thetable{S\arabic{table}}

\setcounter{lem}{0}
\renewcommand{\thelem}{S\arabic{lem}}

\begin{center}
{\Large Supplementary Material for}

{\Large ``On Irreversible Metropolis Sampling Related to Langevin Dynamics"}

\vspace{.1in} {\large Zexi Song \& Zhiqiang Tan}
\end{center}

\section{Metropolis-adjusted sampling algorithms}
\label{sec:MA-algrithms}

\subsection{Metropolis-adjusted BAOAB}
\label{sec:MA-BAOAB}

The BAOAB update is given as follows:
\begin{align}
    x^* & = x_0 - (1+c)\frac{\epsilon^2}{4} \nabla U(x_0) + (1 + c)\frac{\epsilon}{2}  u_0 + \frac{\epsilon}{2}\sqrt{1 - c^2}  Z_0,  \label{eq:newBAO1}\\
    u^* & = c u_0 - \frac{\epsilon}{2}c \nabla U(x_0) - \frac{\epsilon}{2}\nabla U(x^*) + \sqrt{1 - c^2} Z_0,\label{eq:newBAO2}
\end{align}
where $Z_0 \sim \mathcal N(\bo, 1)$, $\eta \geq 0$ is a friction coefficient, and  $c = e^{-\eta \epsilon}$.

We first derive the acceptance probability for using BAOAB as a proposal scheme in the framework of generalized Metropolis--Hastings sampling \citep{Song2020}.
Using (\ref{eq:newBAO2}), the noise $Z_0$ can be expressed as
\[
  Z_0 = (1 - c^2)^{-1/2}\left[u^* - cu_0 + \frac{\epsilon}{2}c\nabla U(x_0)  + \frac{\epsilon}{2}\nabla U(x^*) \right].
\]
Suppose that the same mapping is applied backward (with reversed momentum) from $(x^*, -u^*)$ to $(x_0, -u_0)$ using another noise $-Z^*$.
Then (\ref{eq:newBAO2}) with $(x_0,u_0,Z_0)$ exchanged with $(x^*,-u^*, -Z^*)$ shows that
\begin{equation}
    \label{eq:newBAO3}
    Z^* = - (1 - c^2)^{-1/2}\left[c u^* - u_0 + \frac{\epsilon}{2}\nabla U(x_0)  + \frac{\epsilon}{2}c\nabla U(x^*) \right].
\end{equation}
Remarkably, it can be verified by direct calculation that (\ref{eq:newBAO1}) is also satisfied with $(x_0,u_0,Z_0)$ exchanged with $(x^*,-u^*, -Z^*)$:
\begin{align*}
  x_0 & = x^* - (1+c)\frac{\epsilon^2}{4} \nabla U(x^*) - (1 + c)\frac{\epsilon}{2}u^* - \frac{\epsilon}{2}\sqrt{1 - c^2} Z^*.
\end{align*}
Then the mapping $\Phi$ from $(x_0,u_0,Z_0)$ to $(x^*, u^*, Z^*)$ given by (\ref{eq:newBAO1})-(\ref{eq:newBAO3})
satisfies the following generalized reversibility:
\begin{equation*}
  \text{forward:  }
  \begin{pmatrix}
    x_0 \\
    u_0 \\
    Z_0
  \end{pmatrix}
  \stackrel{\Psi}{\longrightarrow}
  \begin{pmatrix}
    x^* \\
    u^* \\
    Z^*
  \end{pmatrix},
  \qquad
  \text{backward: }
  \begin{pmatrix}
    x^* \\
    -u^* \\
    -Z^*
  \end{pmatrix}
  \stackrel{\Psi}{\longrightarrow}
  \begin{pmatrix}
    x_0 \\
    -u_0 \\
    -Z_0
  \end{pmatrix}.
\end{equation*}
The forward and backward proposal densities are respectively,
\[
  Q(x^*, u^*|x_0, u_0) \propto \exp\left(- \frac{1}{2} Z_0^{\T} Z_0\right),\quad Q(x_0, -u_0|x^*, -u^*) \propto \exp\left(- \frac{1}{2} Z^{*\T} Z^*\right).
\]
By generalized Metropolis--Hastings sampling, we set $(x_1,u_1) = (x^*, u^*)$ with probability $\alpha = \min(1,r)$
or $(x_1, u_1) = (x_0, -u_0)$ with the remaining probability,  where
\begin{align}
    & r = \frac{\pi(x^*, -u^*)Q(x_0, -u_0|x^*, -u^*) }{\pi(x_0, u_0)Q(x^*, u^*|x_0, u_0) }  \nonumber  \\
    & = \exp\left\{ H(x_0, u_0) - H(x^*, u^*) + \frac{1}{2} Z_0^{\T} Z_0 - \frac{1}{2} Z^{*\T} Z^* \right \}, \nonumber \\
    & = \exp\left\{ - [G(x^*, u^*, Z^*) - G(x_0, u_0, Z_0)]\right\} = \exp\{-\Delta G\}. \label{eq:newBAO4}
\end{align}
To make $\exp(-G)$ a proper density function, we define
\begin{equation*}
  G(x,u,Z) = H(x,u) + \frac{1}{2} Z^\T Z + \log(2\pi)^k. 
\end{equation*}
The expression for $\Delta G$ can be calculated as
\begin{align}
    \Delta G  = U(x^*) - U(x_0) & - \left[\frac{\epsilon}{2}u^*  + \frac{\epsilon^2}{8}\nabla U(x^*)\right]^{\T} \nabla U(x^*) \nonumber \\
    & - \left[\frac{\epsilon}{2}u_0 - \frac{\epsilon^2}{8}\nabla U(x_0)\right]^{\T} \nabla U(x_0).  \label{eq:newBAO5}
\end{align}

Next, we extend Proposition~\ref{prop:accept} to Metropolized BAOAB.\
The transition defined in (\ref{eq:newBAO1})--(\ref{eq:newBAO3}), along with the acceptance probability (\ref{eq:newBAO4}), satisfy
all the conditions used in the proof of Lemma~\ref{lem:accept}. Hence when the chain is stationary, the expected acceptance rate is
\begin{equation}
    \label{eq:newBAO6}
    \EE[\alpha] = 2 \PP[\Delta G <0],
\end{equation}
where the $\PP[\Delta = 0]$ term is always 0.
Consider univariate Gaussian target density $\pi(x) = \mathcal N(0,\gamma^{-1})$.
Then $\Delta G$ in (\ref{eq:newBAO5}) evaluated at any $(x,u,Z)$ can be simplified as
\begin{align}
   & \Delta G(x, u, Z)  = \frac{\gamma\epsilon^2}{128} B_1 \cdot B_2,    \label{eq:BAO-dG}
\end{align}
where
\begin{align*}
& B_1 = (2 + 2c)u - \gamma \epsilon (1 + c)x + 2 \sqrt{1 - c^2} Z, \\
& B_2 = [8(1 - c) + 2\gamma \epsilon^2(1 + c)] u - (\gamma^2 \epsilon^3 - 4\gamma\epsilon)(1 + c) x + 2\sqrt{1 - c^2}(\gamma\epsilon^2 - 4) Z.
\end{align*}
In stationarity, $x\sim \mathcal N(0,\gamma^{-1}), u\sim \mathcal N(0,1), Z\sim \mathcal N(0,1)$ independently. Therefore $(B_1,B_2)$ are bivariate normal with
mean 0 and variance matrix
\begin{equation*}
  \begin{pmatrix}
      (1+c)(8 + (1+c)\gamma\epsilon^2) & (1+c)[4-4c + (1+c)\gamma\epsilon^2]\gamma\epsilon^2 \\
      (1+c)[4-4c + (1+c)\gamma\epsilon^2]\gamma\epsilon^2 & [4 - 4c + (1+c)\gamma\epsilon^2][32 + ((1+c) \gamma \epsilon^2 - 4 - 4 c)\gamma\epsilon^2]
    \end{pmatrix}  .
  \end{equation*}
The correlation coefficient between $B_1,B_2$ is
\begin{align*}
    \rho = \frac{(1+c)[4 - 4c + (1+c)\gamma\epsilon^2]\gamma\epsilon^2}{\sqrt{(1+c)[8 + (1+c)\gamma\epsilon^2][4 - 4c +(1+c)\gamma\epsilon^2][32 + (1+c)(\gamma\epsilon^2 -4)\gamma\epsilon^2]}}.
\end{align*}
Because $\rho >0$, using Lemma \ref{lem:binorm} (Section \ref{subsec:proof:prop:accept}), we have
\begin{align*}
    \PP[\Delta G < 0] = \PP[B_1B_2< 0 ]= \frac{1}{2} - \frac{1}{\pi}\arcsin(\rho).
\end{align*}
Combining this with (\ref{eq:newBAO6}) shows that the expected acceptance rate is
\begin{align}
    \EE[\alpha] = 2 \PP[\Delta G < 0] = 1 - \frac{2}{\pi}\arcsin(\rho). \label{eq:newBAO7}
\end{align}
Notice that $ \EE[\Delta G] = \frac{\gamma\epsilon^2}{128}\EE[B_1 B_2] $.
By direct calculation using the expressions from the variance matrix of $(B_1,B_2)$, it can be shown that
\begin{align*}
    \rho^2 = \frac{\EE^2 [B_1B_2]}{\Var(B_1)\Var(B_2)} = \frac{\EE[\Delta G]}{2 + \EE[\Delta G]} .
\end{align*}
Because $\rho$ is always positive, this implies
\begin{equation*}
    \rho = \sqrt{\frac{\EE[\Delta G]}{2 + \EE[\Delta G]}}.
\end{equation*}
By this relation, $\EE[\alpha]$ in (\ref{eq:newBAO7}) can be expressed in terms of $\EE[\Delta G]$ as
\begin{align}
    \EE[\alpha] =  1 - \frac{2}{\pi}\arcsin\left(\sqrt{\frac{\EE[\Delta G]}{2 + \EE[\Delta G]}}\right) = 1 - \frac{2}{\pi} \arctan\left(\sqrt{\frac{\EE[\Delta G]}{2}}\right). \label{eq:newBAO8}
\end{align}
Letting $c = e^{-\eta \epsilon}$ and taking the series expansion of (\ref{eq:newBAO8}) in $\epsilon$, we find that
\[
    \EE[\alpha] = 1 - \frac{\sqrt{2}}{4\pi}\gamma\sqrt{\eta} \epsilon^{5/2} + \mO(\epsilon^{7/2}).
\]

\subsection{Metropolis-adjusted ABOBA}
\label{sec:MA-ABOBA}

The ABOBA update can be stated as
\begin{align}
  \tilde{x} = x_0 + \frac{\epsilon}{2}u_0,\quad  x^* & = x_0 - (1+c)\frac{\epsilon^2}{4} \nabla U(\tilde{x}) + (1 + c)\frac{\epsilon}{2} u_0 + \frac{\epsilon}{2}\sqrt{1 - c^2} Z_0,  \label{eq:newABO1}\\
   u^* & = c u_0 - (1+c)\frac{\epsilon}{2}\nabla U(\tilde{x})+ \sqrt{1 - c^2} Z_0,\label{eq:newABO2}
 \end{align}
where $Z_0 \sim \mathcal N(\bo, 1)$, $\eta \geq 0$ is a friction coefficient, and  $c = e^{-\eta \epsilon}$.

We first derive the acceptance probability for using ABOBA as a proposal scheme in generalized Metropolis-Hastings sampling. The process is similar to that in Section~\ref{sec:MA-BAOAB} for BAOAB.\
Using (\ref{eq:newABO2}), the noise $Z_0$ can be expressed as
\[
  Z_0 = (1 - c^2)^{-1/2}\left[u^* - cu_0 +(1+c)\frac{\epsilon}{2}\nabla U(\tilde{x}) \right].
\]
Then the noise $Z^*$ required for the backward version of (\ref{eq:newABO2}) after momentum reversal is
\begin{equation}
  \label{eq:newABO3}
  Z^* = -(1 - c^2)^{-1/2}\left[cu^* - u_0 +(1+c)\frac{\epsilon}{2}\nabla U(\tilde{x}) \right].
\end{equation}
The gradients in $Z_0$ and $Z^*$ are evaluated at the same value because $x_0 + \frac{\epsilon}{2}u_0 = x^* - \frac{\epsilon}{2}u^* = \tilde{x}$.
While the $Z^*$ expression is derived by the $u$ update, it can be verified by direct calculation that $Z^*$ also gives the backward version of (\ref{eq:newABO1}) from $x^*$ to $x_0$:
\begin{align*}
  x_0 & = x^* - (1+c)\frac{\epsilon^2}{4} \nabla U(\tilde{x}) - (1 + c)\frac{\epsilon}{2}u^* - \frac{\epsilon}{2}\sqrt{1 - c^2} Z^*.
\end{align*}
The acceptance probability is then $\alpha = \min(1,r)$ where $r$ has the same form as (\ref{eq:newBAO4}). For ABOBA, $\Delta G$
can be calculated as
\begin{equation}\label{eq:newABO4}
\Delta G = U(x^*) - U(x_0)  - \frac{\epsilon}{2}\left[u^*  + u_0\right]^{\T} \nabla U(\tilde{x}).
\end{equation}

Next, we extend Proposition~\ref{prop:accept} to Metropolized ABOBA.\
The transition defined in (\ref{eq:newABO1})--(\ref{eq:newABO3}), along with the acceptance probability (\ref{eq:newABO4}), satisfy
all the conditions used in the proof of Lemma~\ref{lem:accept}. Hence when the chain is stationary, the expected acceptance rate is
\[
    \EE[\alpha] = 2 \PP[\Delta G <0].
\]
Consider univariate Gaussian target density $\pi(x) = \mathcal N(0,\gamma^{-1})$.
Then $\Delta G$ in (\ref{eq:newABO4}) evaluated at any $(x,u,Z)$ can be simplified as
\begin{align*}
   & \Delta G(x, u, Z)  = \frac{\gamma\epsilon^2}{128} B_1 \cdot B_2,
\end{align*}
where
\begin{align*}
   & B_1 = (1+c)(4-\gamma\epsilon^2)u - 2\gamma \epsilon (1 + c)x + 4\sqrt{1 - c^2} Z, \\
   & B_2 = (4c - 4 - \gamma\epsilon^2(1+c))u - 2\gamma \epsilon (1 + c)x + 4\sqrt{1 - c^2} Z.
\end{align*}
In stationarity, $x\sim \mathcal N(0,\gamma^{-1}), u\sim \mathcal N(0,1), Z\sim \mathcal N(0,1)$ independently. Therefore, $(B_1,B_2)$ are bivariate normal with mean 0
and variance matrix
\begin{equation*}
    \begin{pmatrix}
      (1+c)[32 + (1+c)\gamma\epsilon^2(\gamma\epsilon^2-4)] & (1+c)[4-4c + (1+c)\gamma\epsilon^2]\gamma\epsilon^2 \\
      (1+c)[4-4c + (1+c)\gamma\epsilon^2]\gamma\epsilon^2 & [8 + (1+c)\gamma\epsilon^2][4 - 4c + (1+c)\gamma\epsilon^2]
    \end{pmatrix} .
  \end{equation*}
The correlation coefficient between $B_1,B_2$ is given by
\[
  \rho = \frac{(1+c)[4 - 4c + (1+c)\gamma\epsilon^2]\gamma\epsilon^2}{\sqrt{(1+c)[8 + (1+c)\gamma\epsilon^2][4 - 4c +(1+c)\gamma\epsilon^2][32 + (1+c)(\gamma\epsilon^2 -4)\gamma\epsilon^2]}}.
\]
Notice that $\Delta G$ is of the same product form as $\Delta G$ in (\ref{eq:BAO-dG}) for BAOAB, with the same coefficient $\frac{\gamma\epsilon^2}{128}$.
Moreover, the expression of the correlation coefficient $\rho$ and $\EE[B_1B_2]$ are also the same as those for BAOAB.
Hence all the calculation in BAOAB directly applies to ABOBA, and the expected acceptance rate for ABOBA is
\[
    \EE[\alpha] = 1 - \frac{2}{\pi} \arctan\left(\sqrt{\frac{\EE[\Delta G]}{2}}\right)  = 1 - \frac{\sqrt{2}}{4\pi}\gamma\sqrt{\eta} \epsilon^{5/2} + \mO(\epsilon^{7/2}).
\]

\subsection{Metropolis-adjusted BP}
\label{sec:MA-BP}

The BP proposal can be written as
\begin{align}
  x^* & = x_0 - \frac{\epsilon^2}{2}\nabla U(x_0) + \epsilon\sqrt{c} u_0 + \epsilon\sqrt{1 - c}Z^{(1)}_0, \label{eq:newBP1} \\
  u^* & = c u_0 - \frac{\epsilon\sqrt{c}}{2}(\nabla U(x_0) + \nabla U(x^*)) + \sqrt{c(1 - c)}Z^{(1)}_0 + \sqrt{1 - c}Z^{(2)}_0. \label{eq:newBP2}
\end{align}
where $c = e^{-\eta\epsilon}$ and $Z^{(1)}_0, Z^{(2)}_0\sim\mathcal{N}(\bo, I)$ independently.
Define
\begin{align}
  Z^{(1)*} & = -(1-c)^{-\frac{1}{2}}\left(\frac{x_0 - x^*}{\epsilon} + \frac{\epsilon}{2}\nabla U(x^*) + \sqrt{c}u^*\right), \nonumber \\
  Z^{(2)*} & = -(1-c)^{-\frac{1}{2}}\left(\frac{\sqrt{c}(x^* - x_0)}{\epsilon} + \frac{\epsilon\sqrt{c}}{2}\nabla U(x_0) - \sqrt{c}u_0\right), \label{eq:newBP3} \\
  Z^* & = (\, (Z^{(1)*})^\T, (Z^{(2)*})^\T \, )^\T, \quad Z_0  = (\, (Z^{(1)}_0)^\T, (Z^{(2)}_0)^\T \, )^\T. \nonumber
\end{align}
Then the mapping from $(x_0,u_0,Z_0)$ to $(x^*,u^*,Z^*)$ satisfies the generalized reversibility. The acceptance probability is
$\alpha = \min(1,r)$, where $r$ has the same form as (\ref{eq:newBAO4}) with $G$ given by
\[
  G(x,u,Z) = H(x,u) + \frac{1}{2}Z^{\T}Z + \log(2\pi)^{\frac{3}{2}k}.
\]
In this case, $\Delta G$ can be calculated as
\begin{align}
  \Delta G 
  & = U(x^*) - U(x_0) - \frac{(x^* - x_0)^{\T}}{2}\left[\nabla U(x^*) + \nabla U(x_0)\right] \nonumber \\
  & \quad + \frac{\epsilon^2}{8}\left[\nabla U(x^*) ^{\T}\nabla U(x^*) - \nabla U(x_0) ^{\T}\nabla U(x_0) \right]. \label{eq:newBP4}
\end{align}
Detailed calculation and discussions on the validity of the BP acceptance probability are included in \cite{Song2020}, where  BP is referred to as UDL.

Next, we extend Proposition~\ref{prop:accept} to Metropolized BP.\
The transition defined in (\ref{eq:newBP1}) through (\ref{eq:newBP3}), along with the acceptance probability
satisfy all the conditions used in the proof of Lemma \ref{lem:accept}. Hence
when the chain is stationary, the expected acceptance rate is
\[
    \EE[\alpha] = 2 \PP[\Delta G <0],
\]
as in Lemma~\ref{lem:accept}.
Consider univariate Gaussian target density $\pi(x) = \mathcal N(0,\gamma^{-1})$.
Then $\Delta G$ in (\ref{eq:newBP4}) evaluated at any $(x,u,Z)$ can be simplified as
\begin{align*}
   & \Delta G(x, u, Z)  = \frac{\gamma^2\epsilon^3}{32} B_1 \cdot B_2, 
\end{align*}
where
\begin{align*}
   & B_1 = 2\sqrt{c} u - \gamma\epsilon x + 2\sqrt{1 - c}Z^{(1)} ,\\
   & B_2 = 2\sqrt{c}\epsilon - (\gamma\epsilon^2 - 4)x + 2\sqrt{1 - c}\epsilon Z^{(1)}.
\end{align*}
Notice that $\Delta G(x, u , Z)$ does not contain $Z^{(2)}$. In stationarity, $x\sim \mathcal N(0,\gamma^{-1}), u\sim \mathcal N(0,1), Z^{(1)}\sim \mathcal N(0,1)$ independently.
Therefore, $(B_1,B_2)$ are bivariate normal with
\begin{equation*}
    \begin{pmatrix}
      B_1 \\
      B_2
    \end{pmatrix}\sim
    \mathcal N\left(\bo, \begin{pmatrix}
      4 + \gamma\epsilon^2 & \gamma\epsilon^3 \\
      \gamma\epsilon^3 & \frac{16}{\gamma} - 4\epsilon^2 + \gamma\epsilon^4
    \end{pmatrix}\right) .
\end{equation*}
The correlation coefficient between $B_1,B_2$ is
\[
    \rho = \sqrt{\frac{\gamma^3\epsilon^6}{64 + \gamma^3\epsilon^6}}.
\]
Because $\rho > 0$, using Lemma \ref{lem:binorm} (Section \ref{subsec:proof:prop:accept}), we have
\[
    \PP[\Delta G < 0] = \PP[B_1B_2< 0 ]= \frac{1}{2} - \frac{1}{\pi}\arcsin(\rho).
\]
Combining this with Lemma~\ref{lem:accept} shows that the expected acceptance is then
\[
    \EE[\alpha] = 2 \PP[\Delta G < 0] = 1 - \frac{2}{\pi}\arcsin(\rho).
\]
Notice that
\[
  \EE[\Delta G] = \frac{\gamma^2\epsilon^3}{32}\EE[B_1B_2] = \frac{\gamma^3\epsilon^6}{32}.
\]
Then apparently,
\[
  \rho^2 = \frac{\EE[\Delta G]}{2 + \EE[\Delta G]} = \frac{\gamma^3\epsilon^6}{64 + \gamma^3\epsilon^6}.
\]
With the relation above,  the expected acceptance rate can be expressed as
\[
    \EE[\alpha] = 1 - \frac{2}{\pi} \arctan\left(\sqrt{\frac{\EE[\Delta G]}{2}}\right),
\]
which does not depend on $c$ or $\eta$ because $\EE[\Delta G]$ dees not. Moreover, $\EE[\alpha]$ can be expanded as
\[
  \EE[\alpha] = 1 - \frac{\gamma^{3/2}}{4\pi}\epsilon^3 + \mO(\epsilon^9).
\]

\section{Preconditioned sampling algorithms}

We present preconditioned versions of Metropolis-adjusted BAOAB, ABOBA, and BP algorithms, in addition to the preconditioned HAMS-A/B taken from \cite{Song2020}.
These algorithms are used in our numerical experiments.

\vspace{.2in}
\begin{algorithm}[!tph] 
  \SetAlgoNoLine
  Initialize $x_0,u_0,\hat{x}_0 = L^\T x_0$ and $\nabla U(\hat{x}_0) = L^{-1}\nabla U(x_0)$.\\
   \For{ $t = 0,1,2,..., N_{iter}$}{
     Sample $w \sim \text{Uniform} [0,1]$ and $\zeta \sim \mathcal N(\bo,I)$ \\
     $\xi = \sqrt{ab}u_t + \sqrt{a(2-a-b)}\zeta, \quad \hat{x}^* = \hat{x}_t - a \nabla U(\hat{x}_t) + \xi$\\
     Propose $x^* = (L^\T )^{-1}\hat{x}^*$\\
     $\nabla U(\hat{x}^*) = L^{-1} \nabla U(x^*), \quad \tilde{\xi} = \nabla U(\hat{x}^*) + \nabla U(\hat{x}_t)$\\
     $\rho = \exp\left\{U(x_t) - U(x^*) + \frac{1}{2-a}(\tilde{\xi})^\T (\xi - \frac{a}{2}\tilde{\xi} )\right\}$ \\
    \eIf{$w < \min(1,\rho)$}{
      $x_{t+1} = x^*$,\quad $\hat{x}_{t+1} = \hat{x}^*$,\quad $\nabla U(\hat{x}_{t+1}) =  \nabla U(\hat{x}^*) $\qquad\# Accept \\
      \If{HAMS-A}
      {
        $u_{t+1} = \left(\frac{2b}{2-a} - 1\right)u_{t} + \frac{2\sqrt{b(2-a-b)}}{2-a}\zeta - \frac{\sqrt{ab}}{2-a}\tilde{\xi}$\\
      }
      \If{HAMS-B}
      {
        $u_{t+1} = u_t - \frac{\sqrt{ab}}{2-a}\tilde{\xi}$ \\
      }

     }{
      $x_{t+1} = x_t, u_{t+1} = -u_t, \hat{x}_{t+1} = \hat{x}_t,\nabla U(\hat{x}_{t+1}) = \nabla U(\hat{x}_t) $ \qquad\# Reject
    }
   }
   \caption{Preconditioned HAMS-A/HAMS-B}   \label{alg:HAMSAB}
  \end{algorithm}

    \begin{algorithm}[!tph] 
      \SetAlgoNoLine
      Initialize $x_0,u_0,\hat{x}_0 = L^\T x_0$ and $\nabla U(\hat{x}_0) = L^{-1}\nabla U(x_0)$.\\
       \For{ $t = 0,1,2,..., N_{iter}$}{
         Sample $w \sim \text{Uniform} [0,1]$ and $\zeta \sim \mathcal N(\bo, I) $ \\
         $\hat{x}^* = \hat{x}_t - (1+c)\frac{\epsilon^2}{4}\nabla U(\hat{x}_t) + (1 + c)\frac{\epsilon}{2} u_t + \frac{\epsilon}{2}\sqrt{1-c^2}\zeta $ \\
         Propose $x^* = (L^\T )^{-1}\hat{x}^*$\\
         $\nabla U(\hat{x}^*) = L^{-1} \nabla U(x^*)$ \\
         Propose $u^* = cu_t - \frac{\epsilon c}{2} \nabla U(\hat{x}_t) - \frac{\epsilon}{2}\nabla U(\hat{x}^*) + \sqrt{1-c^2}\zeta$ \\
         $\rho = \exp\left\{U(x_t) - U(x^*) + \left[\frac{\epsilon}{2}u^* + \frac{\epsilon^2}{8}\nabla U(\hat{x}^*)\right]^{\T} \nabla U(\hat{x}^*)
         + \left[\frac{\epsilon}{2}u_t - \frac{\epsilon^2}{8}\nabla U(\hat{x}_t)\right]^{\T} \nabla U(\hat{x}_t) \right\}$ \\
        \eIf{$w < \min(1,\rho)$}{
          $x_{t+1} = x^*$,\quad $u_{t+1} = u^*$, \quad  $\hat{x}_{t+1} = \hat{x}^*$,\quad $\nabla U(\hat{x}_{t+1}) =  \nabla U(\hat{x}^*) $\qquad\# Accept \\
         }{
          $x_{t+1} = x_t, u_{t+1} = -u_t, \quad \hat{x}_{t+1} = \hat{x}_t,\nabla U(\hat{x}_{t+1}) = \nabla U(\hat{x}_t) $ \qquad\# Reject
        }
       }
       \caption{Preconditioned  BAOAB}   \label{alg:BAO}
    \end{algorithm}

 \begin{algorithm}[!tph] 
    \SetAlgoNoLine
    Initialize $x_0,u_0,\hat x_0 = L^\T x_0$ and $\nabla U(\tilde{x}_0) = L^{-1}\nabla U\left(x_0 + \frac{\epsilon}{2}(L^\T )^{-1}u_0\right)$.\\
     \For{ $t = 0,1,2,..., N_{iter}$}{
       Sample $w \sim \text{Uniform} [0,1]$ and $\zeta \sim \mathcal N(\bo, I) $ \\
       $\hat{x}^* = \hat{x}_t - (1+c)\frac{\epsilon^2}{4}\nabla U(\tilde{x}_t) + (1 + c)\frac{\epsilon}{2} u_t + \frac{\epsilon}{2}\sqrt{1-c^2}\zeta $ \\
       Propose $x^* = (L^\T )^{-1}\hat x^*$\\
       Propose $u^* = cu_t - (1 + c)\frac{\epsilon}{2}\nabla U(\tilde{x}_t) + \sqrt{1-c^2}\zeta$ \\
       $\rho = \exp\left\{U(x_t) - U(x^*) + \frac{\epsilon}{2}(u^* + u_0)^{\T}\nabla U(\tilde{x}_t) \right\}$ \\
      \eIf{$w < \min(1,\rho)$}{
        $x_{t+1} = x^*$,$u_{t+1} = u^*$, $\hat{x}_{t+1} = \hat{x}^*$,\,$\nabla U(\tilde x_{t+1}) =  L^{-1} \nabla U\left(x^* + \frac{\epsilon}{2}(L^\T )^{-1}u^*\right)$\qquad\# Accept \\
       }{
        $x_{t+1} = x_t, u_{t+1} = -u_t,  \hat{x}_{t+1} = \hat{x}_t,\, \nabla U(\tilde x_{t+1}) =  L^{-1} \nabla U\left(x_t - \frac{\epsilon}{2}(L^\T )^{-1}u_t\right) $ \quad\# Reject \\
      }
     }
     \caption{Preconditioned  ABOBA}   \label{alg:ABO}
  \end{algorithm}

    \begin{algorithm}[!tph] 
      \SetAlgoNoLine
      Initialize $x_0,u_0,\hat{x}_0 = L^\T x_0$ and $\nabla U(\hat{x}_0) = L^{-1}\nabla U(x_0)$.\\
       \For{ $t = 0,1,2,..., N_{iter}$}{
         Sample $w \sim \text{Uniform} [0,1]$ and $\zeta_1, \zeta_2 \sim \mathcal N(\bo, I) $ \\
         $u^{+} = \sqrt{c}u_t + \sqrt{1-c}\zeta_1$, \quad $\hat{x}^* = \hat{x}_t + \epsilon u^{+} - \frac{\epsilon^2}{2}\nabla U(\hat{x}_t)$\\
         Propose $x^* = (L^\T )^{-1}\hat{x}^*$\\
         $\nabla U(\hat{x}^*) = L^{-1} \nabla U(x^*), \quad \tilde{\xi} = \nabla U(\hat{x}^*) + \nabla U(\hat{x}_t)$ \\
         $\rho = \exp\left\{U(x_t) - U(x^*) + \frac{\epsilon}{4}\tilde{\xi}^{\T}( 2 u^{+} - \frac{\epsilon}{2}\tilde{\xi}) \right\}$ \\
        \eIf{$w < \min(1,\rho)$}{
          $x_{t+1} = x^*$,\quad  $\hat{x}_{t+1} = \hat{x}^*$,\quad $\nabla U(\hat{x}_{t+1}) =  \nabla U(\hat{x}^*) $\qquad\# Accept \\
          $u_{t+1} = \sqrt{c}(u^{+} - \frac{\epsilon}{2}\tilde{\xi}) + \sqrt{1-c}\zeta_2$
         }{
          $x_{t+1} = x_t, u_{t+1} = -u_t, \quad \hat{x}_{t+1} = \hat{x}_t,\nabla U(\hat{x}_{t+1}) = \nabla U(\hat{x}_t) $ \qquad\# Reject
        }
       }
       \caption{Preconditioned BP}   \label{alg:BP}
    \end{algorithm}

\section{Technical details}
\label{sec:proofs}

\subsection{Proof of Proposition~\ref{prop:sde}}
\label{subsec:proof:prop:sde}

First, consider the case $\phi = 0$. For notational simplicity, assume that the target density $\pi(x)$ is univariate. The proof can be easily extended to multivariate density $\pi(x)$.
Then HAMS proposal given by (\ref{eq:proposal1-mat})--(\ref{eq:proposal2-mat}) becomes
\begin{align}
  \label{eq:proposal5}
  \begin{pmatrix}
    x^* \\
    u^*
  \end{pmatrix} & =
  \begin{pmatrix}
    x_0 \\
    u_0
  \end{pmatrix}
  - \tilde{A}
  \begin{pmatrix}
    \nabla U(x_0) \\
    u_0
  \end{pmatrix}
  +  \begin{pmatrix}
    Z_{0}^{(1)} \\
    Z_{0}^{(2)}
  \end{pmatrix},
\end{align}
where
\begin{align*}
  \tilde{A} &= \begin{pmatrix}
    a_1 & -a_2 \\
    a_2 & 2-a_3
  \end{pmatrix},\qquad
  \begin{pmatrix}
    Z_{0}^{(1)} \\
    Z_{0}^{(2)}
  \end{pmatrix}\sim\mathcal N(\bo, 2A - A^2).
\end{align*}
By the parametrization (\ref{eq:sdeparam1}) for $A$, we have
\begin{align}
  &\tilde{A}  = \begin{pmatrix}
    2 - c_1(1+\sqrt{1-\epsilon^2}) & -\epsilon\sqrt{c_1c_2} \\
    \epsilon\sqrt{c_1c_2} & 2 - c_2(1+\sqrt{1-\epsilon^2})
  \end{pmatrix}, \nonumber\\
  &\Var(Z_{0}^{(1)}) = c_1[2 - c_2\epsilon^2 + 2\sqrt{1-\epsilon^2} + c_1\{\epsilon^2 - 2(1+\sqrt{1-\epsilon^2})\}], \nonumber \\
  &\Var(Z_{0}^{(2)}) = c_2\{2(1+\sqrt{1-\epsilon^2}) - c_1\epsilon^2 - c_2(1+\sqrt{1-\epsilon^2} )^2\}, \nonumber \\
  &\Cov(Z_{0}^{(1)}, Z_{0}^{(2)}) = (c_1 - c_2)\sqrt{c_2c_2}(1+\sqrt{1-\epsilon^2})\epsilon.
\end{align}
Moreover, using (\ref{eq:sdeparam2}) and taking Taylor expansions with respect to $\epsilon$ around $0$ lead to
\begin{align}
   & \tilde{A} = \begin{pmatrix}
    \eta_1\epsilon + \left(\frac{1}{2} - \frac{\eta_1^2}{4}\right)\epsilon^2 + \mO(\epsilon^3)
    & -\epsilon + \frac{1}{4}(\eta_1 + \eta_2)\epsilon^2 + \mO(\epsilon^3) \\
    \epsilon - \frac{1}{4}(\eta_1 + \eta_2)\epsilon^2 + \mO(\epsilon^3) &
    \eta_2\epsilon + \left(\frac{1}{2} - \frac{\eta_2^2}{4}\right)\epsilon^2 + \mO(\epsilon^3)
  \end{pmatrix}, \nonumber \\
  &\Var(Z_{0}^{(1)}) = 2\eta_1\epsilon - \frac{3}{2}\eta_1^2\epsilon^2 + \mO(\epsilon^3), \nonumber \\
  &\Var(Z_{0}^{(2)}) = 2\eta_2\epsilon - \frac{3}{2}\eta_2^2\epsilon^2 + \mO(\epsilon^3), \nonumber \\
  &\Cov(Z_{0}^{(1)}, Z_{0}^{(2)}) = (\eta_2 - \eta_1) \epsilon^2 + \mO(\epsilon^3). \label{eq:sde2}
\end{align}
From (\ref{eq:sde2}) with all $\mO(\epsilon^2)$ terms as remainders, we obtain
\begin{equation*}
    \tilde{A} = \begin{pmatrix}
      \eta_1\epsilon & - \epsilon \\
      \epsilon & \eta_2 \epsilon
    \end{pmatrix} + \mO(\epsilon^2) ,\quad
    \Var\begin{pmatrix}
      Z_{0}^{(1)} \\
      Z_{0}^{(2)}
    \end{pmatrix}
    = \begin{pmatrix}
      2\eta_1\epsilon & 0 \\
      0  & 2\eta_2\epsilon
    \end{pmatrix}  + \mO(\epsilon^2).
  \end{equation*}
Using this approximation, the update (\ref{eq:proposal5}) becomes
\begin{equation}
    \label{eq:sde3}
    \begin{pmatrix}
      x^* \\
      u^*
    \end{pmatrix}  =
    \begin{pmatrix}
      x_0 \\
      u_0
    \end{pmatrix}
    - \begin{pmatrix}
      \eta_1 & -1 \\
      1 & \eta_2
    \end{pmatrix}\begin{pmatrix}
      \nabla U(x_0) \\
      u_0
    \end{pmatrix}\epsilon
    +   \begin{pmatrix}
      \sqrt{2\eta_1}\zeta_1 \\
      \sqrt{2\eta_2}\zeta_2
    \end{pmatrix},
    \quad \zeta_1,\zeta_2 \stackrel{i.i.d.}{\sim} \mathcal N(0,\epsilon),
  \end{equation}
which is Euler's discretization, hence solving SDE (\ref{eq:sde1}) as $\epsilon \to 0$.

Next, we handle the case of nonzero $\phi$, which appears in the HAMS update only through
$\phi(\tilde{Z}^{(1)} + \nabla U(x_0) - \nabla U(x^*))$ in the update (\ref{eq:proposal2-u}) for $u^*$. The term $\tilde{Z}^{(1)}$ is of order $\mO_p(\sqrt{\epsilon})$:
\[
  \tilde{Z}^{(1)} = Z_{0}^{(1)} - a_1 \nabla U(x_0) + a_2 u_0 = \mO_p(\sqrt{\epsilon}),
  \]
because by (\ref{eq:sde2}), $a_1 = \mO(\epsilon)$, $a_2 = \mO(\epsilon)$, $\Var(Z_{0}^{(1)}) = \mO(\epsilon)$, and hence $Z_{0}^{(1)} = \mO_p(\sqrt{\epsilon})$.
Moreover, by the assumption that $\|\nabla^2 U(x)\|\le M$ for a constant $M$ and the mean value theorem,
\[
|\nabla U(x_0) - \nabla U(x^*)| \leq M |x_0 - x^*| = M |\tilde{Z}^{(1)}| = \mO_p (\sqrt{\epsilon}).
\]
For $\phi = \mO(\epsilon)$, combining the preceding results yields $\phi(\tilde{Z}^{(1)} + \nabla U(x_0) - \nabla U(x^*))= \mO_p (\epsilon^{3/2})$.
Hence the additional term depending on $\phi$ does not affect the approximation (\ref{eq:sde3}) for (\ref{eq:proposal5}).
This completes the proof of Proposition~\ref{prop:sde}.\par

\subsection{Parametrization for HAMS-B}
\label{subsec:proof:param-HAMSB}

We discuss the relationship between the parametrization (\ref{eq:HAMSB-param}) and that used in \cite{Song2020}, Section 3.4. The latter,
with $(\epsilon,c)$ renamed $(\delta,d)$, can be stated as
\begin{align}
a_1 = 1 - \sqrt{1 -\delta^2}, \quad a_2 =  \delta \sqrt{d}, \quad a_3 = 2-  d(1-\sqrt{1-\delta^2}) , \quad \phi = \frac{ \delta\sqrt{d}}{1+\sqrt{1-\delta^2}} ,  \label{eq:HAMSB-param-old}
\end{align}
where $a_1$, $a_2$, and $\phi$ would be the same as in (\ref{eq:HAMSA-param}) with $(\epsilon,c_2)$ replaced by $(\delta, d)$.
By matching the expressions in (\ref{eq:HAMSB-param}) and (\ref{eq:HAMSB-param-old}), $(\epsilon, c_1)$   and $(\delta, d)$ are related as follows:
\begin{align*}
& 1 - \sqrt{1-\delta^2} = 2-  c_1 (1+\sqrt{1-\epsilon^2}) ,  \\
& 2 - d (1-\sqrt{1-\delta^2}) = 1+\sqrt{1-\epsilon^2} .
\end{align*}
Solving for $(\delta, d)$ from the above equations gives
\begin{align*}
& \delta^2 = 1- \left[ c_1 (1 + \sqrt{1-\epsilon^2}) -1 \right]^2, \\
& d = \frac{1- \sqrt{1-\epsilon^2}}{ 2-  c_1 (1+\sqrt{1-\epsilon^2})} .
\end{align*}
For $c_1 = \me^{- \eta_1 \epsilon/2}$ in (\ref{eq:sdeparam2}), taking Taylor expansions  yields
\begin{align*}
& \delta^2 = 2 \eta_1 \epsilon + (1- \frac{3}{2} \eta_1^2) \epsilon^2 + \mO (\epsilon^2) ,\\
& d = \frac{ \epsilon + \mO(\epsilon^3)} { 2\eta_1 + (1 -\frac{\eta_1^2}{2}) \epsilon + \mO(\epsilon^2) }.
\end{align*}
For any fixed $\eta_1 >0$, it follows that $\delta^2 = \mO(\epsilon)$ and $d = \mO (\epsilon)$.
Hence a $\delta$ value translates into a much smaller value for $\epsilon$ in the new parametrization, and
the $d$ value also tends to be much smaller than 1.
Nevertheless, if $\eta_1=0$, then, by the leading terms,
$\delta \approx \epsilon$ and $d \approx 1$, which are expected for the corresponding Hamiltonian dynamics.

\subsection{Proof of Proposition~\ref{prop:phichoice}}
\label{subsec:proof:prop:phichoice}
When $\pi(x) = \mathcal N(0,\gamma^{-1})$, we have $\nabla U(x) = \gamma x$ and the HAMS proposal becomes
\begin{align}
    x^* & = (1 - a_1\gamma)x_0 + a_2u_0 + Z_{0}^{(1)}, \label{eq:proposal6} \\
    u^* & =\{a_1\phi \gamma(\gamma - 1) - a_2\gamma\}x_0 + \{a_3 - a + \phi a_2(1-\gamma)\}u_0 + \phi(1-\gamma)Z_{0}^{(1)} + Z_{0}^{(2)},  \\
    Z^{(1)*} & = Z_{0}^{(1)} - a_1\gamma(x_0 + x^*) + a_2(u_0 - u^*), \\
     Z^{(2)*}& = Z_{0}^{(2)} - a_2\gamma(x_0 + x^*) + a_3(u_0 - u^*).
\end{align}
We can also simplify $\Delta G$ as
\begin{equation}
  \label{eq:dG3}
  \Delta G = \frac{\gamma}{2}((x^*)^2 - x_0^2) + \frac{1}{2}((u^*)^2 - u^2_0) + \frac{1}{2}Z^{*\T} (2A - A^2)^{-1} Z^* - \frac{1}{2}Z_0^\T (2A - A^2)^{-1} Z_0.
\end{equation}
Combining (\ref{eq:proposal6}) -- (\ref{eq:dG3}) shows that $\Delta G$ can be expressed as a quadratic form of $(x_0,u_0,Z_{0}^{(1)},Z_{0}^{(2)})^\T$
stated in Proposition~\ref{prop:phichoice}:
\[
\Delta G =  (x_0, u_0, Z_{0}^{(1)}, Z_{0}^{(2)})\, D(\gamma) \, (x_0, u_0, Z_{0}^{(1)}, Z_{0}^{(2)})^\T.
\]
The entries of $D(\gamma)$ are then computed using \texttt{Mathematica}. We find that $\Delta G$ does not contain
any $[Z_{0}^{(2)}]^2$ terms, therefore $d_{44}(\gamma) = 0$. For the other diagonal entries,
the order of $d_{11}(\gamma)$ is $\gamma^4$ and the orders of $d_{22}(\gamma), d_{33}(\gamma)$ are both $\gamma^2$.
For  $d_{11}(\gamma)$, the coefficient of $\gamma^4$ is
\begin{equation}
    \label{eq:d1}
    \frac{a_1^2  h(\phi)}{2a_1(a_3 -2) - 2(a_2^2 + 2a_3 - 4)} .
\end{equation}
For  $d_{22}(\gamma)$, the coefficient of $\gamma^2$ is
  \begin{equation}
    \label{eq:d2}
    \frac{a_2^2  h(\phi)}{2a_1(a_3 -2) - 2(a_2^2 + 2a_3 - 4)} .
  \end{equation}
For  $d_{33}(\gamma)$, the coefficient of $\gamma^2$ is
  \begin{equation}
    \label{eq:d3}
    \frac{h(\phi)}{2a_1(a_3 -2) - 2(a_2^2 + 2a_3 - 4)}.
  \end{equation}
The same function of $\phi$ that appears in (\ref{eq:d1}), (\ref{eq:d2}) and (\ref{eq:d3}) is quadratic in $\phi$,
  \begin{equation*}
    h(\phi) = \phi^2(4 - 2a_1) - \phi 4a_2 +2a_1 +a_2^2 - a_1a_3.
  \end{equation*}
This quadratic function $h(\phi)$ is always $\ge 0$, with the discriminant
\begin{align*}
& \quad 16 a_2^2 - 4 (4-2a_1) (2 a_1 + a_2^2 - a_1 a_3) \\
& = - 8a_1 ( 4 - 2 a_1 - 2 a_3 + a_1 a_3 - a_2^2 ) \le 0,
\end{align*}
because $a_1\ge 0$, and $(2-a_1)(2-a_3)\geq a_2^2$ due to the constraint $\bo \leq A \leq 2I$.
Therefore $|h(\phi)|$ is minimized when $\phi = \frac{a_2}{2-a_1}$.

\subsection{Proof of Corollary~\ref{cor:phichoice}}
\label{subsec:proof:cor1}
When $\phi = a2/(2-a_1)$, we have
\begin{align}
  u^* & = \frac{a_1 + a_2^2 + 2a_3 - a_1a_3 - 2}{2-a_1} u_0 - \frac{a_2}{2-a_1}\nabla U(x_0) - \frac{a_2}{2-a_1}\nabla U(x^*) + \frac{a_2}{2-a_1} Z^{(1)}_0 + Z^{(2)}_0\label{eq:cor1proof1} \\
  Z^{(1)*} & = a_2\left(2 - a_3  - \frac{a_2^2}{2-a_1}\right) u_0 + \left(\frac{a_2^2}{2-a_1} - a_1\right) \nabla U(x_0) + \left(\frac{a_2^2}{2-a_1} - a_1\right) \nabla U(x^*)  \nonumber \\
           & \qquad + (1 - \frac{a_2^2}{2-a_1})Z^{(1)}_0 - a_2 Z^{(2)}_0 \\
  Z^{(2)*} & = a_3\left(2 - a_3  - \frac{a_2^2}{2-a_1}\right) u_0 + \frac{a_2(a_1+ a_3 - 2)}{2-a_1} \nabla U(x_0) + \frac{a_2(a_1+ a_3 - 2)}{2-a_1} \nabla U(x^*) \nonumber \\
           & \qquad - \frac{a_2 a_3}{2-a_1}Z^{(1)}_0 + (1- a_3)Z^{(2)}_0.
\end{align}
Furthermore, because $A$ matrix is block diagonal, $(2A - A^2)^{-1}$ is also block diagonal given by
\begin{equation}
  \label{eq:cor1proof2}
  (2A - A^2)^{-1} =  [(a_1a_3 - a_2^2)(4 + a_1a_3 - a_2^2 - 2a_1 - 2a_3)]^{-1}\begin{pmatrix}
    ( 2a_3 -a_3^2 - a_2^2) I & a_2(a_1 + a_3 - 2) I \\
    a_2(a_1 + a_3 - 2) I  & ( 2a_3 -a_1^2 - a_2^2) I
  \end{pmatrix}.
\end{equation}
Recall that $\Delta G$ is defined as
\[
  \Delta G = U(x^*) - U(x_0) + \frac{1}{2} \left\{ (u^*)^{\T} u^* - u_0^\T u_0 + \frac{1}{2}Z^{*\T} (2A - A^2)^{-1} Z^* - \frac{1}{2}Z_0^\T (2A - A^2)^{-1} Z_0\right\}.
\]
Substituting (\ref{eq:cor1proof1}) -- (\ref{eq:cor1proof2}) into the above, we obtain results in Corollary~\ref{cor:phichoice}.

\subsection{Proof of Proposition~\ref{prop:variance}}
\label{subsec:proof:prop:variance}
When the target density is $\mathcal N(0,\gamma^{-1})$, the HAMS proposal can be equivalently
written in the form of (\ref{eq:proposal4}) with
\begin{align}
& \Phi = \begin{pmatrix}
  1 - a_1\gamma & a_2 \\
  a_1\phi\gamma(\gamma - 1) - a_2\gamma & a_3 - 1 + \phi a_2(1-\gamma)
\end{pmatrix}, \quad \mathbf{\zeta} \sim \mathcal N(\bo, W), \label{eq:proposal60} \\
& W = \begin{pmatrix}
  2a_1 - a_1^2 - a_2^2 & \phi(1-\gamma)(2a_1 - a_1^2 -a_2^2) + 2a_2 -a_1a_2 - a_2a_2 \\
  \phi(1-\gamma)(2a_1 - a_1^2 -a_2^2)  & \phi^2(1-\gamma)^2(2a_1 - a_1^2 - a_2^2) + 2a_3 - a_3^2 -a_2^2 \\
  + 2a_2 -a_1a_2 - a_2a_2 & + 2\phi(1-\gamma)(2a_2 - a_1a_2 - a_2a_3)
\end{pmatrix}.\nonumber
\end{align}
Similarly as in \cite{Burrage2007}, Section 2, taking the variance of both sides in (\ref{eq:proposal4}) shows that the stationary variance $V$ satisfies the following equation,
\begin{equation}
  \label{eq:var5}
  V = \Phi V \Phi^\T + W.
\end{equation}
Substituting the default choice $\phi = a_2/(2-a_1)$ in $\Phi$ and $W$ and solving for $V$ in (\ref{eq:var5}), we
obtain (\ref{eq:var1}) stated in Proposition~\ref{prop:variance}.

Note that without using the default choice  $\phi$,
a general expression of $V$ in terms of $a_1,a_2,a_3$ and $\phi$ can also be obtained from (\ref{eq:var5}).
But the expression is complicated and not informative, hence not presented here.

\subsection{Proof of Lemma~\ref{lem:accept}}
\label{subsec:proof:lem:accept}
Denote  as $\Psi$  the mapping from $(x_0,u_0,Z_0)$ to $(x^*, u^*, Z^*)$ given by (\ref{eq:proposal1})--(\ref{eq:proposal3}).
Then $\Psi$ satisfies the following generalized reversibility:
\begin{equation}
  \label{eq:genrev1}
  \text{forward:  }
  \begin{pmatrix}
    x_0 \\
    u_0 \\
    Z_0
  \end{pmatrix}
  \stackrel{\Psi}{\longrightarrow}
  \begin{pmatrix}
    x^* \\
    u^* \\
    Z^*
  \end{pmatrix},
  \qquad
  \text{backward: }
  \begin{pmatrix}
    x^* \\
    -u^* \\
    -Z^*
  \end{pmatrix}
  \stackrel{\Psi}{\longrightarrow}
  \begin{pmatrix}
    x_0 \\
    -u_0 \\
    -Z_0
  \end{pmatrix}.
\end{equation}
Let $S$ be the mapping that changes the signs of $u$ and $Z$, that is, $S(x,u,Z) = (x, -u, -Z)$.
Define the composite $\hat{\Psi} = S \circ \Psi$. Then (\ref{eq:genrev1}) can be equivalently stated as
\begin{equation}
  \label{eq:genrev2}
  \Psi(\hat{\Psi}(x,u,Z)) = S(x,u,Z).
\end{equation}
Moreover, because the function $G$ is even in $u$ and $Z$, we have
\begin{equation}
  \label{eq:accept3}
  G\circ S = G.
\end{equation}

According to (\ref{eq:mr}), the acceptance rate at any current value $(x,u,Z)$ is
\begin{align*}
  \alpha(x,u,Z) & = \min[1, \exp(-\Delta G(x,u,Z))], 
\end{align*}
where $\Delta G(x,u,Z) = G(\Psi(x,u,Z) ) -  G(x,u,Z)$, and $G$ is redefined as
\[
  G(x,u,Z) = H(x,u) + \frac{1}{2} Z^\T (2A - A^2)^{-1} Z + \frac{1}{2}\log(2\pi)^k + \frac{1}{2}\log \left[(2\pi)^{2k} |Det(2A - A^2)|\right].
\]
The determinant terms are included to make $\exp(-G)$ a valid density function.
If the target density is $\mathcal{N}(\bo, I)$, then HAMS is rejection free with
$\Delta G \equiv 0$ and hence Lemma~\ref{lem:accept} trivially holds: $\EE[\alpha] = \PP[\Delta G = 0] = 1$.
If the target density is not
$\mathcal{N}(\bo, I)$, then $\PP[\Delta G = 0] = 0$. Hence it suffices to show that
$\EE[\alpha] = 2\PP[\Delta G < 0]$.

At stationarity, the density of $(x,u,Z)$ is $\exp(-G(x,u,Z))$. Then
\begin{align*}
  \EE[\alpha] & = \int \min[1, \exp(-\Delta G(x,u,Z))]\cdot \exp(-G(x,u,Z)) \, \dif x\, \dif u\, \dif Z \\
  & = \underbrace{\int_{\Delta G<0} \exp(-G(x,u,Z))\, \dif x\, \dif u\, \dif Z  }_{(I)} \\
  & \quad + \underbrace{\int_{\Delta G>0} \exp(-\Delta G(x,u,Z))\exp(-G(x,u,Z))\, \dif x\, \dif u\, \dif Z  }_{(II)}.
\end{align*}
Apparently $(I) = \PP[\Delta G <0]$. In the following, we show that $(I) = (II)$.
On one hand, $(II)$ can be directly calculated as
\begin{align}
  (II) &= \int_{\Delta G>0} \exp[- (\Delta G(x,u,Z) + G(x,u,Z) )]\, \dif x\, \dif u\, \dif Z\nonumber\\
  &= \int_{\Delta G>0} \exp[-G(\Psi(x,u,Z))] \dif x\, \dif u\, \dif Z \nonumber\\
   \text{(by (\ref{eq:accept3}))} \quad &= \int_{\Delta G>0} \exp[-G(\hat{\Psi}(x,u,Z))] \dif x\, \dif u\, \dif Z. \label{eq:accept4}
\end{align}
On the other hand, $(I)$ can be shown to be
\begin{align}
  \label{eq:accept5}
 (I) & = \int_{\Delta G<0} \exp(-G(\hat x,\hat u,\hat Z))\, \dif \hat x\, \dif \hat u\, \dif\hat Z \nonumber \\
  & = \int_{\Delta G >0} \exp(-G(\hat{\Psi}(x,u,Z)))\, \dif  x\, \dif u\, \dif Z.
\end{align}
The first step follows by replacing $(x,u,Z)$ with $(\hat x,\hat u,\hat Z)$ in the notation.
The second step involves a change of variables in the integration: $(\hat x, \hat u, \hat Z) = \hat{\Psi}(x,u, Z)$.
The mapping $\hat \Psi$ can be expressed as a series of shear mappings and sign changes, and hence has a unit Jacobian.
Moreover, there is a one-to-one correspondence
between $(x,u,Z)$ with $\Delta G(x,u,Z) > 0$ and $(\hat x,\hat u,\hat Z)$ with $\Delta G(\hat x,\hat u,\hat Z) < 0$ under the change of variables:
\begin{align*}
  \Delta G(\hat x,\hat u,\hat Z)&  =  G(\Psi(\hat x,\hat u,\hat Z) ) -  G(\hat x,\hat u,\hat Z)  \\
  \text{(by definition)}\quad & = G[ \Psi ( \hat{\Psi}( x, u, Z) )] - G(\hat{\Psi}( x, u, Z) ) \\
  \text{(by (\ref{eq:genrev2}))} \quad & = G(S(x,u,Z))  - G(\hat{\Psi}( x, u, Z) ) \\
  \text{(by (\ref{eq:accept3}))} \quad & = G(x,u,Z)  - G(\Psi( x, u, Z) ) = - \Delta G(x,u,Z).
\end{align*}
Comparing (\ref{eq:accept4}) and (\ref{eq:accept5}) completes the proof.

\subsection{Proof of Proposition~\ref{prop:accept}}
\label{subsec:proof:prop:accept}

First, we prove the following result about bivariate normal random variables.
\begin{lem}
  \label{lem:binorm}
  For $\tau \in [-1,1]$, let
  \[
    \begin{pmatrix}
      X \\
      Y
    \end{pmatrix}\sim
    \mathcal N\left(\bo, \begin{pmatrix}
      1 & \tau \\
      \tau & 1
    \end{pmatrix}\right).
  \]
Then
 $$   \PP[X>0 \text{ and } Y>0] = \frac{1}{4} + \frac{1}{2\pi}\arcsin(\tau) .$$
\end{lem}
\textbf{Proof of Lemma~\ref{lem:binorm}.}\; Define $\zeta = (Y - \tau X)/\sqrt{1 - \tau^2}$. Then $\zeta$ and $X$
are jointly normal with $(X, \zeta)^\T \sim \mathcal{N}(\bo, I)$. Using the fact that $Y>0$ is equivalent to $\zeta > -\frac{\tau}{\sqrt{1-\tau^2}}X$
and making a change to polar coordinates, we have
\begin{align*}
  & \quad \PP[X>0 \text{ and } Y>0] = \PP\left[ X>0 \text{ and } \zeta > -\frac{\tau}{\sqrt{1-\tau^2}} X\right] \\
  & = \int_{x = 0}^{\infty} \int_{\zeta =  -\frac{\tau}{\sqrt{1-\tau^2}} x}^{\infty}  \frac{1}{2\pi}\exp\left( - \frac{x^2}{2} - \frac{\zeta^2}{2}\right) \, d\zeta\, \dif x
  = \int_{\theta = \arctan\left( - \frac{\tau}{\sqrt{1-\tau^2}} \right)}^{\pi/2} \int_{r = 0}^{\infty} \frac{1}{2\pi} e^{-r^2/2} r \, \dif r\, d\theta \\
 & = \int_{\theta = \arctan\left( - \frac{\tau}{\sqrt{1-\tau^2}} \right)}^{\pi/2} \frac{d\theta}{2\pi} = \frac{1}{4} + \frac{1}{2\pi}\arcsin(\tau).
\end{align*}

Next, to apply Lemma~\ref{lem:accept} for $\EE[\alpha]$, we calculate $\PP[\Delta G< 0]$ for a univariate normal target. According to Corollary~\ref{cor:phichoice},
\begin{align*}
  \Delta G(x,u,Z) & = \frac{a_1\gamma(\gamma-1)}{2(2-a_1)} B_1 \cdot B_2,
\end{align*}
where $B_1 =  (a_2u + Z^{(1)} - a_1\gamma x) $ and $B_2 = (a_2u + Z^{(1)} + (2 - a_1\gamma)x)$.
At stationarity, we have $x\sim \mathcal N(0, \gamma^{-1}), u\sim\mathcal N(0,1)$ and $Z^{(1)} \sim \mathcal N(0, 2a_1 -a_1^2 -a_2^2)$.
Then $(B_1,B_2)$ are jointly normal with
\begin{equation}
  \begin{pmatrix}
    B_1 \\
    B_2
  \end{pmatrix}\sim
  \mathcal N\left(\bo, \begin{pmatrix}
     a_1^2(\gamma - 1) + 2a_1  & a_1^2(\gamma - 1) \\
    a_1^2(\gamma - 1) &  a_1^2(\gamma - 1) - 2a_1 + \frac{4}{\gamma}
  \end{pmatrix}\right) . \label{eq:dG4}
\end{equation}
The correlation coefficient between $B_1$ and $B_2$ is
\begin{equation*}
  \rho = \frac{a_1^2(\gamma - 1)\sqrt{\gamma}}{\sqrt{a_1(a_1^3(\gamma - 1)^2\gamma - 4a_1 + 8)}}.
\end{equation*}
Using Lemma~\ref{lem:binorm}, we have
\begin{align*}
  & \PP(B_1 > 0 \text{ and }B_2 > 0 ) 
  = \frac{1}{4} + \frac{1}{2\pi}\arcsin(\rho),
\end{align*}
which leads to
\begin{align*}
 & \quad \PP(B_1B_2 > 0 ) = \PP(B_1 > 0 \text{ and }B_2 > 0 ) + \PP(B_1 < 0 \text{ and }B_2 < 0 )  \\
 & = 2  \PP(B_1 > 0 \text{ and }B_2 > 0 )  = \frac{1}{2} + \frac{1}{\pi}\arcsin(\rho),
\end{align*}
and $ \PP(B_1B_2 < 0 ) = \frac{1}{2} - \frac{1}{\pi}\arcsin(\rho)$.
Clearly $\mathrm{Sign}(\rho) = \mathrm{Sign}(\gamma - 1)$.
Then depending on whether $\gamma >1$, we have
\begin{equation*}
  \PP[\Delta G < 0] =
  \begin{cases}
    \PP[B_1B_2< 0] =  \frac{1}{2} - \frac{1}{\pi}\arcsin(\rho) & \text{if }  \gamma > 1, \\
    \PP[B_1B_2> 0] =  \frac{1}{2} + \frac{1}{\pi}\arcsin(\rho) & \text{if } 0<\gamma< 1, \\
    0 & \text{if } \gamma = 1.
 \end{cases}
\end{equation*}
Combining this with Lemma~\ref{lem:accept} shows that the expected acceptance rate is
\begin{equation}
  \label{eq:accept6}
  \EE[\alpha] =
  \begin{cases}
    1 - \frac{2}{\pi}\arcsin(\rho) & \text{if }  \gamma > 1 \\
    1 + \frac{2}{\pi}\arcsin(\rho) & \text{if } 0<\gamma< 1 \\
    1 & \text{if } \gamma = 1.
 \end{cases}
\end{equation}

Finally we relate the above expression to the expected change $\EE[\Delta G]$. Because
$\EE[B_1 B_2] = a_1^2(\gamma -1)$ by (\ref{eq:dG4}), we have
\begin{equation*}
  \EE[\Delta G] = \frac{a_1\gamma(\gamma - 1)}{2(2-a_1)} \EE[B_1B_2] = \frac{a_1^3\gamma (\gamma - 1)^2}{2(2-a_1)},
\end{equation*}
which, by direct calculation, is related to $\rho$ as follows:
\begin{align}
  \label{eq:accept7}
  \rho^2 = \frac{\EE[\Delta G]}{2 + \EE[\Delta G]}\, \Longleftrightarrow  \,\rho = \mathrm{Sign}(\gamma - 1)\sqrt{\frac{\EE[\Delta G]}{2 + \EE[\Delta G]}}.
\end{align}
Substituting (\ref{eq:accept7}) into  (\ref{eq:accept6}), we obtain the unified expression in Proposition~\ref{prop:accept}:
\begin{equation*}
  \EE[\alpha] = 1 - \frac{2}{\pi}\arcsin\left( \sqrt{\frac{\EE[\Delta G]}{2 + \EE[\Delta G]}} \right) = 1 - \frac{2}{\pi} \arctan\left(\sqrt{\frac{\EE[\Delta G]}{2}}\right).
\end{equation*}

\subsection{Proof of Proposition \ref{prop:eigen1}}

The two eigenvalues of $\Phi$ are given by
$$
  \frac{1}{2}\left(a_3 - a_1 + \sqrt{(a_1+a_3 - 2)^2 - 4a_2^2}\right) .
$$
The spectral radius (i.e., maximum modulus of the two eigenvalues) is
\[
    \rho(\Phi) = \begin{cases}
        \sqrt{a_1 + a_3 -a_1a_3 + a_2^2 - 1}, & \text{ if } 4 a_2^2 \geq (a_1 + a_3 - 2)^2, \\
        \frac{1}{2}\left(|a_3 - a_1| + \sqrt{(a_1+a_3 - 2)^2 - 4a_2^2}\right), & \text{ if } 4 a_2^2 < (a_1 + a_3 - 2)^2.
     \end{cases}
\]
For fixed $a_1$ and $\nu = a_2^2/a_3$, we write the spectral radius of $\Phi$ as a function of $a_3$
\begin{equation}
    \label{eq:eigenproof1}
    \begin{aligned}
        \rho(a_3) & = \begin{cases}
            \sqrt{a_1 + a_3 - a_1a_3 + \nu a_3 - 1}, & \text{ if } R_1 \leq a_3 \leq R_2, \\
            \frac{1}{2}\left(|a_3 - a_1| + \sqrt{(a_1+a_3 - 2)^2 - 4\nu a_3}\right), & \text{ if } a_3 < R_1 \text{ or } a_3 > R_2,
         \end{cases}
    \end{aligned}
\end{equation}
where $R_1$ and $R_2$ are the roots of $(a_1+a_3 - 2)^2 - 4 \nu a_3$ as a function of $a_3$, given by
\begin{align*}
R_1  = 2\nu + 2 - a_1 - 2\sqrt{ \nu (\nu + 2 - a_1)} = (\sqrt{\nu + 2 - a_1} - \sqrt{\nu})^2 , \\
R_2  = 2\nu + 2 - a_1 + 2\sqrt{ \nu (\nu + 2 - a_1)} = (\sqrt{\nu + 2 - a_1} + \sqrt{\nu})^2 .
\end{align*}
Assume that $0<a_1<2$ and $\nu \leq a_1 \leq 1 + \nu$. Then as we show later,
the function $\rho(a_3)$ is nonincreasing when $a_3 < R_1$ and nondecreasing where $a_3\geq R_2$. In the intermediate case $R_1 \leq a_3 \leq R_2$, the function $\rho(a_3)$ is nondecreasing because
by the condition $a_1\leq \nu + 1$,
\begin{align*}
    & \rho(a_3) = \sqrt{a_1 + a_3 - a_1a_3 + \nu a_3 - 1}  \\
  \Longrightarrow  & \frac{d\rho}{d a_3} = \frac{1+\nu -a_1}{2\sqrt{a_1 + a_3 - a_1a_3 + a_3\nu - 1}}\geq 0.
\end{align*}
Consequently, the spectral radius of $\Phi$ is minimized at
\[
    a_3^* = R_1 = (\sqrt{\nu + 2 - a_1} - \sqrt{\nu})^2,
\]
with the minimum spectral radius
\[
    \frac{|a_3^* - a_1|}{2}
\]
and the implied choice of $a_2$
\[
    a_2^* = \pm \sqrt{\nu a_3^*}.
\]
The condition $\bo \leq A \leq 2 I$ holds if and only if $0 \le a_1, a_3 \le 2$, $a_1a_3 \ge a_2^2$, and $(2-a_1) (2-a_3) \ge a_2^2$.
With $a_2^2 = \nu a_3$ and $\nu \le a_1$, this condition dictates that
\[
    0\leq a_3 \leq \frac{2(2-a_1)}{\nu+2 - a_1},
\]
which is always satisfied by $a_3^*$, because
\begin{align*}
a_3 \le  (\sqrt{\nu + 2 - a_1} - \sqrt{\nu})  (\sqrt{\nu + 2 - a_1} + \sqrt{\nu}) = 2- a_1 \le  \frac{2(2-a_1)}{\nu+2 - a_1}.
\end{align*}
The last inequality follows with $\nu\le a_1$ and $\nu +2 -a_1 \ge 2$.

In the remainder of this section, we demonstrate the monotonicity of $\rho(a_3)$ for $a_3 < R_1$ or $a_3 >R_2$ as mentioned above. We distinguish four cases.

\textbf{Case 1:}  $a_3 \geq a_1, a_3 > R_2$. Then $\rho(a_3)$ becomes
\begin{equation}
    \label{eq:eigenproof2}
    \rho(a_3) = \frac{1}{2}\left(a_3 - a_1 + \sqrt{(a_1+a_3 - 2)^2 - 4\nu a_3}\right),
\end{equation}
with the derivative
\begin{equation}
    \label{eq:eigenproof3}
    \frac{d\rho}{d a_3} = \frac{1}{8}((a_1+a_3 - 2)^2 - 4\nu a_3)^{-1/2}\left(a_1 + a_3 - 2 - 2\nu + \sqrt{(a_1+a_3 - 2)^2 - 4\nu a_3}\right).
\end{equation}
Then $\frac{d\rho}{d a_3} > 0$ because
\begin{align*}
    & \quad a_1 + a_3 - 2 - 2\nu + \sqrt{(a_1+a_3 - 2)^2 - 4\nu a_3} \\
     & > a_1 + R_2 - 2 - 2\nu + \sqrt{(a_1+a_3 - 2)^2 - 4\nu a_3} \\
     &  = 2\sqrt{\nu^2 + 2\nu - a_1\nu} +  \sqrt{(a_1+a_3 - 2)^2 - 4\nu a_3} \ge  0.
\end{align*}

\textbf{Case 2:} $a_1 \leq a_3 < R_1$. Then $\rho(a_3)$ and $d\rho/d a_3$ are the same as in (\ref{eq:eigenproof2}) and (\ref{eq:eigenproof3}).
For $0<a_1<2$, it holds that $\nu \le \sqrt{\nu(2-a_1 + \nu )}$. Then $ \frac{d\rho}{d a_3} \leq 0$ because
\begin{align*}
   &  a_1 + a_3 < a_1 + R_1 = 2 + 2(\nu - \sqrt{\nu(2-a_1 + \nu )}) \le 2  \\
    \Longrightarrow & 2 - a_1 - a_3 + 2\nu \geq 2 - a_1 - a_3    \geq \sqrt{(a_1 + a_3 - 2)^2 - 4\nu a_3}  \\
    \Longrightarrow & a_1 + a_3 - 2 - 2\nu + \sqrt{(a_1 + a_3 - 2)^2 - 4\nu a_3}\leq 0 .
\end{align*}

\textbf{Case 3:} $a_3 < a_1, a_3 < R_1$. Then $\rho(a_3)$ becomes
\begin{equation}
    \label{eq:eigenproof4}
    \rho(a_3) = \frac{1}{2}\left(a_1 - a_3 + \sqrt{(a_1+a_3 - 2)^2 - 4\nu a_3}\right),
\end{equation}
with the derivative
\begin{equation}
    \label{eq:eigenproof5}
    \frac{d\rho}{d a_3} = \frac{1}{8}((a_1+a_3 - 2)^2 - 4\nu a_3)^{-1/2}\left(a_1 + a_3 - 2 - 2\nu - \sqrt{(a_1+a_3 - 2)^2 - 4\nu a_3}\right).
\end{equation}
Then $\frac{d\rho}{d a_3} < 0$ because
\begin{align*}
    & \quad a_1 + a_3  - 2 - 2\nu - \sqrt{(a_1+a_3 - 2)^2 - 4\nu a_3} \\
     & < a_1 + R_1 - 2 - 2\nu - \sqrt{(a_1+a_3 - 2)^2 - 4\nu a_3} \\
     &  = -2\sqrt{\nu^2 + 2\nu - a_1\nu} -  \sqrt{(a_1+a_3 - 2)^2 - 4\nu a_3} \le 0.
\end{align*}

\textbf{Case 4:} $2 > a_1 \geq a_3 > R_2$. Then $\rho(a_3) $ and $d\rho/d a_3$ are the same as in (\ref{eq:eigenproof4}) and (\ref{eq:eigenproof5}).
Notice that
\begin{align*}
    & a_1 + a_3 > a_1 + R_2 = 2 + 2\nu + 2 \sqrt{\nu(2-a_1 + \nu)} \\
    \Longrightarrow & a_1 + a_3 - 2 - 2\nu >0.
\end{align*}
By the condition $a_1\leq 1 + \nu$, we have $2+\nu - a_1 >0$. Then $\frac{d\rho}{d a_3} \ge  0$ because
\begin{align*}
    & 4\nu(2+\nu - a_1 ) \ge 0 \\
    \Longrightarrow & (a_1 + a_3 -2)^2 - 4\nu a_3 + 4\nu(2+\nu - a_1 ) \ge (a_1 + a_3 -2)^2 - 4\nu a_3  \\
    \Longrightarrow & (a_1 + a_3 - 2 - 2\nu)^2 \ge (a_1 + a_3 -2)^2 - 4\nu a_3 \\
    \Longrightarrow & a_1 + a_3 - 2 - 2\nu \ge \sqrt{(a_1 + a_3 -2)^2 - 4\nu a_3} .
\end{align*}

Combining all four cases shows that
\[
 \frac{d\rho}{a_3}\leq 0 \text{ if } a_3 < R_1, \quad \text{ and } \quad \frac{d\rho}{a_3}\geq 0 \text{ if } a_3 \geq R_2.
\]
This completes the proof of Proposition 5.

\subsection{Proof of Proposition \ref{prop:eigen1-b}}

Consider a transformation $\tilde a_1 = 2-a_1$, $\tilde a_2= a_2$, and $\tilde a_3 = 2-a_3$. Then the $\Phi$ matrix becomes
$$ \Phi = \begin{pmatrix}
    1- a_1 & a_2 \\
    - a_2 & a_3-1
\end{pmatrix}  =
 \begin{pmatrix}
    \tilde a_1 -1 & \tilde a_2 \\
    - \tilde a_2 & 1 -\tilde a_3
\end{pmatrix} .$$
The eigenvalues of $\Phi$, hence also the spectral radius, depend on $(\tilde a_3, \tilde a_2, \tilde a_1)$ in the same way
as $\Phi$ depends on $(a_1, a_2, a_3)$. Moreover,
fixed $a_3$ and $\tilde \nu = a_2^2 / (2-a_1)$ translate into fixed $\tilde a_3$ and $\tilde \nu = \tilde a_2^2 / \tilde a_1$.
The condition $\tilde \nu \le 2-a_3 \le 1 + \tilde \nu$ translates into
$\tilde \nu \le \tilde a_3 \le 1 + \tilde \nu$.
Hence Proposition~\ref{prop:eigen1} can be applied to obtain that for fixed $\tilde a_3$ and $\tilde \nu$,
the spectral radius of $\Phi$ is minimized over $(\tilde a_1, \tilde a_2)$ by the choice
$$ {\tilde a_1}^* = ( \sqrt{ \tilde \nu + 2- \tilde a_3} - \sqrt{\tilde \nu} )^2 ,$$
which leads to $ 2- a_1^* = ( \sqrt{ \tilde \nu +a_3} -\sqrt{\tilde \nu})^2$ as stated in Proposition~\ref{prop:eigen1-b}.

\subsection{Proof of Corollary \ref{cor:eta-ab}}
To obtain HAMS-A, we set $\nu = a_1 = 1- \sqrt{1-\epsilon^2}$ in (\ref{eq:eigen1}). Then
\[
    a_3^* = (\sqrt{2} - \sqrt{1-\sqrt{1-\epsilon^2}})^2.
\]
In the SDE parameterization, we also have
\[
    a_3^* = \me^{-\frac{\eta_2\epsilon}{2}}(1+\sqrt{1-\epsilon^2}).
\]
Therefore
\begin{align*}
 &  \me^{-\frac{\eta_2\epsilon}{2}}(1+\sqrt{1-\epsilon^2}) = (\sqrt{2} - \sqrt{1-\sqrt{1-\epsilon^2}})^2 \\
  \Longrightarrow & \eta_2 = \frac{2}{\epsilon}\log\left[\frac{1 + \sqrt{1-\epsilon^2}}{(\sqrt{2} - \sqrt{1-\sqrt{1-\epsilon^2}})^2}\right].
\end{align*}
Similarly for HAMS-B, let $\tilde{\nu} = 2 - a_3 = 1 - \sqrt{1 - \epsilon^2}$ in (\ref{eq:eigen1-b}). Then
\[
    2 - a_1^* = (\sqrt{2} - \sqrt{1-\sqrt{1-\epsilon^2}})^2.
\]
According to SDE parameterization, we have
\[
 a_1^* = 2 - \me^{-\frac{\eta_1\epsilon}{2}}(1+\sqrt{1-\epsilon^2}) \Rightarrow 2 - a_1^* = \me^{-\frac{\eta_1\epsilon}{2}}(1+\sqrt{1-\epsilon^2}).
\]
Hence
\begin{align*}
    &  \me^{-\frac{\eta_1\epsilon}{2}}(1+\sqrt{1-\epsilon^2}) = (\sqrt{2} - \sqrt{1-\sqrt{1-\epsilon^2}})^2 \\
    \Longrightarrow & \eta_1 = \frac{2}{\epsilon}\log\left[\frac{1 + \sqrt{1-\epsilon^2}}{(\sqrt{2} - \sqrt{1-\sqrt{1-\epsilon^2}})^2}\right].
\end{align*}
Thus $\eta_1 = \eta_2$. Taking expansions we have
\[
 \eta_1 = \eta_2 = 2 + \frac{5}{12}\epsilon^2 + \mO(\epsilon^4).
\]

\subsection{Proofs of Propositions~\ref{prop:matching}--\ref{prop:matching3}}
\label{sec:det-matching}

For a univariate target density $\pi(x)$, the HAMS updates (\ref{eq:proposal1-mat})--(\ref{eq:proposal3-mat}) can be equivalently stated as follows:
\begin{align}
  & x^* = x_0 - a_1 \nabla U(x_0) + a_2 u_0 + Z_0^{(1)}, \label{eq:match1} \\
  & u^* = (a_3 + \phi a_2 - 1) u_0 + (\phi - \phi a_1 - a_2) \nabla U(x_0)
  - \phi \nabla U(x^*) + \phi Z_0^{(1)} + Z_0^{(2)}, \label{eq:match2} \\
  & Z_0^{(1)},Z_0^{(2)} \text{ are zero mean Gaussian with } \Var(Z_0^{(1)}) = 2a_1 - a_1^2 - a_2^2, \nonumber \\
  & \Var(Z_0^{(2)}) = 2a_3 - a_3^2 - a_2^2 \text{  and  } \Cov(Z_0^{(1)},Z_0^{(2)}) = 2a_2 - a_1a_2 - a_2a_3.\label{eq:match3}
\end{align}
The variance and covariance of $(x^*,u^*)$ in (\ref{eq:match1}) and (\ref{eq:match2}) given $(x_0,u_0)$ are then
\begin{align}
  &\Var(x^*) = \Var(Z_0^{(1)}), \quad \Cov(x^*, u^*) = \phi \Var(Z_0^{(1)}) + \Cov(Z_0^{(1)},Z_0^{(2)}) ,\nonumber \\
  &\Var(u^*) = \phi^2\Var(Z_0^{(1)}) + 2 \phi \Cov(Z_0^{(1)},Z_0^{(2)}) + \Var(Z_0^{(2)}).\label{eq:match4}
\end{align}
Throughout this section, $\Var(x^*),\Var(u^*)$ and $\Cov(x^*, u^*)$ are understood to be conditional on $(x_0,u_0)$.
For matching between HAMS and existing algorithms, we first identify $a_1,a_2,a_3,\phi$ to match the coefficients for $\nabla U(x_0)$ and $ u_0$ in (\ref{eq:match1}) and
those of $\nabla U(x_0)$, $\nabla U(x^*)$, and $u_0$ in (\ref{eq:match2})
and then we compare $\Var(x^*),\Var(u^*)$ and $\Cov(x^*, u^*)$.
Even with modification to existing algorithms, the matching is nontrivial because
there are five coefficients of $\nabla U(x_0)$, $\nabla U(x^*)$, and $u_0$, but only four tuning parameters $a_1,a_2,a_3,\phi$.

\vspace{.1in}
\textbf{Rescaled GJF.}\; The rescaled GJF update is
\begin{align}
  x^* & = x_0 - \frac{\epsilon^2}{2+\eta\epsilon}\nabla U(x_0) +
  \frac{\epsilon\sqrt{4-\epsilon^2}}{2+\eta\epsilon}  u_0  + \frac{\epsilon}{2+\eta\epsilon} W, ,\label{eq:GJF5} \\
  u^* & = \frac{2-\eta\epsilon}{2+\eta\epsilon} u_0 +
  \frac{\eta\epsilon^2 - 2\epsilon}{\sqrt{4 - \epsilon^2}(2+\eta\epsilon)} \nabla U(x_0) - \frac{\epsilon}{\sqrt{4-\epsilon^2}}\nabla U(x^*)
  +\frac{4}{\sqrt{4 - \epsilon^2}(2 + \eta \epsilon )} W, \label{eq:GJF6}
\end{align}
where $W \sim \mathcal N(0, 2\eta \epsilon)$. The coefficients of $\nabla U(x_0)$, $\nabla U(x^*)$, and $u_0$ between (\ref{eq:match1})--(\ref{eq:match2}) and (\ref{eq:GJF5})--(\ref{eq:GJF6})
are matched, remarkably, by setting
\begin{equation}
  \label{eq:GJF7}
  a_1 = \frac{\epsilon^2}{2+\eta\epsilon},\, a_2 = \frac{\epsilon\sqrt{4 - \epsilon^2}}{2 + \eta\epsilon},\,
  a_3 = \frac{4 - \epsilon^2}{2 + \eta\epsilon},\,\phi = \frac{\epsilon}{\sqrt{4 - \epsilon^2}}.
\end{equation}
Then using the HAMS formulas in (\ref{eq:match3})--(\ref{eq:match4}), we find
\begin{align*}
  & \Var(x^*) = \frac{2\eta \epsilon^3}{(2 + \eta\epsilon)^2},
  \quad \Var(u^*) = \frac{32\eta\epsilon}{(4-\epsilon^2)(2 + \eta\epsilon)^2},\\
  & \Cov(x^*, u^* ) = \frac{8\eta\epsilon^2}{(2+\alpha\epsilon)^2\sqrt{4 - \epsilon^2}},
\end{align*}
which match  exactly the variances and covariance of $(x^*,u^*)$ in (\ref{eq:GJF5})--(\ref{eq:GJF6}).
Moreover, the choices of $(a_1,a_2,a_3)$ in (\ref{eq:GJF7}) satisfy $a_1a_3 = a_2^2$, corresponding to a singular matrix $A$ in HAMS-A.\ Thus
the rescaled GJF is identical to HAMS-A except for the choice of $\phi$.
As stated by Proposition~\ref{prop:phichoice}, HAMS-A uses $\phi = a_2/(2-a_1)$, which by the values in (\ref{eq:GJF7}) leads to
\[
  \phi = \frac{\epsilon}{\sqrt{4 - \epsilon^2} + \frac{2\epsilon\eta}{\sqrt{4 - \epsilon^2}}}
  \neq \frac{\epsilon}{\sqrt{4 - \epsilon^2} }  .
\]
The difference in $\phi$ is $\mO(\epsilon^2)$.\par

\vspace{.1in}
\textbf{Rescaled BAOAB.}\; The rescaled BAOAB update is
\begin{align}
  x^* &= x_0 - \frac{\epsilon^2}{4}(1+e^{-\eta\epsilon})\nabla U(x_0) + \frac{\epsilon\sqrt{4-\epsilon^2}}{4}(1+e^{-\eta\epsilon}) u_0
  + \frac{\epsilon \sqrt{1-e^{-2\eta\epsilon}}}{2} W,  \label{eq:BAO5} \\
  u^* &= e^{-\eta\epsilon}u_0 - \frac{\epsilon e^{-\eta\epsilon}}{\sqrt{4-\epsilon^2}}\nabla U(x_0) - \frac{\epsilon}{\sqrt{4-\epsilon^2}}\nabla U(x^*)
  + 2\sqrt{\frac{1-e^{-2\eta\epsilon}}{4-\epsilon^2}} W, \label{eq:BAO6}
\end{align}
where $W\sim \mathcal N(0,1)$. The coefficients of $\nabla U(x_0)$, $\nabla U(x^*)$, and $u_0$  between (\ref{eq:match1})--(\ref{eq:match2}) and (\ref{eq:BAO5})--(\ref{eq:BAO6})
are matched, remarkably, by setting
\begin{equation}
  \label{eq:BAO7}
  a_1 = \frac{\epsilon^2}{4}(1+e^{-\eta\epsilon}),\, a_2 =  \frac{\epsilon\sqrt{4-\epsilon^2}}{4}(1+e^{-\eta\epsilon}) ,
  \, a_3 = (1+e^{-\eta\epsilon})\left(1-\frac{\epsilon^2}{4}\right), \phi = \frac{\epsilon}{\sqrt{4-\epsilon^2}}.
\end{equation}
Using the HAMS formulas in (\ref{eq:match3})--(\ref{eq:match4}), we find
\begin{align*}
  &\Var(x^*) = \frac{\epsilon^2(1 - e^{-2\eta\epsilon})}{4},
  \quad \Var(u^*) = \frac{4 - 4e^{-2\eta\epsilon}}{4 - \epsilon^2},
  \quad \Cov(x^*, u^* ) = \frac{(1 - e^{-2\eta\epsilon})\epsilon}{\sqrt{4 - \epsilon^2}},
\end{align*}
which agrees with (\ref{eq:BAO5}) and (\ref{eq:BAO6}). By (\ref{eq:BAO7}), the rescaled BAOAB corresponds to
a singular $A$ matrix and only differs from HAMS-A by $\mO(\epsilon^2)$ in $\phi$. The $\phi$ value implied by HAMS-A
is
\[
\phi = \frac{\epsilon}{\sqrt{4 - \epsilon^2} + \frac{4(e^{\epsilon\eta - 1})}{\sqrt{4 - \epsilon^2}(1 + e^{\eta\epsilon})}}
\neq \frac{\epsilon}{\sqrt{4-\epsilon^2}}.
\]
The difference in $\phi$ is $\mO(\epsilon^2)$.

\vspace{.1in}
\textbf{IL with full-step momentum.}\;
The IL update (\ref{eq:IL1})--(\ref{eq:IL2}) can be rewritten as
\begin{align}
    x^* & = x_0 - \epsilon^2\left(1-\frac{c}{2}\right) \nabla U(x_0) + \epsilon\left(1-\frac{c}{2}\right)u_{-\frac{1}{2}} + \frac{\epsilon}{2}\sqrt{c(2-c)} W,  \label{eq:Goga3}\\
    u_{\frac{1}{2}} & = (1-c) u_{-\frac{1}{2}} - \epsilon(1-c)\nabla U(x_0) + \sqrt{c(2-c)}W, \label{eq:Goga4}
  \end{align}
where $W \sim \mathcal N(0,1)$.
With the full-step momentum in Proposition~\ref{prop:matching},  (\ref{eq:Goga3})--(\ref{eq:Goga4}) leads to
\begin{align}
x^* & = x_0 - \frac{\epsilon^2}{2}\left(1-\frac{\tilde{c}}{2}\right) \nabla U(x_0) + \frac{\epsilon\sqrt{4-\epsilon^2}}{2}\left(1-\frac{\tilde{c}}{2}\right) u_0 + \frac{\epsilon}{2}\sqrt{\tilde{c}(2-\tilde{c})}W \label{eq:Goga5}.\\
u^* & = (1-\tilde{c}) u_0 - \frac{\epsilon(1-\tilde{c})}{\sqrt{4-\epsilon^2}}\nabla U(x_0)  - \frac{\epsilon}{\sqrt{4-\epsilon^2}} \nabla U(x^*) + \frac{2\sqrt{\tilde{c}(2-\tilde{c})}}{\sqrt{4-\epsilon^2}} W, \label{eq:Goga6}
\end{align}
where $W \sim \mathcal N(0,1)$.
By substituting $\tilde{c} = 1 - e^{-\eta\epsilon}$, we see that (\ref{eq:Goga5})-(\ref{eq:Goga6}) becomes identical to (\ref{eq:BAO5}) and (\ref{eq:BAO6}). Hence
IL with full-step momentum is equivalent to rescaled BAOAB and matches HAMS-A in the same manner.\par

\vspace{.1in}
\textbf{Modified BP.}\; The modified BP update is
\begin{align}
  x^* & = x_0 - (1 - \sqrt{1 - \epsilon^2})\nabla U(x_0) + \epsilon\sqrt{c} u_0 + \epsilon \sqrt{1-c} W_1,\label{eq:UDL3} \\
  u^* & = cu_0 - \frac{\sqrt{c}\epsilon}{1 + \sqrt{1 - \epsilon^2}}\nabla U(x_0) -\frac{\sqrt{c}\epsilon}{1 + \sqrt{1 - \epsilon^2}}\nabla U(x^*) + \sqrt{c(1-c)} W_1 + \sqrt{1-c} W_2, \label{eq:UDL4}
\end{align}
where $ W_1,W_2 \sim \mathcal N(0,1)$ independently.
The coefficients of $\nabla U(x_0)$, $\nabla U(x^*)$, and $u_0$  between (\ref{eq:match1})--(\ref{eq:match2}) and (\ref{eq:UDL3})--(\ref{eq:UDL4})
are matched, remarkably, by setting
\begin{align}
  a_1 = 1- \sqrt{1-\epsilon^2},\quad  a_2 = \epsilon\sqrt{c}, \quad
  a_3 =1 + c\sqrt{1-\epsilon^2}, \quad \phi = \frac{\sqrt{c}\epsilon}{1+\sqrt{1-\epsilon^2}}. \label{eq:UDL5}
\end{align}
Using the HAMS formulas in (\ref{eq:match3})--(\ref{eq:match4}), we find
\begin{align*}
  & \Var(x^*) = \epsilon^2(1 - c), \quad  \Var(u^*) = 1 - c^2, \quad \Cov(x^*, u^*) = \epsilon(1 - c)\sqrt{c},
\end{align*}
which match exactly the variances and covariance of $(x^*,u^*)$ in (\ref{eq:UDL3}) and (\ref{eq:UDL4}).
The $\phi$ choice in (\ref{eq:UDL5}) also agrees with the default value $\phi = a_2/(2-a_1)$ for HAMS.

\vspace{.1in}
\textbf{Modified VEC.}\;
The modified VEC update is
\begin{align}
  x^* & = x_0 - \frac{\epsilon^2}{2} \nabla U(x_0) + \frac{2\epsilon - \eta\epsilon^2}{2} u_0
  + \frac{\sqrt{2\eta}\epsilon^{3/2}}{2} W_1 + \frac{\sqrt{6\eta}\epsilon^{3/2}}{6} W_2 \label{eq:mVanden1} \\
  u^* & = \left(1-\eta\epsilon + \frac{\eta^2\epsilon^2}{2}\right) u_0 + \left(\frac{\eta\epsilon^2 - \epsilon}{2} - \frac{\epsilon^3}{4} \right) \nabla U(x_0)
  -\frac{\epsilon}{2}\nabla U(x^*) \nonumber \\
  & \quad  + \frac{\sqrt{2\eta\epsilon}}{2}(2-\eta\epsilon) W_1 - \frac{\sqrt{6}}{6}(\eta\epsilon)^{3/2}W_2,  \label{eq:mVanden2}
\end{align}
where $W_1, W_2 \sim \mathcal N(0,1)$, independently.
The coefficients of $\nabla U(x_0)$, $\nabla U(x^*)$, and $u_0$  between (\ref{eq:match1})--(\ref{eq:match2}) and (\ref{eq:mVanden1})--(\ref{eq:mVanden2})
are matched, remarkably, by setting
\begin{align}
  a_1 = \frac{\epsilon^2}{2},\quad a_2 = \epsilon - \frac{\eta\epsilon^2}{2},\quad a_3 = 2 - \frac{\epsilon}{4}(2-\eta\epsilon)(2\eta+\epsilon), \quad \phi = \frac{\epsilon}{2}. \label{eq:Vanden3}
\end{align}
Using the HAMS formulas in (\ref{eq:match3})--(\ref{eq:match4}), we find
\begin{align*}
  & \Var(x^*) = \eta\epsilon^3 - \frac{\eta^2\epsilon^4 + \epsilon^4}{4}, \\
  &  \Var(u^*) = 2\eta\epsilon - 2 \eta^2\epsilon^2 + \frac{\epsilon}{16}[8\eta\epsilon^2(1+2\eta^2) - 4\epsilon^4(1+\eta^2 + \eta^4) + 4\eta\epsilon^4 - \epsilon^5], \\
  & \Cov(x^*, u^*) =  \eta\epsilon^2 - \eta^2\epsilon^3 + \frac{\epsilon^2}{4}\eta(1+\eta^2) - \frac{1}{8}\epsilon^3,
\end{align*}
whereas according to (\ref{eq:mVanden1}) and (\ref{eq:mVanden2}) in the modified VEC update,
\begin{align*}
  & \Var(x^*) =  \frac{2\eta}{3}\epsilon^3,
  \quad \Var(u^*) =  2 \eta\epsilon - 2 \eta^2\epsilon^2 + \frac{2}{3} \eta^3 \epsilon^3,
  \quad \Cov(x^*, u^*) = \eta\epsilon^2 - \frac{2}{3} \eta^2\epsilon^3.
\end{align*}
The differences between the corresponding variances and covariances are $\mO(\epsilon^3)$.
The $\phi$ choice in (\ref{eq:Vanden3}) only differs from the default value $\phi = a_2/(2-a_1)$ by $\mO(\epsilon^2)$.

\vspace{.1in}
\textbf{Shifted HAMS.}\;
The update (\ref{eq:sHAMS1}) in shifted HAMS can be rewritten as
\begin{align}
  x^* & = x_0 - a_1\nabla U(\tilde{x}) + (ba_1 + a_2)u_0 + Z_0^{(1)}, \label{eq:sHAMS2} \\
  u^* & = (a_3 + ba_2 -1)u_0 - a_2 \nabla U(\tilde{x}) + Z_0^{(2)}. \label{eq:sHAMS3}.
\end{align}
Our matching approach using shifted HAMS is similar that using original HAMS.\ For each method, we first identify
 $a_1, a_2, a_3, b$ to match the coefficients for $\nabla U(\tilde x)$ and $u_0$ in (\ref{eq:sHAMS2})--(\ref{eq:sHAMS3})
and then we compare $\Var(x^*),\Var(u^*)$ and $\Cov(x^*, u^*)$.
While the first step is relatively straightforward with four coefficients of  $\nabla U(\tilde x)$ and $u_0$
and four tuning parameters $a_1,a_2,a_3,b$, the close matching in the variances and covariance remains nontrivial.

\vspace{.1in}
\textbf{Modified ABOBA.}\;
Consider modified ABOBA update with $b$ to be determined:
\begin{align}
  x^* & = x_0 - b\frac{\epsilon}{2}(1+e^{-\eta\epsilon}) \nabla U(\tilde{x})
  + b(1+e^{-\eta\epsilon}) u_0 + b \sqrt{1-e^{-2\eta\epsilon}}W,  \label{eq:mABO1} \\
  u^* & = e^{-\eta\epsilon}u_0 - \frac{\epsilon}{2}(1+e^{-\eta\epsilon})\nabla U(\tilde{x}) + \sqrt{1-e^{-2\eta\epsilon}} W, \label{eq:mABO2}
\end{align}
where $W \sim \mathcal N(0,1)$.
Matching the coefficients of $\nabla U(\tilde x)$ and $u_0$ between (\ref{eq:sHAMS2})--(\ref{eq:sHAMS3}) and (\ref{eq:mABO1})--(\ref{eq:mABO2}), we obtain
\[
a_1 = \frac{1}{2}(1+e^{-\eta\epsilon})(1-\sqrt{1-\epsilon^2}),\, a_2 = \frac{\epsilon}{2}(1+e^{-\eta\epsilon}),
\, a_3 = \frac{1}{2}(1+e^{-\eta\epsilon})(1+\sqrt{1-\epsilon^2}),\, b = \frac{1-\sqrt{1-\epsilon^2}}{\epsilon}.
\]
The variances and covariance implied by shifted HAMS are
\begin{align*}
  &  \Var(x^*)  =  \frac{1}{2}(1-\sqrt{1-\epsilon^2})(1-e^{-2\eta\epsilon}) = \frac{\eta\epsilon^3}{2} - \frac{\eta^2\epsilon^4}{2} + \left(\frac{\eta}{8} + \frac{\eta^2}{3}\right)\epsilon^5 + O(\epsilon^6), \\
  & \Var(u^*) = \frac{1}{2}(1+\sqrt{1-\epsilon^2})(1-e^{-2\eta\epsilon}) = 2\eta\epsilon - 2\eta^2\epsilon^2 + \left(\frac{4}{3}\eta^3 - \frac{\eta}{2}\right)\epsilon^3 + O(\epsilon^4), \\
  & \Cov(x^*, u^*) = \frac{\epsilon}{2}(1-e^{-2\eta\epsilon}) = \eta^2\epsilon^2 - \eta^2\epsilon^3 + \frac{2\eta^3\epsilon^4}{3} + O(\epsilon^5) ,
\end{align*}
whereas those computed from (\ref{eq:mABO1})--(\ref{eq:mABO2}) are
\begin{align*}
  &\Var(x^*) = \frac{(1-e^{-2\eta\epsilon})(1-\sqrt{1-\epsilon^2})^2 }{\epsilon^2}
  = \frac{\eta\epsilon^3}{2} - \frac{\eta^2\epsilon^4}{2} + \left(\frac{\eta}{4} + \frac{\eta^2}{3}\right)\epsilon^5 + O(\epsilon^6), \\
  &\Var(u^*) = 1 - e^{-2\eta\epsilon} = 2\eta\epsilon - 2\eta^2\epsilon^2 + \frac{4}{3}\eta^3\epsilon^3 + O(\epsilon^4), \\
  &\Cov(x^*, u^*) = \frac{1-\sqrt{1-\epsilon^2}}{\epsilon}(1-e^{-2\eta\epsilon}) =
  \eta^2\epsilon^2 - \eta^2\epsilon^3 +\left(\frac{2}{3}\eta^3 + \frac{\eta}{4}\right)\epsilon^4 + O(\epsilon^5).
\end{align*}
The differences between the corresponding variances and covariances are $\mO(\epsilon^3)$.

\vspace{.1in}
\textbf{Modified SPV.}\; Consider modified SPV update with $b$ to be determined:
\begin{align}
  x^* & = x_0 - \frac{b(1-e^{-\eta\epsilon})}{\eta} \nabla U(\tilde{x}) + b(1+e^{-\eta\epsilon}) u_0
  + b\sqrt{1-e^{-2\eta\epsilon}} W_1, \label{eq:mSPV1} \\
  u^* & = e^{-\eta\epsilon}u_0 - \frac{1-e^{-\eta\epsilon}}{\eta}\nabla U(\tilde{x}) + \sqrt{1 - e^{-2\eta\epsilon}} W_1. \label{eq:mSPV2}
\end{align}
where $W \sim \mathcal N(0,1)$.
Matching the coefficients of $\nabla U(\tilde x)$ and $u_0$ between (\ref{eq:sHAMS2})--(\ref{eq:sHAMS3}) and (\ref{eq:mSPV1})--(\ref{eq:mSPV2}), we obtain
\begin{align*}
  & a_1 = \frac{1}{2}\left\{1+e^{-\eta\epsilon} - \sqrt{(1+e^{-\eta\epsilon})^2 - \frac{4(1-e^{-\eta\epsilon})^2}{\eta^2}} \right\},\quad a_2 = \frac{1-e^{-\eta\epsilon}}{\eta}, \\
  & a_3 = \frac{1}{2}\left\{1+e^{-\eta\epsilon} + \sqrt{(1+e^{-\eta\epsilon})^2 - \frac{4(1-e^{-\eta\epsilon})^2}{\eta^2}} \right\}, \\
  & b = \frac{\eta\left(1 + e^{-\eta\epsilon} - \sqrt{(1+e^{-\eta\epsilon})^2 - \frac{4(1-e^{-\eta\epsilon})^2}{\eta^2}}\right)}{2(1-e^{-\eta\epsilon})}.
\end{align*}
The variances and covariance implied by shifted HAMS are
\begin{align*}
  & \Var(x^*) = \frac{\eta^3\epsilon^3}{2} - \frac{\eta^2\epsilon^4}{2} + \left(\frac{\eta}{8} + \frac{\eta^3}{4}\right)\epsilon^5 + O(\epsilon^6), \\
  & \Var(u^*) = 2\eta\epsilon - 2\eta^2\epsilon^2 + \left(\frac{4\eta^3}{3} - \frac{\eta}{2}\right)\epsilon^3 + O(\epsilon^4), \\
  & \Cov(x^*, u^*) = \frac{(1-e^{-\eta\epsilon})^2}{\eta} = \eta\epsilon^2 - \eta^2\epsilon^3 + \frac{7\eta^3\epsilon^4}{12} + O(\epsilon^5),
\end{align*}
whereas according to (\ref{eq:mSPV1}) and (\ref{eq:mSPV2}) in the modified SPV update,
\begin{align*}
  & \Var(x^*) = \frac{\eta^3\epsilon^3}{2}  - \frac{\eta^2\epsilon^4}{2} + \left(\frac{\eta}{4} + \frac{\eta^3}{4}\right)\epsilon^5 + O(\epsilon^6), \\
  & \Var(u^*) = 2\eta\epsilon - 2\eta^2\epsilon^2 + \frac{4\eta^3}{3} \epsilon^3 + O(\epsilon^4), \\
  & \Cov(x^*, u^*) =   \eta\epsilon^2 - \eta^2\epsilon^3 + \frac{7\eta^3\epsilon^4}{12} + \frac{\eta\epsilon^4}{4} + O(\epsilon^5).
\end{align*}
The differences between the corresponding variances and covariances are $\mO(\epsilon^3)$.

\vspace{.1in}
\textbf{Modified Mannella's leapfrog.}
Consider modified Mannella's leapfrog update with $b$ to be determined:
\begin{align}
  x^* & = x_0 - b \frac{2\epsilon}{2+\eta\epsilon}\nabla U(\tilde{x}) + b\frac{4}{2+\eta\epsilon} u_0
   + b \frac{2\sqrt{2\eta}}{2+\eta\epsilon} W, \label{eq:MLeap1} \\
  u^* & = \frac{2-\eta\epsilon}{2+\eta\epsilon} u_0 - \frac{2\epsilon}{2+\eta\epsilon}\nabla U(\tilde{x})
  + \frac{2\sqrt{2\eta}}{2+\eta\epsilon} W \label{eq:MLeap2},
\end{align}
where $W \sim \mathcal N(0, \epsilon)$.
The coefficients of $\nabla U(\tilde x)$ and $u_0$ between (\ref{eq:sHAMS2})--(\ref{eq:sHAMS3}) and (\ref{eq:MLeap1})--(\ref{eq:MLeap2}) are matched by setting
\begin{align*}
  & a_1 = \frac{2(1-\sqrt{1-\epsilon^2} )}{2+\eta\epsilon},\quad  a_2 = \frac{2\epsilon}{2+\eta\epsilon}, \\
  & a_3 = \frac{2(1+\sqrt{1-\epsilon^2} )}{2+\eta\epsilon} ,\quad
  b =  \frac{1-\sqrt{1-\epsilon^2}}{\epsilon}.
 \end{align*}
The variances and covariance given by shifted HAMS are
\begin{align*}
  & \Var(x^*)  = \frac{4\eta\epsilon(1-\sqrt{1-\epsilon^2})}{(2+\epsilon\eta)^2}  = \frac{\eta\epsilon^3}{2} - \frac{\eta^2\epsilon^4}{2} + \frac{\eta}{8}(1+3\eta^2)\epsilon^5 + O(\epsilon^6), \\
  & \Var(u^*)  =  \frac{4\eta\epsilon(1+\sqrt{1-\epsilon^2})}{(2+\epsilon\eta)^2} = 2\eta\epsilon - 2\eta^2\epsilon^2  + \frac{\eta}{2}(3\eta^2 - 1)\epsilon^3 + O(\epsilon^4), \\
  & \Cov(x^*, u^*) =  \frac{4\eta\epsilon^2}{(2+\epsilon\eta)^2} = \eta\epsilon^2 - \eta^2\epsilon^3 + \frac{3\eta^3\epsilon^4}{4} + O(\epsilon^5) .
\end{align*}
The variances and covariance given by (\ref{eq:MLeap1}) and (\ref{eq:MLeap2}) are
\begin{align*}
  & \Var(x^*) = \frac{\eta\epsilon^3}{2} - \frac{\eta^2\epsilon^4}{2} + \frac{\eta}{8}(2+3\eta^2)\epsilon^5 + O(\epsilon^6), \\
  & \Var(u^*) =  2\eta\epsilon - 2\eta^2\epsilon^2  + \frac{3\eta^3\epsilon^3}{2} + O(\epsilon^4), \\
  & \Cov(x^*, u^*) = \eta\epsilon^2 - \eta^2\epsilon^3 + \frac{\eta}{4}(1 + 3\eta^2)\epsilon^4 + O(\epsilon^5).
\end{align*}
The differences between the corresponding variances and covariances are $\mO(\epsilon^3)$.

\section{Details and additional results for numerical experiments} \label{sec:sim-detail}

\subsection{Double well} \label{sec:sim-detail-DW}
In the double well experiment, there is no preconditioning. In Algorithms~\ref{alg:HAMSG}--\ref{alg:BP} we take $L \equiv I$.
For HAMS-A, we set $a = 1 - \sqrt{1-\epsilon^2}, b = e^{-\eta_2\epsilon/2}(1+\sqrt{1 - \epsilon^2})$ in Algorithm~\ref{alg:HAMSAB}. For HAMS-B, we first set $\tilde{a} = e^{-\eta_1\epsilon/2}(1+\sqrt{1-\epsilon^2}), \tilde{b} =1 - \sqrt{1-\epsilon^2} $ and then use the transformation
$a = 2 - \tilde{a}, b = (\tilde a \tilde b)/(2 - \tilde a)$ in Algorithm~\ref{alg:HAMSAB}. For HAMS-1/2/3, we set $c_1 = e^{-k\epsilon^2/2}$ for $k = 1,2,3$ respectively and $c_2 = e^{-\eta_2\epsilon/2}$, define $a_1,a_2,a_3$ by (\ref{eq:sdeparam1}), and then apply Algorithm~\ref{alg:HAMSG}. For BAOAB, ABOBA and BP we set $c = e^{-\eta\epsilon}$ in Algorithms~\ref{alg:BAO}--\ref{alg:BP} with $\eta$ set to $\eta_2$ in HAMS.\par

\vspace{.1in}
\textbf{Equivalence of temperatures.} We show that $T_{C1} = T_{C2} = T_K$. By the definition (\ref{eq:temp1}), $x$ and $u$ are independent and $u\sim \mathcal N(0,T)$.
Thus $T_K =\EE[u^2] = T$. For the configurational temperatures, we use Stein's identity \citepsup{Ley2017}, which states that
for any differentiable function $f(x)$ such that $f(x) \pi(x) \to 0$ as $x \to \pm\infty$,
\begin{align*}
\EE \left[ f(x) \frac{\nabla\pi(x)}{\pi(x)} \right] = - \EE [ f^\prime (x) ].
\end{align*}
Notice that $\nabla U(x) = - T \frac{\nabla\pi(x)}{\pi(x)}$.
Taking $f(x)=x$ shows that
\begin{align*}
 & T_{C1} = \EE[x\cdot\nabla U(x)] = -T\EE\left[ x \frac{\nabla \pi(x)}{\pi(x)}\right] = T.
\end{align*}
Moreover, taking $f(x) = \nabla U(x)$ shows that
\begin{align*}
  &  \EE[(\nabla U(x))^2] =  -T\EE\left[ \nabla U(x) \frac{\nabla \pi(x)}{\pi(x)}\right] = T \EE [ \nabla^2 U(x) ].
\end{align*}
and hence
\begin{align*}
  & T_{C2} = \frac{\EE[(\nabla U(x))^2]}{\EE[\nabla^2 U(x)]} =T.
\end{align*}

\begin{figure}[!tph]  
  \centering
  \begin{subfigure}{.49\linewidth}
      \centering
      \includegraphics[width = \textwidth]{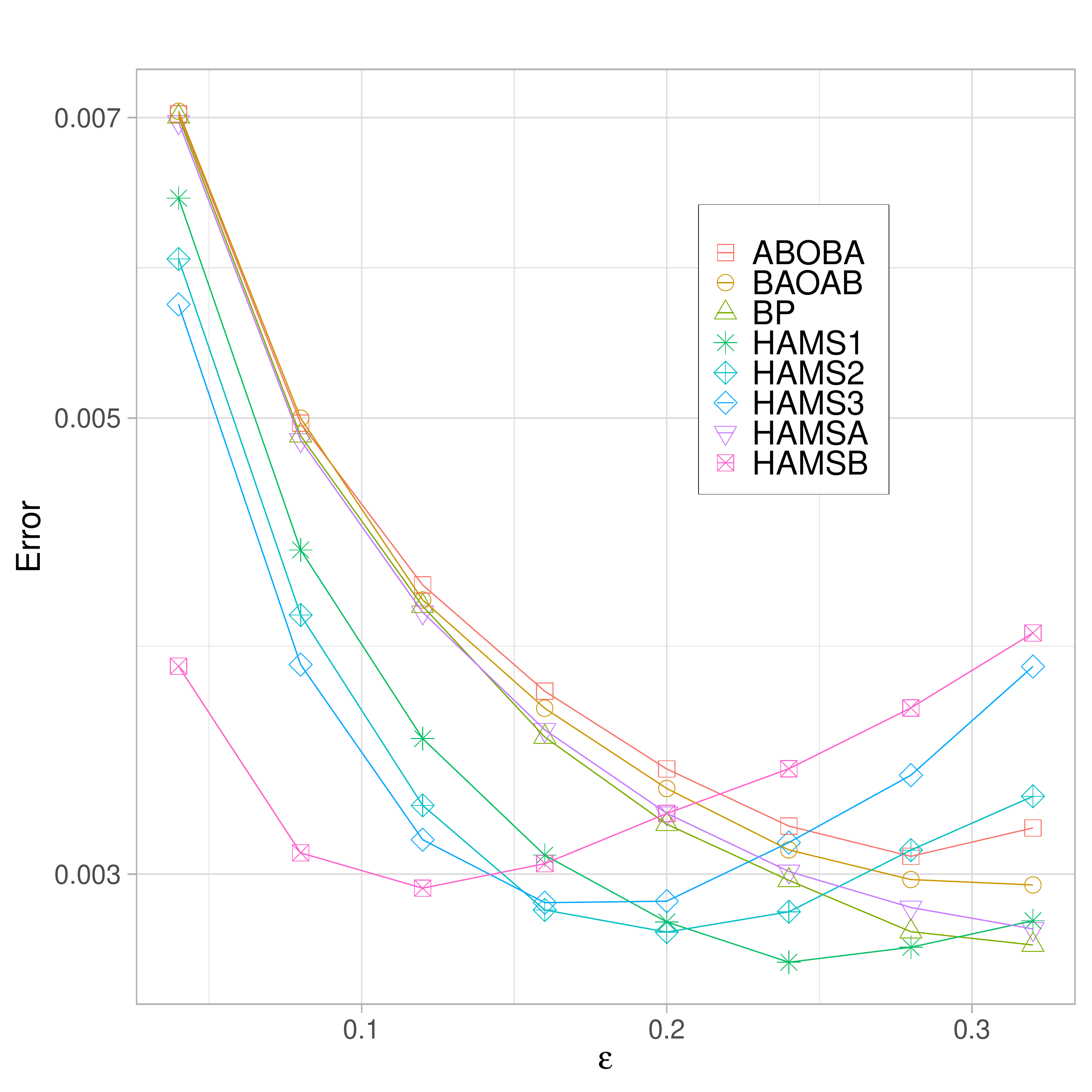}
      \caption{}
  \end{subfigure}
      \hfill
  \begin{subfigure}{.49\linewidth}
      \centering
      \includegraphics[width = \textwidth]{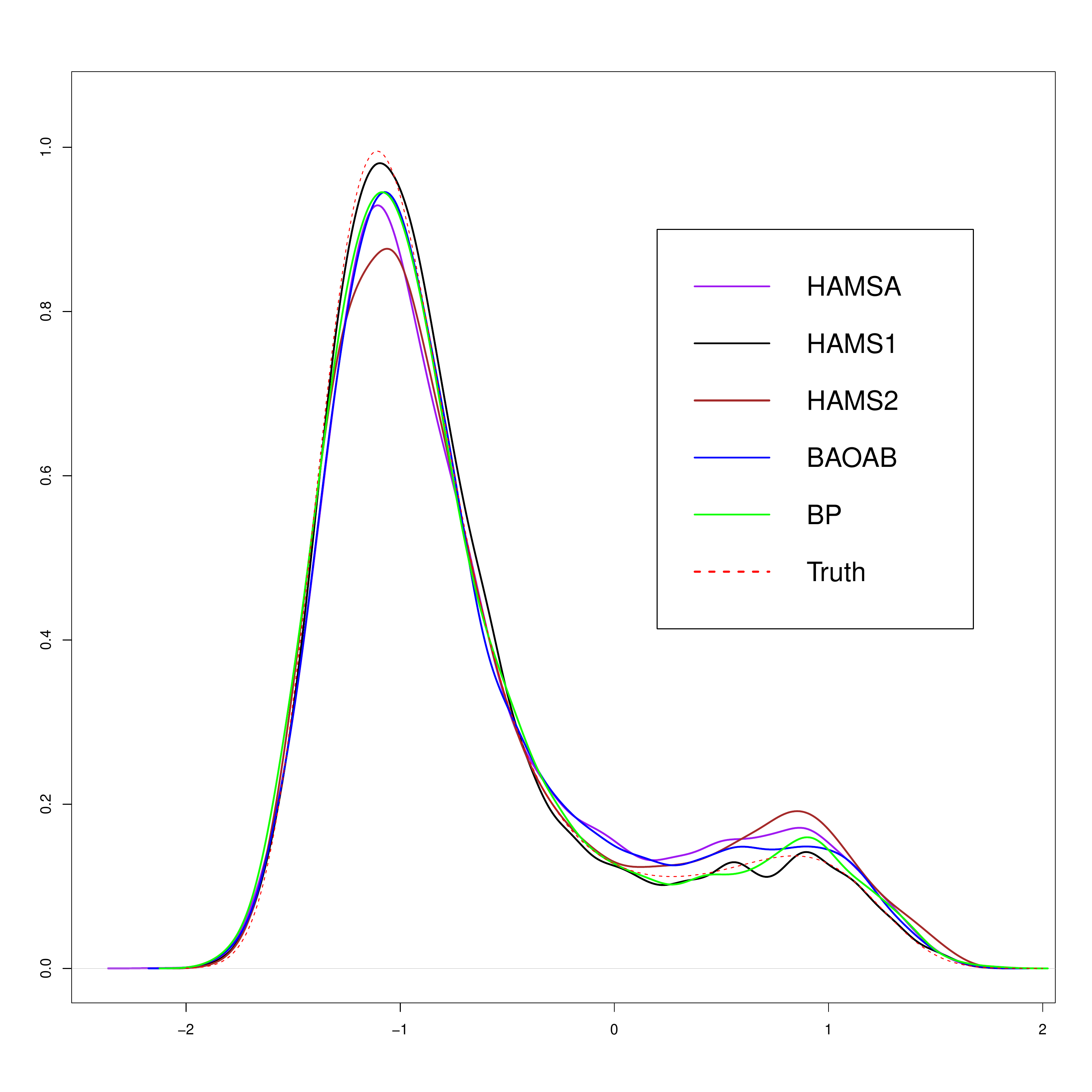}
      \caption{}
  \end{subfigure}
 \vspace{-.2in}  
  \RawCaption{\caption{(Left) Root mean squared errors in densities based on $3000$ repeated experiments. (Right) A sample density plot from an individual run with $\epsilon = 0.24$.}
  \label{fig:DWDens}}
\end{figure}

\vspace{.1in}
\textbf{Density estimation.} In addition to the temperatures, we report the performance of density estimation.
Following  \cite{Leimkuhler2013}, the error in density estimation is computed by dividing the interval $[-2,2]$ into $16$
equal sized bins and compare the empirical density with the truth obtained from numerical integration. The left panel of Figure~\ref{fig:DWDens} shows the errors on log scale.
Comparison between the methods is consistent with that in temperature estimation.
When $\epsilon$ is small, HAMS-$k$ has better performance as $k$ increases (including HAMS-A with $k = 0$). Moreover, HAMS-A, BAOAB, ABOBA and BP have comparable performance for small $\epsilon$.
The error of HAMS-B is the smallest for $\epsilon \leq 0.12$ but quickly increases afterwards. The overall best performance is achieved by HAMS-1 at $\epsilon = 0.24$.

Figure~\ref{fig:DWDens} also shows density plots, produced using \texttt{density()} in R,
from an individual run when $\epsilon = 0.24$. This confirms that HAMS-1 best tracks the shape of the true density.

\vspace{.1in}
\textbf{Error calculation.} We describe how the errors are calculated in Figure~\ref{fig:DWError} and Figure~\ref{fig:DWDens}.
Let $\{x_{ij}\}$ and $\{u_{ij}\}$ be the samples collected, indexed by $i = 1,...,N$ draws and $j = 1,...,J$ repetitions. Let $T$ be the true temperature.
Then
\begin{align*}
  & T_{C1}^{(j)} = \frac{1}{N}\sum_{i = 1}^N x_{ij}\cdot \nabla U(x_{ij}),\quad \text{Error in }T_{C1} = \sqrt{\frac{1}{J}\sum_{j = 1}^J(T_{C1}^{(j)} - T)^2} \, , \\
  & T_{C2}^{(j)} = \frac{\sum_{i = 1}^N(\nabla U(x_{ij}))^2}{\sum_{i = 1}^N \nabla^2 U(x_{ij})},\quad \text{Error in }T_{C2} = \sqrt{\frac{1}{J}\sum_{j = 1}^J(T_{C2}^{(j)} - T)^2} \, , \\
  & T_{K}^{(j)} = \frac{1}{N}\sum_{i = 1}^N u_{ij}^2, \qquad \text{Error in }T_{K} = \sqrt{\frac{1}{J}\sum_{j = 1}^J(T_{K}^{(j)} - T)^2} \, .
\end{align*}
For errors in densities, let $\omega^*_k$ be the area under the true density curve in the $k$th bin, $k=1,\ldots, M(=16)$. Then
\[
  \omega_{kj} = \frac{1}{N} \sum_{i = 1}^N \mathbb I \{x_{ij} \in k\text{th bin}\}, \, e_j = \frac{1}{M}\sum_{k = 1}^M |\omega_{kj} - \omega^*_k|,
  \, \text{Error in density } = \sqrt{\frac{1}{J}\sum_{j = 1}^J e_j^2 }.
\]
The normalizing constants needed to evaluate $\omega^*_k$s are obtained from \texttt{Mathematica}.

\subsection{Stochastic volatility} \label{sec:sim-detail-SV}

\textbf{Detailed expressions.}
We apply preconditioning and use the default parameter choices implied by Proposition~\ref{prop:eigen1} and \ref{prop:eigen1-b}. For HAMS-A, we set $a = 1 - \sqrt{1-\epsilon^2}, b = (\sqrt{2} - \sqrt{a})^2$ in Algorithm~\ref{alg:HAMSAB}.
For HAMS-B, we first set $\tilde{b} =1 - \sqrt{1-\epsilon^2}, \tilde{a} = (\sqrt{2} - \sqrt{\tilde{b}})^2 $ and then use the transformation
$a = 2 - \tilde{a}, b = (\tilde a \tilde b)/(2 - \tilde a)$ in Algorithm~\ref{alg:HAMSAB}. For HAMS-1/2/3, we set  for $k = 1,2,3$
\[
  c_1 = \exp(-k\epsilon^2/2), \quad c_2 = \max\left[\frac{1}{2}, \left\{ \frac{3 - \sqrt{1-\epsilon^2}}{1+\sqrt{1-\epsilon^2}} - 2\sqrt{2}\epsilon(1+\sqrt{1-\epsilon^2})^{-3/2} \right\} c_1\right],
\]
define $a_1,a_2,a_3$ by (\ref{eq:sdeparam1}), and then apply Algorithm~\ref{alg:HAMSG}. Here we restrict $c_2\geq \frac{1}{2}$ to ensure the condition $a_1\leq 1 + \nu$ in Proposition~\ref{prop:eigen1}
is satisfied. For BAOAB, ABOBA and BP, we set $c$ as below to be consistent with the choice in HAMS-A,
\[
  c = \left\{ \frac{3 - \sqrt{1-\epsilon^2}}{1+\sqrt{1-\epsilon^2}} - 2\sqrt{2}\epsilon(1+\sqrt{1-\epsilon^2})^{-3/2} \right\}.
\]

Next, we give the detailed expressions for stochastic volatility model. Using (\ref{eq:sveq1}) and (\ref{eq:sveq2}), we derive the conditional density of $\bx$ given $\by$,
\begin{align*}
  p(\bx|\by,\beta,\sigma,\varphi) & \propto \underbrace{ p(x_1)\prod_{t=2}^T p(x_t|x_{t-1},\varphi,\sigma) }_{\mathcal N(\bx|\bo,C)} \cdot
  \overbrace{\prod_{t = 1}^T p(y_t| x_t,\beta)}^{\mathcal N(\by| \bo, \beta^2 \exp(\bx))} \\
  & \propto \exp\left\{-\frac{1}{2}\bx^\T C^{-1}\bx\right\} \beta^{-T} \exp\left\{-\frac{1}{2}\sum_{t=1}^T(x_t + \beta^{-2}y_t^2\exp(-x_t)) \right\}.
\end{align*}
where the $(i,j)$ position of $C$ matrix is given by $\varphi^{|i - j|}\sigma^2/(1 - \varphi^2)$. After suppressing the dependency on $\by$ and the parameters,
the negative log density or potential function of $\bx$ is
\[
  U(\bx) = \frac{1}{2} \bx^\T C^{-1}\bx + \frac{1}{2}\sum_{t=1}^T(x_t + \beta^{-2}y_t^2\exp(-x_t)).
\]
The gradient is
\[
  \nabla U(\bx) = C^{-1}\bx  - \frac{1}{2}\beta^{-2}\by\exp(-\bx) + \frac{1}{2}\mathbf{1},
\]
where $\mathbf{1}$ is a vector of all $1$'s. The Hessian is
\begin{align*}
\nabla^2 U(\bx) & = C^{-1} + \frac{1}{2} \diag[ \beta^{-2}\by^2\exp(-\bx) ].
\end{align*}
The square $\by^2$ is taken component-wise. Using the relation between $\by$ and $\bx$,
the diagonal elements in the second term can be expressed as
\[
\beta^{-2}\by^2\exp(-\bx)  = \beta^{-2}\exp(-\bx) \bz^2 \beta^2 \exp(\bx) = \bz^2.
\]
Hence $\EE [ \nabla^2 U(\bx)] = C^{-1} + \frac{1}{2}I$. For preconditioning, we use $\hat{\Sigma} = \left(C^{-1} + \frac{1}{2}I \right)^{-1}$.

\begin{figure}[!tph]  
  \centering
  \begin{subfigure}{.49\linewidth}
      \centering
      \includegraphics[width = \textwidth]{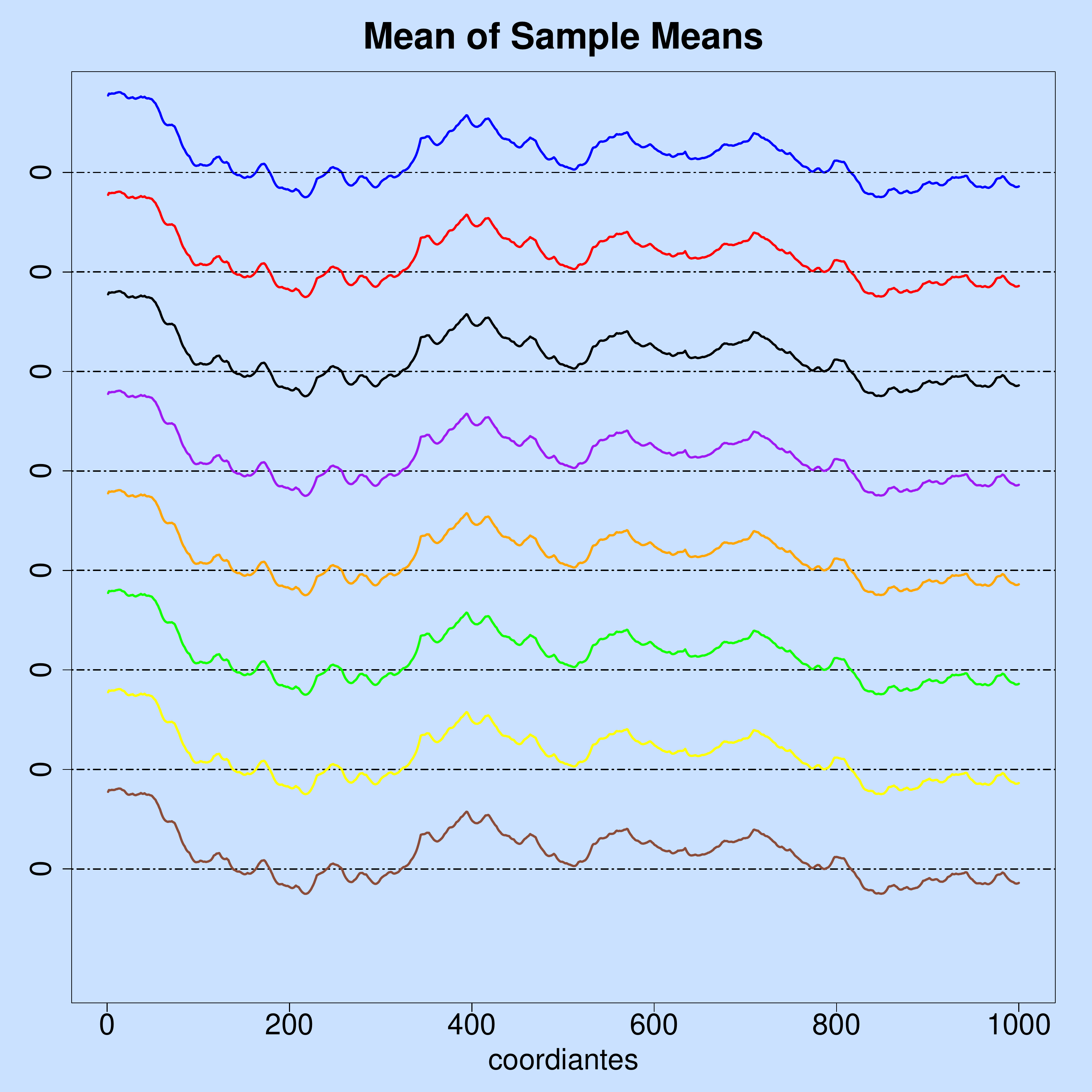}
      \caption{}
  \end{subfigure}
      \hfill
  \begin{subfigure}{.49\linewidth}
      \centering
      \includegraphics[width = \textwidth]{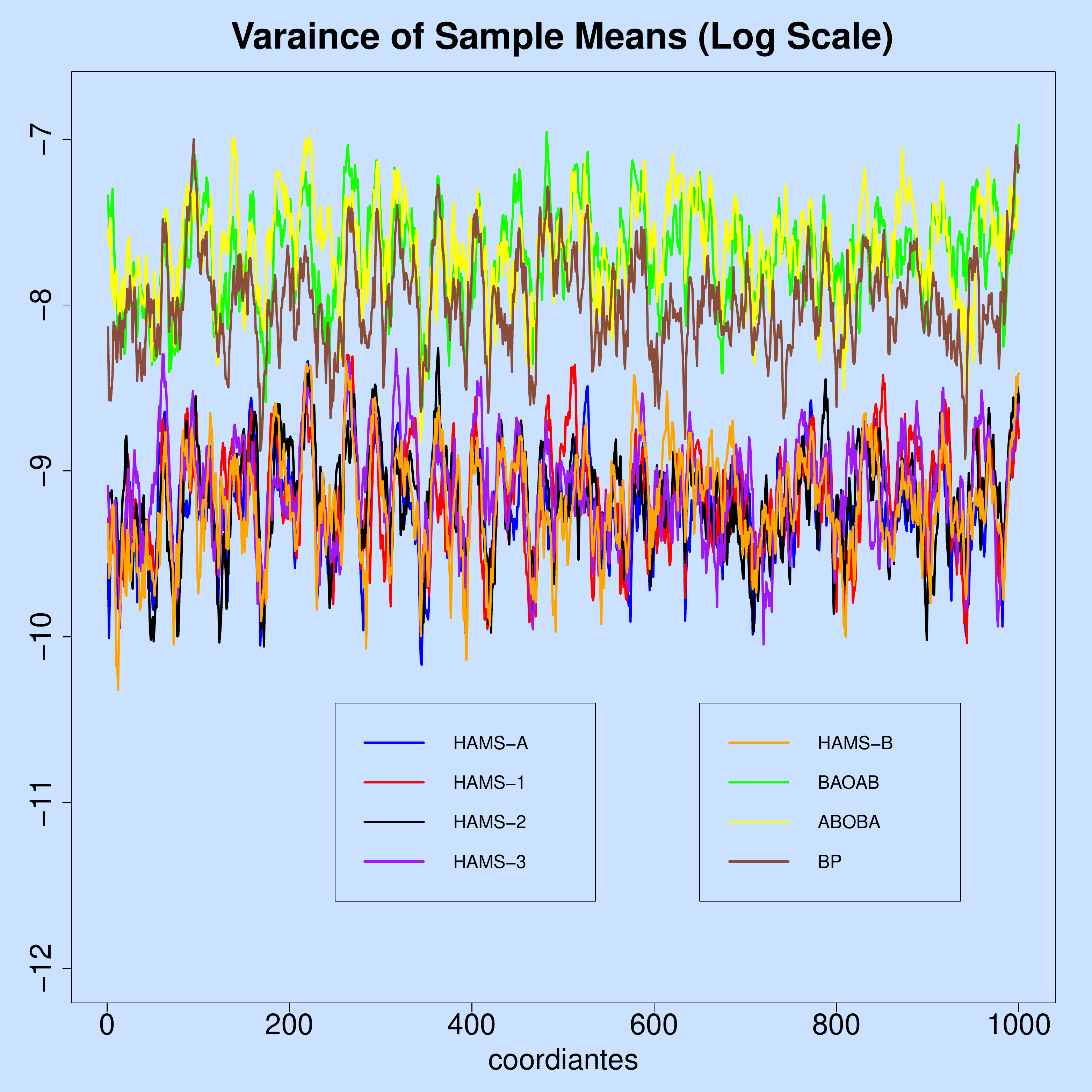}
      \caption{}
  \end{subfigure}
 \vspace{-.2in}  
  \RawCaption{  \caption{Averages of sample means (shifted) and variances of sample means (log-scale)
  of all latent variables over $50$ repetitions for sampling latent variables in the stochastic volatility model.
   \label{fig:svmeanvar}}}
\end{figure}

\vspace{.1in}
\textbf{Means and variance of sample means.} In the left panel of Figure~\ref{fig:svmeanvar}, we plot the
average of sample means of all the 1000 latent coordinates across $50$ repetitions. The curves are shifted by a constant to be aligned with zero. We see that
the overall shapes of average sample means are similar among all methods. However, the variances of sample means as shown in the right panel of Figure~\ref{fig:svmeanvar}
indicate that HAMS methods are more consistent than the remaining methods.\par

\subsection{Log-Gaussian Cox model} \label{sec:sim-detail-Cox}

\textbf{Detailed expressions.}
The tuning parameters $(a,b)$, $(c_1,c_2)$, and $c$ are set in the same manner as in Section~\ref{sec:sim-detail-SV}. We now provide the necessary expressions
for the log-Gaussian Cox model. From (\ref{eq:coxeq1}), we obtain the potential function of $\bx$ (given $\by$),
\[
  U(\bx) = \frac{1}{2}x^\T C^{-1}x - \sum_{i,j}(y_{ij}x_{ij} - n^{-1}\exp(x_{ij}+\mu)).
\]
The gradient is
\[
  \nabla U(\bx) = C^{-1}\bx - \by + n^{-1}\exp(\bx + \mu).
\]
The Hessian is
\[
  \nabla^2 U(\bx) = C^{-1} + n^{-1}\diag[\bx + \mu].
\]
Because marginally $\bx \sim \mathcal N(0,C)$, we take the expectation $\EE[\nabla^2 U(\bx)] = C^{-1} + n^{-1}\diag[\sigma^2/2 + \mu]$. We
take $\hat{\Sigma} = \left\{ C^{-1} + n^{-1}\diag[\sigma^2/2 + \mu] \right\}^{-1}$ for preconditioning.

\begin{figure}[t]  
  \centering
  \begin{subfigure}{.49\linewidth}
      \centering
      \includegraphics[width = \textwidth]{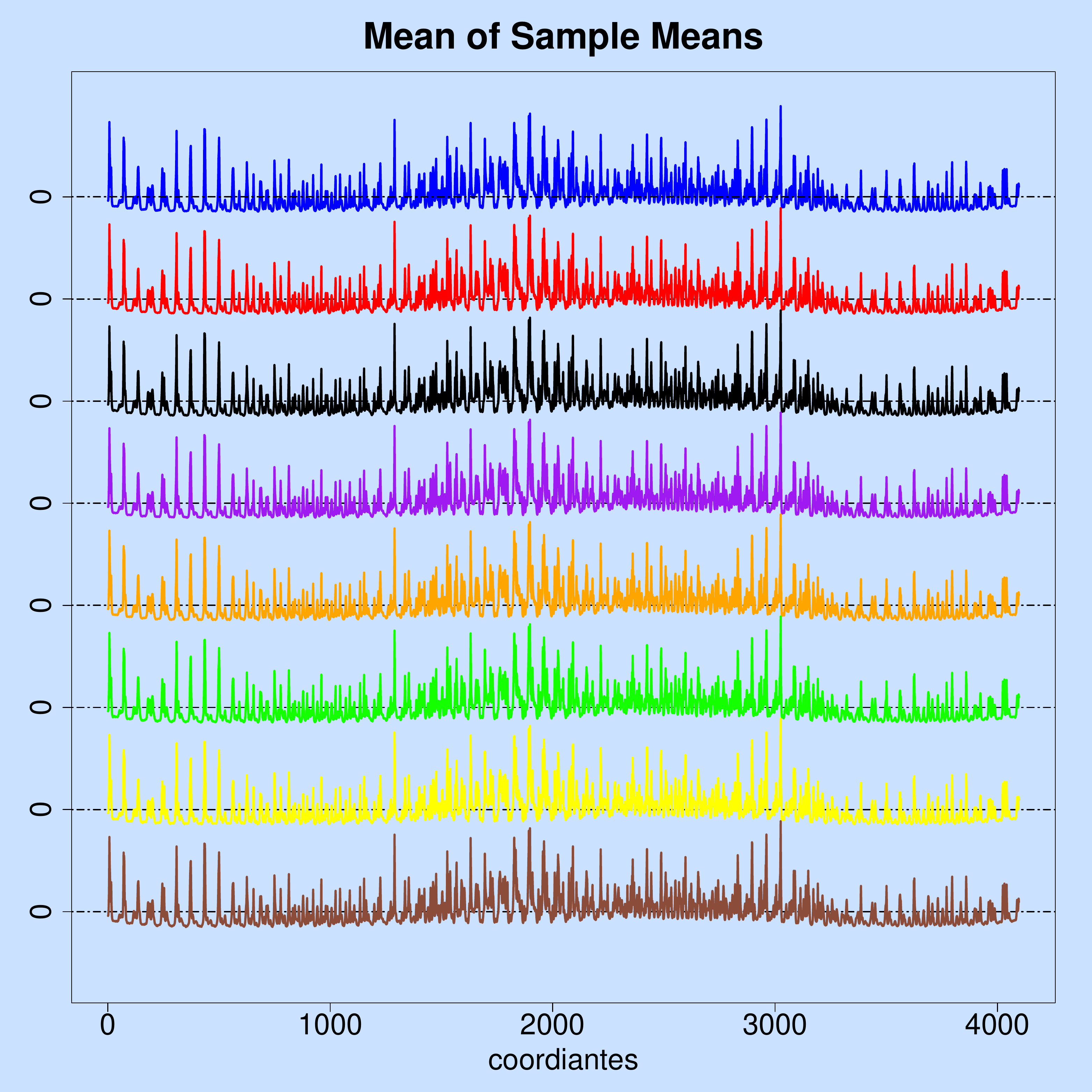}
      \caption{}
  \end{subfigure}
      \hfill
  \begin{subfigure}{.49\linewidth}
      \centering
      \includegraphics[width = \textwidth]{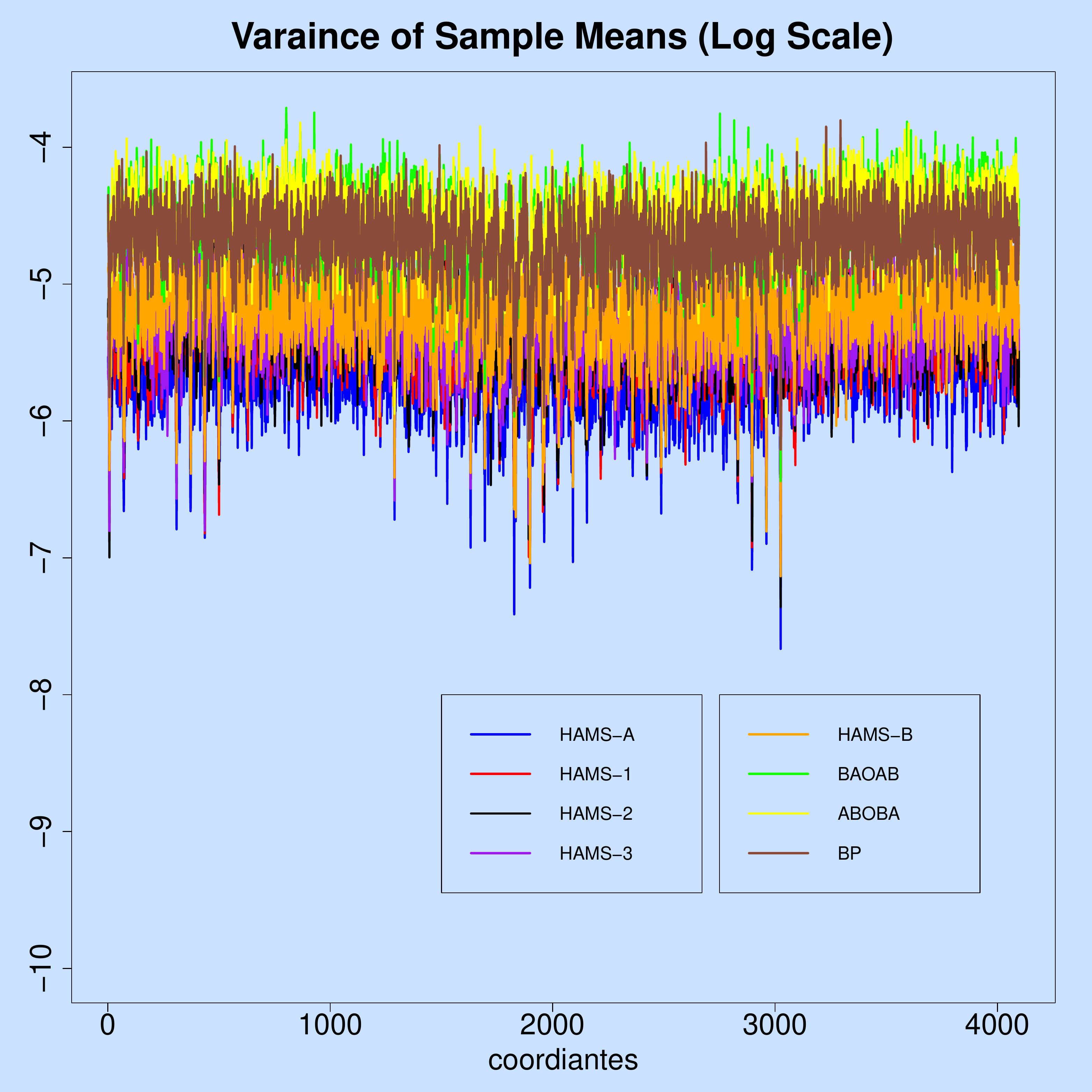}
      \caption{}
  \end{subfigure}
 \vspace{-.2in}  
  \RawCaption{  \caption{Averages of sample means (shifted) and variances of sample means (log-scale)
  of all latent variables over $50$ repetitions for sampling latent variables in the log-Gaussian Cox model.
   \label{fig:coxmeanvar}}}
\end{figure}

\vspace{.1in}
\textbf{Means and variance of sample means.}
The average sample means for the Cox model line up consistently among different methods as shown by Figure~\ref{fig:coxmeanvar}. But the variances of sample means
can be differentiated in the right panel of Figure~\ref{fig:coxmeanvar}, in spite of the overlaps.
We see that HAMS-A has the smallest variation across repeated simulations, followed by HAMS-1, HAMS-2, HAMS-3, HAMS-B, BP, ABOBA and BAOAB.

\bibliographystylesup{apalike}
\bibliographysup{hams2ref}

\end{document}